\documentclass[letterpaper]{article}
\usepackage{fullpage}
\usepackage{ktmacros}
\usepackage[backend=bibtex,maxnames=30,maxbibnames=30,maxcitenames=30,maxalphanames=30,giveninits=true,doi=false,url=true]{biblatex}
\newcommand*{\citet}[1]{\AtNextCite{\AtEachCitekey{\defcounter{maxnames}{2}}} \textcite{#1}}

\newcommand*{\citep}[1]{\cite{#1}}

\bibliography{refs}

\newcommand{\agg}{\ensuremath{\mathcal{A}}}
\newcommand{\minbatch}{\ensuremath{B}}
\newcommand{\Dec}{\ensuremath{Dec}}
\newcommand{\sa}{\ensuremath{\mathcal{SA}}}
\newcommand{\SAA}{\ensuremath{\textsc{SA}_2}}
\newcommand{\gf}{\mathbb{F}}
\newcommand{\rappork}{{\sc Rappor}$_K$}

\newif\ifbiblatex
\biblatextrue

\newcommand{\Description}[1]{}

\title{Samplable Anonymous Aggregation for Private Federated Data Analysis}
\author{Kunal Talwar\thanks{Apple.} \footnote{Corresponding Author. \texttt{ktalwar@apple.com}}
\and Shan Wang\footnotemark[1]
\and Audra McMillan\footnotemark[1]
\and Vojta Jina\thanks{Research performed while at Apple.}
\and Vitaly Feldman\footnotemark[1]
\and	Pansy Bansal\footnotemark[1]
\and	Bailey Basile\footnotemark[1]
\and	Aine Cahill\footnotemark[1]
\and	Yi Sheng Chan\footnotemark[1]
\and	Mike Chatzidakis\footnotemark[1]
\and	Junye Chen\footnotemark[1]
\and	Oliver Chick\footnotemark[1]
\and	Mona Chitnis\footnotemark[1]
\and	Suman Ganta\footnotemark[1]
\and	Yusuf Goren\footnotemark[1]
\and	Filip Granqvist\footnotemark[1]
\and	Kristine Guo\footnotemark[1]
\and	Frederic Jacobs\footnotemark[1]
\and	Omid Javidbakht\footnotemark[1]
\and	Albert Liu\footnotemark[1]
\and	Richard Low\footnotemark[1]
\and	Dan Mascenik\footnotemark[1]
\and	Steve Myers\footnotemark[1]
\and	David Park\footnotemark[1]
\and	Wonhee Park\footnotemark[1]
\and	Gianni Parsa\footnotemark[1]
\and	Tommy Pauly\footnotemark[1]
\and	Christian Priebe\footnotemark[1]
\and	Rehan Rishi\footnotemark[1]
\and	Guy N. Rothblum\footnotemark[1]
\and	Michael Scaria\footnotemark[3]
\and	Linmao Song\footnotemark[1]
\and	Congzheng Song\footnotemark[1]
\and	Karl Tarbe\footnotemark[1]
\and	Sebastian Vogt\footnotemark[1]
\and	Luke Winstrom\footnotemark[1]
\and	Shundong Zhou\footnotemark[1]
}

\begin{document}
\maketitle

\begin{abstract}
  We revisit the problem of designing scalable protocols for private statistics and private federated learning when each device holds its private data. Locally differentially private algorithms require little trust but are (provably) limited in their utility. Centrally differentially private algorithms can allow significantly better utility but require a trusted curator. This gap has led to significant interest in the design and implementation of simple cryptographic primitives, that can allow central-like utility guarantees without having to trust a central server.

  Our first contribution is to propose a new primitive that allows for efficient implementation of several commonly used algorithms, and allows for privacy accounting that is close to that in the central setting without requiring the strong trust assumptions it entails.  {\em Shuffling} and {\em aggregation} primitives that have been proposed in earlier works enable this for some algorithms, but have significant limitations as primitives.
  We propose a {\em Samplable Anonymous Aggregation} primitive, which computes an aggregate over a random subset of the inputs and show that it leads to better privacy-utility trade-offs for various fundamental tasks.
  Secondly, we propose a system architecture that implements this primitive and perform a security analysis of the proposed system. Our design combines additive secret-sharing  with anonymization and authentication infrastructures.
 \end{abstract}

 \section{Introduction}
 Learning aggregate population trends can allow for better data-driven decisions, and machine learning can improve user experience. Compared to learning from public curated datasets, learning from the user population offers several benefits. As an example, a next-word prediction model trained on words typed by users (a) can better fit the actual distribution of language used on devices, (b) can adapt faster to shifts in distribution, and (c) can more faithfully represent smaller sub-populations that may not be well-represented in curated datasets. At the same time, training such models may involve sensitive user data. This tension has led to increasing interest in cross-device federated learning  and analytics\footnote{Federated data analytics (such as collecting statistics and telemetry), and federated learning (such as training ML models) are the two most common forms of federated data analyses. While everything we say in this work applies to both kinds of analyses, for brevity we will often use federated learning when we mean federated learning and federated analytics.}, and such systems have been deployed at scale~\cite{ErlingssonPK14, Apple2017, DingKY17, ENPA:2021, ZhangRXZZK23}.

 In federated learning, the data of a user stays on their device, and only model updates are sent to a server. This reduction in the data sent to the servers may however not be sufficient to ensure user privacy.
 A long line of work~\cite{ShokriSSS17,NasrSH19, CarliniLEKS19, Feldman20, FeldmanZ20, CTWJ+21, CIJL+23}, has shown that trained models can memorize training data. Differential privacy~\cite{Dwork:2006, DworkR14} can address the privacy concerns by provably preventing memorization of data that is unique to one or a few users. Private gradient-based learning algorithms add sufficient noise to gradients (aggregated over a batch) to mask individual contributions. In the setting of a trusted curator, one can use centrally differentially private algorithms for data analysis (e.g. ~\cite{DworkR14,Kasiviswanathan:2008}) and machine learning (e.g. ~\cite{ChaudhuriMS11,BassilyST14,DLDP}). It is natural to aim to combine differential privacy with federated learning, and indeed private federated learning (PFL) and analytics systems aim to compute differentially private (DP) models and statistics in a federated setup.

 \begin{figure*}
       \centering
   \includegraphics[width=0.3\linewidth]{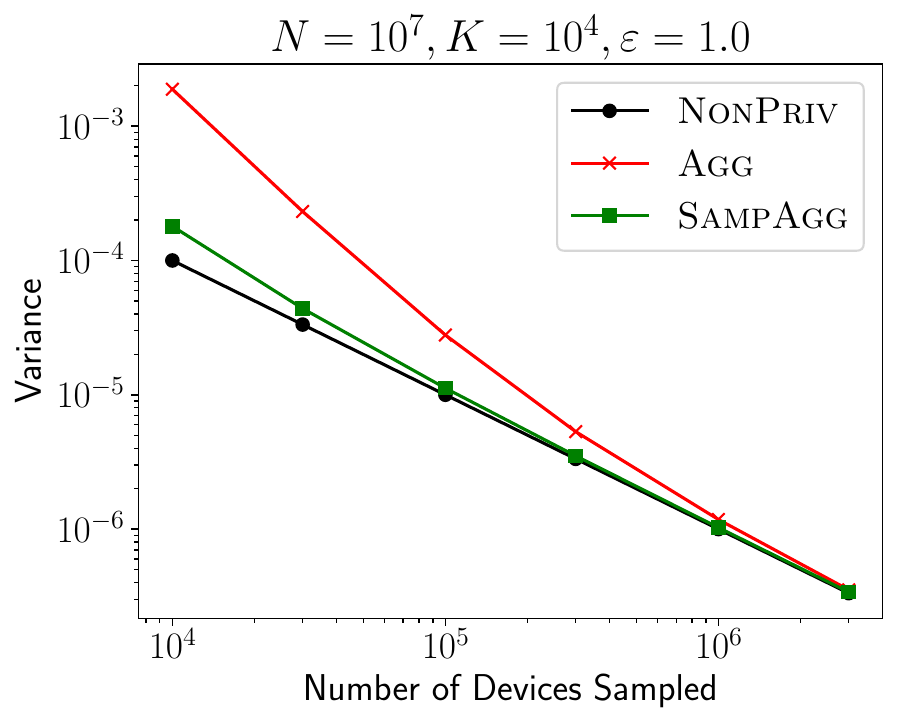}
   \includegraphics[width=0.3\linewidth]{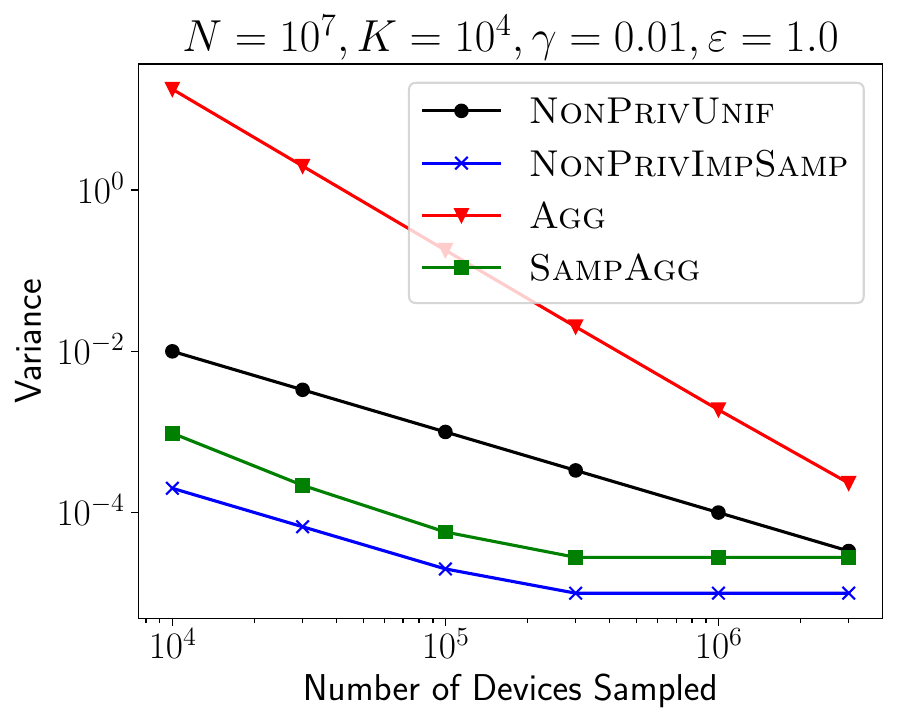}
   \includegraphics[width=0.3\linewidth]{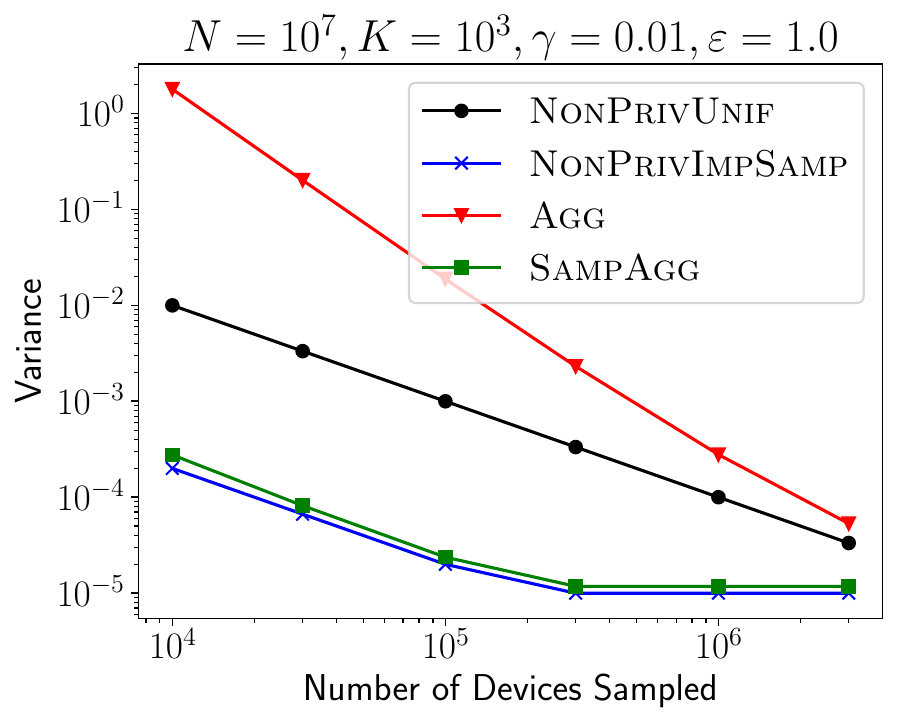}
   \caption{Expected Squared Error of a non-private baseline ({\sc NonPriv}), Aggregation model ({\sc Agg}) and Samplable Aggregation ({\sc SampAgg}) on a histogram task, for a histogram on uniform ground truth distribution (Left), and Skewed ground truth (Middle, Right), with data-dependent sampling (also showing non-private Importance Sampling). The population size ($N$), the support size ($K$), the privacy parameter ($\varepsilon$), and the fraction of non-default values ($\gamma$) are shown. We plot the Variance (expected squared error) of the algorithm agains the number of devices sampled $M$. More details and additional plots are in \cref{sec:experiments}.}\label{fig:histogram}
   \Description{Plots showing the benefits of Samplable Aggregation on Histograms.}
 \end{figure*}

 Natural implementations of private learning algorithms in the federated setup, however, require placing a fair amount of trust in the server enabling this computation. Indeed a long and influential line or research~\cite{MelisSCS19,FowlGCGG22,GeipingBDM20, WangSZSWQ19, YinMVAKM21,ZhaoMB20,ZhuLH19,WenGFGG22,PasquiniFA22,BoenischDSSSP23a, BoenischDSSSP23b} shows that an adversarial server can compromise individual privacy in some commonly-used setups.
 The desire to reduce the trust assumptions in PFL has led to significant interest in the design and implementation of specific primitives, that are useful enough for a variety of DP algorithms, yet simple enough for secure and scalable implementations. Secure aggregation~\cite{ODOpaper, Bonawitz17,unlynx,crypteps,honeycrisp,orchard} and shuffling~\cite{Bittau17,ErlingssonFMRTT19,CheuSUZZ19} are two prominent examples of such primitives.

 Our first contribution is to propose a new, more powerful primitive that we call Samplable Anonymous Aggregation. We show that this primitive is powerful enough to allow for strong differential privacy guarantees for some of the commonly-used algorithms for private learning and statistics. For private histograms, it allows us to get privacy-utility trade-offs that are close to those achievable centrally with a trusted curator. For private learning, it allows us to get privacy bounds close to those achieved by the moments accountant~\cite{DLDP} in the central setting; indeed it is the first federated learning primitive that is able to achieve this.

 As our second contribution, we describe the architecture of a scalable system that can implement this primitive under suitable trust assumptions. Our proposed architecture builds on the Prio system~\cite{Corrigan-GibbsB17}, where clients use additive secret-sharing to share their data to two or more servers. As long as at least one of the servers is honest, Prio ensures relevant security properties. Our proposed architecture combines Prio with additional elements including client-side sampling, and anonymization and device authentication infrastructures.  We critically analyze the security aspects of the architecture and propose additional hardenings that can make it difficult to violate the trust assumptions of the system. We now give additional details of our contributions.

 \medskip\noindent{\bf New Primitive:} Our proposed primitive of {\em Samplable Anonymous Aggregation} specifies the ideal functionality that we desire. It being an Aggregation means that we compute an aggregate of (some) client contributions. Samplability means that this aggregate can be computed over a random subset of the clients. Anonymity here means that the subset we aggregate over stays hidden: the adversary only learns the sum over a random subset, but does not know whether any specific client is in this random subset. We defer the precise definition to \cref{sec:sa2}.

 Compared to the shuffling primitive~\cite{Bittau17}, aggregation can allow for better privacy guarantees, particularly for private federated learning where there is a large asymptotic gap~\cite{CheuSUZZ19, BalleBGN19a} in the privacy-utility tradeoffs of aggregation vs. shuffling for the problem of mean estimation. Compared to the previously studied primitive of Aggregation~\cite{ODOpaper,Bonawitz17, BellBGLR20}, samplability allows for better privacy analysis due to {\em Privacy Amplification by Sampling}~\cite{Kasiviswanathan:2008}.
 These privacy gains are significant. As we show in~\cref{exa:pfl-sampling}, for typical parameters used in cross-device settings, privacy amplification by sampling can help improve the differential privacy guarantees from $\eps \approx 100$ to $\eps \approx 1$.

 When communication constraints are present, the utility benefit of samplable anonymous aggregation is non-trivial even for the simpler tasks. In \cref{fig:histogram}, we consider the task of learning a distribution over an alphabet of size $K$ (for example, learning the usage distribution of language tokens, or learning the distribution of error codes for an application). While the population $N$ may often be large, we may want to learn a good approximation of this distribution by collecting reports from a smaller number $M$ of samples, while preserving differential privacy. In the left Figure, we plot the expected squared error against the number of samples without privacy ({\sc NonPriv}), when using an aggregator or a shuffler alone ({\sc Agg}), and when using a samplable aggregator ({\sc SampAgg}). Since sampling improves privacy, a samplable aggregator allows us to add lower variance noise, compared to an aggregator, for the same final privacy parameter. Thus we see that {\sc SampAgg} improves on {\sc Agg}. This gap becomes even larger when we have a very skewed distribution where a non-default value occurs only with a small probability $\gamma$ and we care about the distribution of the non-default values (e.g. an error code occurs for a user only if there is an error). In this case, absent privacy concerns, one would send a report only when there is a non-default value or more broadly, use some kind of importance sampling. With a normal aggregator, importance sampling is not private as it reveals that a user has a non-default value. With a samplable anonymous aggregator, we can additionally benefit from this importance sampling and thus see larger gains for some parameters (\cref{fig:histogram} middle and right).

 \medskip\noindent{\bf Implementation:} Our proposed implementation of this primitive (detailed in \cref{sec:architecture}) uses a split-trust model, where we rely on two or more non-colluding servers. Effectively, the clients secret share their data to two servers, which can then run a multi-party computation to evaluate the aggregate.
 Our protocol needs a single message from each client, thus avoiding the complications due to churn and dropout that are inherent in multi-round protocols.

 In addition to non-colluding servers, our approach relies on an anonymization infrastructure that ensures that the messages received at the server cannot be linked to their sender. Using such an infrastructure allows us to get {\em strong anonymity} that ensures that even the {\em set} of clients contributing to a collection remains hidden from the server. To ensure that contributions are coming from genuine clients, we propose using an anonymous rate-limited device authentication method such as Privacy Pass~\cite{ietf-privacypass-rate-limit-tokens-01}.

 A large-scale deployment of a PFL system may be subject to attacks from adversarial client devices, that may seek to poison the computation, or attempt to manipulate the learnt statistics or the learnt model to behave in a certain way~\cite{BagdasaryanVHES20,BhagojiCMC19, SunKSM19,BaruchBG19,WangSRVASLP20,CheuSU19}.
 A robust system should limit the potential impact of an adversarial client.
 Prio allows devices to send zero-knowledge proofs of {\em validity} that can be verified by the server to guarantee this robustness to poisoning. Recent advances in efficient proofs in this framework~\cite{BBCGI19,BonehBCGI21, BellGGKMRS22, CastroP22,Talwar22,AddankiGJOP22,RatheeSWP22,BonehBCGI23}, enable this robustness with small computational and communication overheads.

 \medskip\noindent{\bf Organization:} The rest of the paper is organized as follows. We start with some differential privacy background in \cref{sec:prelims}, and present the formal definition of our primitive in \cref{sec:sa2}. We describe the proposed architecture for implementing the primitive, along with its security analysis and additional hardenings in~\cref{sec:architecture}. \cref{sec:discussion} discusses properties of the primitive and additional enhancements.
 We discuss additional related work in \cref{sec:related} and conclude with some open directions in \cref{sec:conclusions}.

 \section{Preliminaries}
 \label{sec:prelims}

 Differential privacy~\cite{Dwork:2006} defines a way to measure the individual privacy loss of a randomized algorithm.
 Intuitively, an algorithm is differentially private if the distribution of outputs is reasonably stable with respect to a single individual changing their data. There are several ways to formalize the notion of closeness of distributions that are commonly used to define variants of differential privacy (DP). The most popular are the hockey-stick divergence, used to define $(\eps,\delta)$-differential privacy, and the R\'enyi divergence, used to R\'enyi differential privacy (RDP).

 \begin{definition}[Hockey-stick divergence]
 The {\em hockey-stick} divergence between two random variables $P$ and $Q$ is defined by:
 \begin{displaymath}
   \dalpha{e^{\eps}}(P\|Q) = \int \max\{0, P(x)-e^{\eps} Q(x)\} dx,
 \end{displaymath}
   where we use the notation $P$ and $Q$ to refer to both the random variables and their probability density functions. We say that $P$ and $Q$ are $(\eps, \delta)$-indistinguishable if $\max\{\dalpha{e^{\eps}}(P\|Q), \dalpha{e^{\eps}}(Q\|P)\}\le\delta$. \end{definition}

 \begin{definition}[R\'enyi divergence]
 For two random variables $P$ and $Q$, the \emph{R\'enyi divergence} of $P$ and $Q$ of order $\alpha>1$ is \[D^{\alpha}(P\|Q) = \frac{1}{\alpha-1}\ln \E_{x\sim Q}\left[\left(\frac{P(x)}{Q(x)}\right)^{\alpha} \right] .\]
 \end{definition}

 We say that two databases are neighboring if one can be obtained from the other by the addition or deletion of the data of a single individual. Another notion is based on replacement, where two databases are neighboring if one can be obtained from the other by replacing the data of one individual. As these are related up to a factor of two, we will not dwell on this distinction and state results for one or the other when appropriate. We can now define differentially private algorithms with respect to both the hockey-stick divergence\footnote{The definition in~\cite{Dwork:2006} is stated slightly differently, but is easily seen to be equivalent~\cite{BartheO13}}, and R\'enyi divergence.

 \begin{definition}[Central DP]\cite{Dwork:2006}\label{centralDP}
 An algorithm $\cA:\cD^n\to\output$ is $(\eps, \delta)$-\emph{differentially private} if for all neighboring databases $X$ and $X'$, $\cA(X)$ and $\cA(X')$ are $(\eps, \delta)$-indistinguishable. If $\delta=0$, we refer to $\cA$ as satisfying $\eps$-DP, and say the algorithm satisfies \emph{pure differential privacy}. If $\delta>0$, then we say $\cA$ satisfies \emph{approximate differential privacy}.

 An algorithm $\cA:\cD^n\to\output$ is $(\alpha, \rho(\alpha))$-\emph{R\'enyi differentially private} if for all neighboring databases $X$ and $X'$, $D^{\alpha}(\cA(X)\| \cA(X'))\le\rho(\alpha)$.
 \end{definition}

 The two variants of differential privacy are related. Any algorithm that satisfies pure DP also satisfies RDP. If an algorithm satisfies RDP, then it also satisfies approximate DP. We give the formal conversion statements in the following theorem.

 \begin{theorem}\cite{Bun:2016, mironov2017renyi, CanonneKS20}
 If $\cA:\cD^n\to\output$ is $\eps$-DP, then for any $\alpha>1$, $\cA$ satisfies $(\alpha, \frac{1}{2}\eps^2\alpha)$-RDP. Conversely, for any $\delta\in(0,1]$, if $\cA$ is $(\alpha, \eps)$-RDP then it is $(\eps~+~\frac{\log(1/\delta)+(\alpha-1)\log(1-1/\alpha)-\log(\alpha)}{\alpha-1}, \delta)$-DP.
 \end{theorem}

 Differential privacy satisfies two key properties that allow for private building blocks to be combined to design more complex algorithms. First, it satisfies {\em post-processing}: applying an arbitrary data-independent map does not increase the privacy cost. Second, it degrades smoothly as we adaptively compose multiple private algorithms. {\em Composition} theorems allow us to analyze the privacy cost of a sequence of private algorithms. There are different versions of these composition results, and other DP results in this work, for different DP variants. While they are typically similar in their aymptotics, their numerical constants can differ. We will aim to present the simplest variant for each DP result in this section. For practical privacy accounting, one would often use other variants of these results.

 \begin{theorem}[RDP Composition \cite{mironov2017renyi}\label{compositionofRDP}]
 Let $\cA_i:\cD^n\to\output_1$ for $i=1,2,\ldots,T$ be a (possibly adaptive) sequence of randomized algorithms such that $\cA_i$ is $(\alpha, \rho_i(\alpha))$-RDP. Then the algorithm defined by $X\mapsto(Y_1,Y_2,\ldots,Y_T)$ where $Y_i=\cA_i(X)$ is $(\alpha, \sum_{i=1}^T \rho_i(\alpha))$-RDP.
 \end{theorem}

 One can easily generalize Theorem~\ref{compositionofRDP} to adaptively compose any number of RDP algorithms. A similar theorem exists for composing approximate DP algorithms.
 \begin{theorem}[Advanced Composition \cite{DRV10}\label{composition}]
 Let $\cA_i:\cD^n\to\output_i$, for $i=1, 2,\ldots, T$ be a (possibly adaptive) sequence of randomized algorithms such that $\cA_i$ is $(\eps, \delta)$-DP. Then for any $\delta'>0$, the algorithm defined by  $X\mapsto(Y_1,Y_2,\ldots,Y_T)$ where $Y_i=\cA_i(X)$  is $(\eps', T\delta+\delta')$-DP where $\eps' = \eps\sqrt{2T\ln \frac 1 \delta'} + T\eps(e^{\eps}-1)$.
 \end{theorem}
 Note that in the expression for $\eps'$ above, the second term is $O(\eps^2 T)$ whenever $\eps < 1$. For $\eps < \frac{1}{\sqrt{T}}$, the first term will dominate, and thus the composition of $T$ $(\eps,\delta)$-DP mechanisms will be $(O(\eps\sqrt{T}), T\delta)$-DP for small $\eps$. Asymptotically, and in practice, the tightest privacy bounds on many practical DP algorithms are derived by composing the RDP guarantees and converting the final RDP guarantee to approximate DP.

 \subsection{Privacy Amplification by Sampling}\label{subsampling}

 Sampling is a common primitive used in data collection systems for a variety of reasons, including balancing communication and computation costs associated with large scale data analysis.
 Choosing to include only a sample of the available users in the differentially private computation provides additional privacy, by giving each user plausible deniability about whether their data was included in the computation or not. This can be formalized for a variety of sampling methods including sampling-with-replacement (selecting a random subset of a fixed size) and Poisson sampling (selecting each data point independently with a certain probability) in both approximate and R\'enyi DP~\cite{Kasiviswanathan:2008, Ullman:2017, Balle:2018, WangLF16b, Wang:2021}. Below we state the privacy amplification by sampling bound for Poisson sampling for approximate differential privacy.

 \begin{theorem}\label{samplingWOR}\cite{Kasiviswanathan:2008}
 Let $\cA$ be an $(\eps,\delta)$-DP mechanism, and $\gamma\in[0,1]$ be a sampling rate. Given a data set $X$, let $\cA_s$ be the algorithm that first samples a subset $X'$ by including each data point in $X$ with probability $\gamma$, then outputs $\cA(X')$. Then $\cA_s$ is $(\eps',\delta')$-DP for $\eps'=\log(1+\gamma(e^\eps -1))$ and
 $\delta'=\gamma\delta$.
 \end{theorem}

 Note that when $\eps$ is small, $\eps'\approx \gamma\eps$, so the privacy guarantee is, roughly, scaled by the sampling rate. Similar bounds exist for sampling-with-replacement, and R\'enyi DP, and numerical bounds for these for specific mechanisms of interest are implemented in various libraries\cite{DLDP,Mironov2019RnyiDP,tfprivacy, yousefpour2021opacus}.
 \begin{example}[Gaussian and Subsampled Gaussian Mechanisms]
 Let us consider a particular differentially private algorithm, and the impact that Poisson sampling has on this mechanism.
 The \emph{Gaussian mechanism} is a DP algorithm that hides the impact of a single individual by adding Gaussian noise to the output of the computation. Given $f:\cD^n\to\mathbb{R}^d$ a deterministic function, define the sensitivity of $f$ to be $\Delta_f = \max_{X,X' \mbox{ neighbors}}\|f(X)-f(X')\|_2$ to be the maximum amount that the output of the function can vary on any pair of neighboring datasets $X$ and $X'$. The Gaussian mechanism on input $X$ outputs $f(X)+\mathcal{N}(0, \Delta_f^2 \sigma^2 \mathbb{I}_d)$.
 It satisfies $(\eps,\delta)$-DP for $\sigma \geq \sqrt{2 \ln 1.25/\delta} / \eps$, and $(\alpha, \alpha/2\sigma^2)$-RDP.

 Further, given $q \in [0,1]$, the algorithm that first samples a subset $X'$ by including each user with probability $q$, then outputs $f(X')+\mathcal{N}(0, \Delta_f^2 \sigma^2 \mathbb{I}_d)$ is $(2q\eps, 2\delta)$-DP and $(\alpha, q^2 \alpha / 2\sigma^2)$-RDP as long as $\alpha$ is suitable bounded~\cite{DLDP,Mironov2019RnyiDP}.

 These analytic bounds can be improved by numerical methods. We can use bounds from the Analytic Gaussian Mechanism~\cite{Balle2018ImprovingTG}, and the moments accountant~\cite{DLDP}, and will use those in subsequent examples in this work.
 \end{example}

 \subsection{Local Differential Privacy}

 Definition~\ref{centralDP} is commonly referred to as the \emph{central model} of differential privacy, where the curator (or server) has direct access to the data. Local DP algorithms constrain the whole communication from each client itself to be differentially private. In the fully adaptive case, they can communicate with the server in an arbitrary order with adaptive interaction. Formally, a protocol satisfies local $(\eps,\delta)$-DP if the transcripts of the interaction on any two pairs of neighbouring datasets are $(\eps,\delta)$-indistinguishable. When there is only a single round of interaction, the condition on transcripts reduces to each user interacting with the server using a mechanism that is differentially private with respect to that user's data. We call such mechanisms for the local reports of a user \emph{local randomizers.}

 \begin{definition}[Local randomizer]\label{localrandomizer}
 An algorithm $\lr\colon \cD\to \output$ is an $(\eps, \delta)$-DP \emph{local randomizer}  if there exists a reference distribution $\mathcal{D}_{\mbox{ref}}$ such that
 for all data points $x\in \cD$, $\lr(x)$ and $\mathcal{D}_{\mbox{ref}}$ are $(\eps, \delta)$-indistinguishable.
 \end{definition}

 Formally, an adaptive single pass $(\eps,\delta)$-DP local protocol can be described by a sequence of local randomizers $\Aldp[i]:\out[1]\times\cdots\times\out[i-1]\times\mathcal{D}\to\out[i]$ for $i\in[n]$, where $\cD$ is the data domain, $\out[i]$ is the range space of $\Aldp[i]$ and the $i$-th user returns $z_i=\Aldp[i](z_{1:i-1}, x_i)$. We require that the local randomizer $\Aldp[i](z_{1:i-1}, \cdot)$ be $(\eps,\delta)$-DP for all values of auxiliary inputs $z_{1:i-1}\in\out[1]\times\cdots\times\out[i-1]$. We can also define local randomizers with respect to R\'enyi DP, however throughout this paper, we will focus on $(\epsilon, \delta)$-DP local randomizers.

 \begin{example}[{\sc Rappor}$_{K}$.] Given an input $x \in [K]$, the mechanism \rappork\ first transforms $x$ to a $K$-dimensional $0$-$1$ vector $e_x$, which is $1$ in the $x$th coordinate and $0$ elsewhere. It then flips each bit of $e_x$, independently, with probability $\frac{1}{e^{\eps_0}+1}$ to get a $0$-$1$ vector. This mechanism satisfies $(\eps_0,0)$-DP in the local model~\cite{ErlingssonPK14}.
 \end{example}

 The aggregate and shuffle models of differential privacy are distributed models of computation where the clients hold their own data and interact with the server in a federated model through an implementation of the appropriate primitive. We will discuss the impact of these primitives in the following section.

 \paragraph{Notation.} $\sum$ always refers to a commutative associative addition operation, such as summation over reals or over a finite field. We abuse notation and use sets to mean {\em multisets}.

 \section{Aggregation Systems and their Implications}
 \label{sec:sa2}
 Towards our final definition, we start by defining a simpler primitive: an aggregation functionality. While such a functionality has been defined in previous work, we define a version that is slightly more general.

 Our aggregator $\agg$  will be parameterized by a {\em cohort size} \minbatch. This parameter specifies the minimum number of contributions that must be aggregated before release. This would typically correspond to the batch size in a private federated learning setting, and we will often use batch and cohort interchangeably. The aggregator is also parameterized by a decoding function \Dec, that prescribes how messages from users are to be transformed before aggregating. It receives messages from a set of users, decodes each of them using \Dec, and outputs the sum of the decodings, as long as at least \minbatch\ messages are received.
 Formally, it receives $m_1,\ldots,m_k$ for some $k$, and returns

 \begin{align*}
     \agg_{\minbatch}^{\Dec}(\{m_1, \ldots, m_k\}) = \left\{\begin{array}{ll} \sum_{i=1}^k Dec(m_i) & \mbox{if } k \geq \minbatch\\ \bot &\mbox{otherwise} \end{array} \right.
 \end{align*}
 When \minbatch\ and \Dec\ are obvious from context, we will often just write \agg\ to mean $\agg_{\minbatch}^{\Dec}$.
 A single-round protocol in the Aggregator model is defined by a randomizer $R$, and the decoding function \Dec\ used in the aggregator. We say that the protocol $(R, \Dec, \minbatch, \agg)$ is $(\eps,\delta)$-DP (resp. $(\alpha, \rho(\alpha))$-RDP) in the aggregator model if for any $k$, $\agg_{\minbatch}^{\Dec}(R(x_1), \ldots, R(x_k))$ is $(\eps,\delta)$-DP (resp. $(\alpha, \rho(\alpha))$-RDP). Note that we treat $k$ as public in this definition.

 A multi-round protocol in the Aggregator model is defined by a sequence of single-round protocols. The privacy cost of a multi-round protocol can be upper bounded by the sequential composition of the privacy cost of each of the single-round protocols. More generally, each user may choose to participate in only a subset of the single-round protocols, in which case better privacy bounds on the composition may be proven by a more careful analysis.

 \begin{example}\label{exa:pfl-agg-only}
     Let $x_i \in \Re^d$ with $\|x_i\|_2 \leq 1$. Let $R(x_i) = x_i + \mathcal{N}(0, \frac{\sigma^2}{\minbatch} \mathbb{I}_d)$ and let $\Dec$ be the identity function. Then the aggregation $\agg_{\minbatch}^{\Dec}(R(x_1), \ldots, R(x_k))$ is distributed as $\sum_i x_i + \mathcal{N}(0, \frac{k\sigma^2}{\minbatch} \mathbb{I}_d)$ whenever $k \geq \minbatch$ and equals $\bot$ otherwise. Then for $\eps = \frac{\sqrt{2\log \frac {1.25} \delta}}{\sigma}$, the protocol is $(\eps,\delta)$-DP in the aggregator model.
     As a concrete example, one can verify that for  $\sigma = 7$, this protocol is $(1, 10^{-8})$-DP in the aggregator model. Using the better numerical bounds from~\cite{Balle2018ImprovingTG}, the same privacy bound holds for $\sigma=5.1$.
 \end{example}

 \begin{example} \label{exa:fedstats}
 Consider any $\eps_0$-local DP mechanism $R^{(\eps_0)}$, an arbitrary $\Dec$ and for $\delta>0$, let $\eps = \eps_{\texttt{shuffle}}(\eps_0, \minbatch, \delta)$ be the bound\footnote{The analytical bound in~\cite{FeldmanMT2020} is $\ln\left(1+(e^{\eps_0}-1)\left(\frac{4\sqrt{2\ln(4/\delta)}}{\sqrt{(e^{\eps_0}+1)\minbatch}}+\frac{4}{n}\right)\right)$, and they also give numerical bounds that are tighter.} on the privacy of shuffled version of the local responses proved in~\cite{FeldmanMT2020}. Since aggregation is commutative and associative, the output of the aggregator
 is a post-processing of the shuffled responses. For any $k \leq \minbatch$, the output is independent of the data and hence private. For $k \geq \minbatch$, the shuffling result applies and gives us $(\eps, \delta)$-DP in the aggregator model.
 As a concrete example, for $\eps_0=4.0$ and $\minbatch= 10,000$, the numerical bounds in~\cite{FeldmanMT2020} imply that this protocol satisfies $(0.61, 1e-10)$-DP. When $R^{\eps_0}$ is the {\sc Rappor}$_K$ algorithm~\cite{ErlingssonPK14}, the aggregation allows for asymptotically optimal reconstruction of histograms~\cite{FeldmanMT2020}.
 \end{example}

 Note that the aggregator model assumes that all the $k$ clients are following the protocol. Any bound in the aggregator model extends to a bound in a robust version of this model, where at least $k\geq \minbatch$ of the clients follow the protocol (see~\cref{subsec:security}).
 Moreover, note that from the point of view of one of the clients, the noise is coming from $(\minbatch-1)$ other clients rather than $\minbatch$. Since we  will deal with large values of $\minbatch$, we ignore this distinction for simplicity.

 We next define the stronger primitive of Samplable Anonymous Aggregation. This primitive takes in an additional {\em sampling} parameter $q$. For inputs $x_1, \ldots, x_k$, the sampled aggregate is the aggregate as before of $\Dec(R(x_i))$s, but now taken only over a randomly chosen subset of indices. Formally, for a multiset $S$, let $S_{\downarrow q}$ be the random multiset defined by selecting each index independently with probability $q$. Then
 \begin{align*}
     \sa_{\minbatch}^{\Dec}(\{m_1, \ldots, m_k\}) &=  \agg_{\minbatch}^{\Dec}(\{m_1, \ldots, m_k\}_{\downarrow q})
 \end{align*}

 Note that in this definition, we have assumed that the aggregator only outputs the aggregate, but not the set of indices that contributed to the sum. This {\em secrecy of the sample} is what enables the privacy amplification by sampling analyses.

 As before, a single round protocol $(R, \Dec, \minbatch, \agg, q)$ satisfies a certain differential privacy guarantee in the Samplable Anonymous Aggregator (\SAA) model if $\sa_{\minbatch}^{\Dec}(R(x_1),\ldots, R(x_k))$ satisfies the differential privacy guarantee for any $k$. A multi-round protocol will be analyzed by sequential (adaptive) composition of single round protocols, possibly taking into account that each user may not participate in each round.

 \begin{example}\label{exa:pfl-sampling}
 Consider a Private Federated Learning process where at each iteration, each user independently with probability $q$ decides whether or not to participate in this iteration. If it decides to participate, it sends its gradient, clipped to norm 1, with additive $\mathcal{N}(0, \frac{\sigma^2}{\minbatch})$ noise added. The aggregator outputs the sum of these noisy gradient updates. Then by privacy amplification by sampling results, it follows that this protocol satisfies $(\eps, \delta)$-DP in the \SAA\ model for $\eps \approx q\frac{\sqrt{2\log \frac {1.25} \delta}}{\sigma}$. These analytic bounds can be improved using numerical analysis techniques. As an example, if $\sigma=5.1$ and $q=0.02$, the protocol satisfies $(0.034,10^{-8})$-DP in the \SAA\ model (using~\cite{Balle2018ImprovingTG} and privacy amplification by sampling). Running this algorithm over $2,500$ iterations leads to an overall privacy cost of $(0.8, 10^{-8})$-DP in the \SAA\ model using the moments accountant. Note that this is a significant improvement over what one would get in the aggregation model.
 Indeed without amplification, the resulting $\eps$ is larger than $100$. A more careful analysis, that exploits the fact that when sampling, each user only participates in a $\approx q$ fraction of the rounds gives an $\eps \approx 8.3$. The parameters above are in the same ballpark as used in large-scale deployments such as~\cite{xu2023federated} that lead to accurate models.

 We also note that this analysis does not make any assumptions about the server generating the models on which the gradients are computed. Those may be chosen adversarially, and the privacy analysis continues to hold, as it only requires that each step is a sampled Gaussian.
 \end{example}

 \begin{example}\label{exa:fedstats-sampling} Consider the aggregation of \rappork\ outputs as in \cref{exa:fedstats}. Recall that with $\eps_0 = 4.0$ and $B=10,000$, the shuffled output satisfies $(0.61, 10^{-10})$-DP. Now suppose that we had $1,000,000$ clients, and each was sampled with probability $q=0.02$, and their outputs were aggregated. With respect to the dataset of $1,000,000$ clients, this output satisfies $(\eps, 10^{-10})$-DP for $\eps < 0.02$. Thus in the \SAA\ model, for these parameters, the algorithm satisfies $(0.02, 10^{-10})$-DP.
 \end{example}

 One can also define a version of this definition where the set is chosen by sampling a random subset of indices of a fixed size $\minbatch$. This distinction is akin to the different flavors of privacy amplification by sampling. These different models are close to each other:  when $\minbatch < qk - \sqrt{qk\log \frac 1 \delta}$, then the Poisson sample is of size at least $\minbatch$ with probability $(1-\delta)$. In this case, the distribution of the set of chosen indices in the Poisson case is a mixture of distributions over random subsets of a fixed size $B'$ for a random $B' \geq \minbatch$.

 \subsection{The Impact of Sampling and Anonymity}
 \label{sec:experiments}
 This is an opportune moment to better understand the impact of sampling.  Sampling is often done for reasons of efficiency: when trying to learn a population average (or a population gradient), a small sample may be sufficient to get a reasonable estimate, and sampling may help both minimize the impact on the client device, and reduce the load on the server.

 Consider an iterative algorithm run for $T$ steps, with each step being run on a $q$ fraction of the data. When the sample is not anonymous, each user participates on average in $qT$ steps. Thus if each step is $(\eps,\delta)$-DP, the advanced composition theorem tells us that the algorithm is $\approx (\eps\sqrt{qT}, q\delta T)$-DP.

 On the other hand, when the sample is in fact random, privacy amplification by sampling implies a bound of $\approx (q\eps, q\delta)$ for each of the $T$ steps. Now composing over all $T$ steps, the overall algorithm is $\approx (q\eps\sqrt{T}, q\delta T)$-DP. This gain of about $\sqrt{q}$ in the privacy cost is borne out in numerical privacy accounting as well, as we saw in~\cref{exa:pfl-sampling}.

 Typical private federated learning and analytics applications run many such algorithms for various reasons, such as hyperparameter tuning, keeping the model current in the face of distribution shift, or getting telemetry over time to evaluate the impact of changes in the system. Running algorithm such as above, the $1/\sqrt{q}$ reduction in the privacy cost per algorithm translates to being able to run $1/q$ times as many algorithms at the same privacy cost.

 This is a significant win. The sampling factor $q$ can often be quite small when dealing with large populations of users. This gain, of being able to run $1/q$ times as many experiments comes at {\em absolutely no cost to utility}. From a statistical point of view, the algorithm being run in the two cases (random known sample vs. random unknown sample) is exactly the same. The gain in privacy cost here is coming {\em entirely} from being able to hide the identity of the client devices! Counter-intuitive at first, this gain is similar to what one achieves from {\em shuffling}. However, while that shuffling gain is applicable to a much narrower class of local DP algorithms, the gain from sampling applies more broadly.

 How large is this $1/q$ factor in practice? In typical PFL applications that have been reported, the batch size is typically in the range of a few thousand to tens of thousands, depending partly on the model size~\cite{xu2023federated,federated-personalization}. The population of devices potentially contributing to these collections varies depending on the task at hand. As an example, for training a keyboard model, ~\cite{xu2023federated} have reported population sizes ranging from ~1M (Portugese keyboard model for Portugal) to ~16M (Portugese keyboard model for Brazil), and other works~\cite{improving-on-device-speaker} have used estimates in the 100M range. Thus for these setups $1/q$ varies from $20$ on the extreme low end to $10000$ on the high end, with values in the hundreds being typical. Similar ranges of $1/q \approx 60$ have been reported for Federated statistics applications in ~\cite{cormode2022sample, zhu2020federated}. Thus making the aggregation samplable can allow us to run many more learning tasks compared to aggregation alone, for the same overall privacy budget.
 \begin{figure*}[h]
   \centering
   \includegraphics[width=0.3\linewidth]{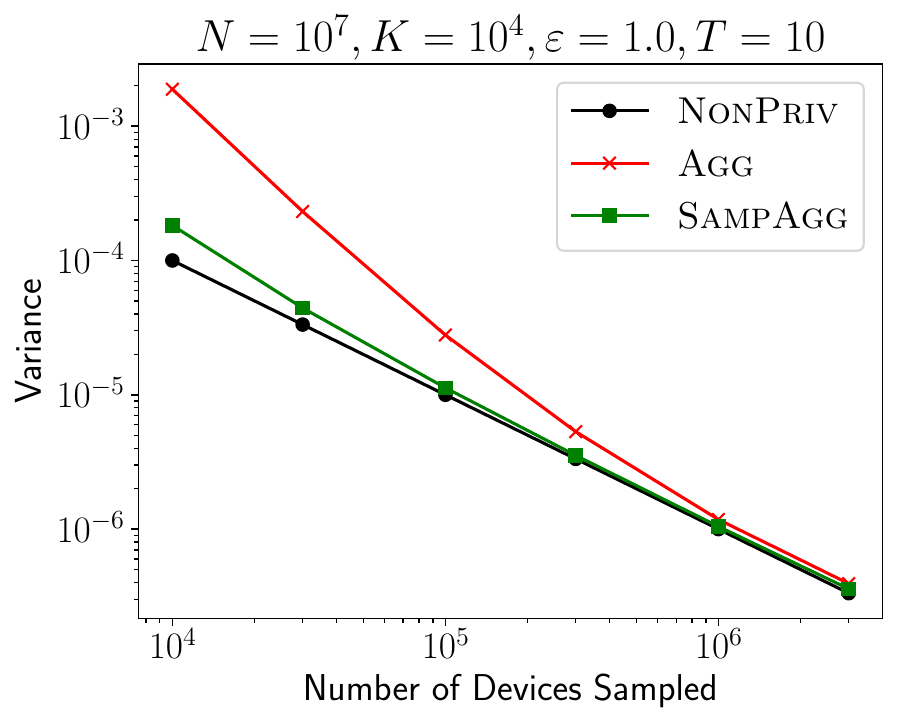}
   \includegraphics[width=0.3\linewidth]{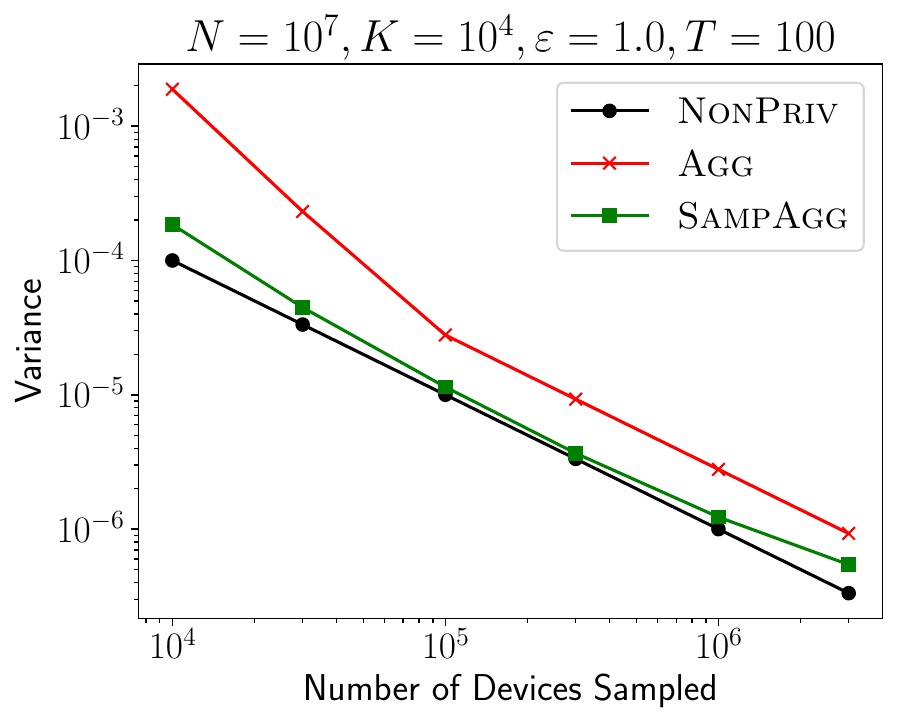}
   \includegraphics[width=0.3\linewidth]{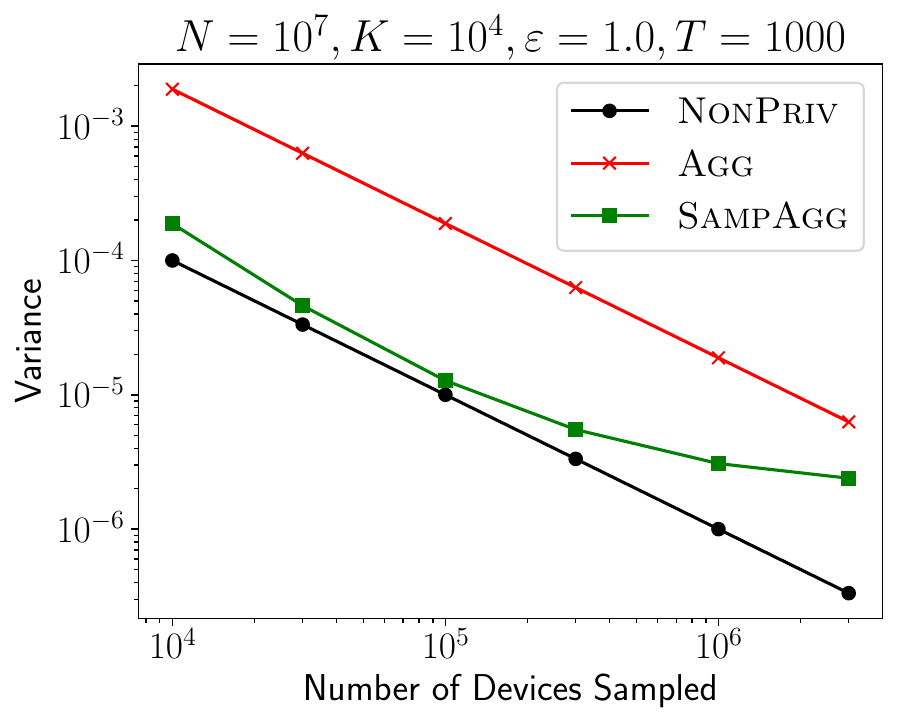}
   \caption{Expected Squared Error of a non-private baseline ({\sc NonPriv}), Aggregation model ({\sc Agg}) and Samplable Aggregation ({\sc SampAgg}) on a histogram task, for varying values of tasks $T$ for a fixed total privacy budget.}\label{fig:hist-vary-T}
   \Description{Plots showing the benefits of Samplable Aggregation on Histograms.}
 \end{figure*}

 \begin{figure*}[h]
   \centering
   \includegraphics[width=0.3\linewidth]{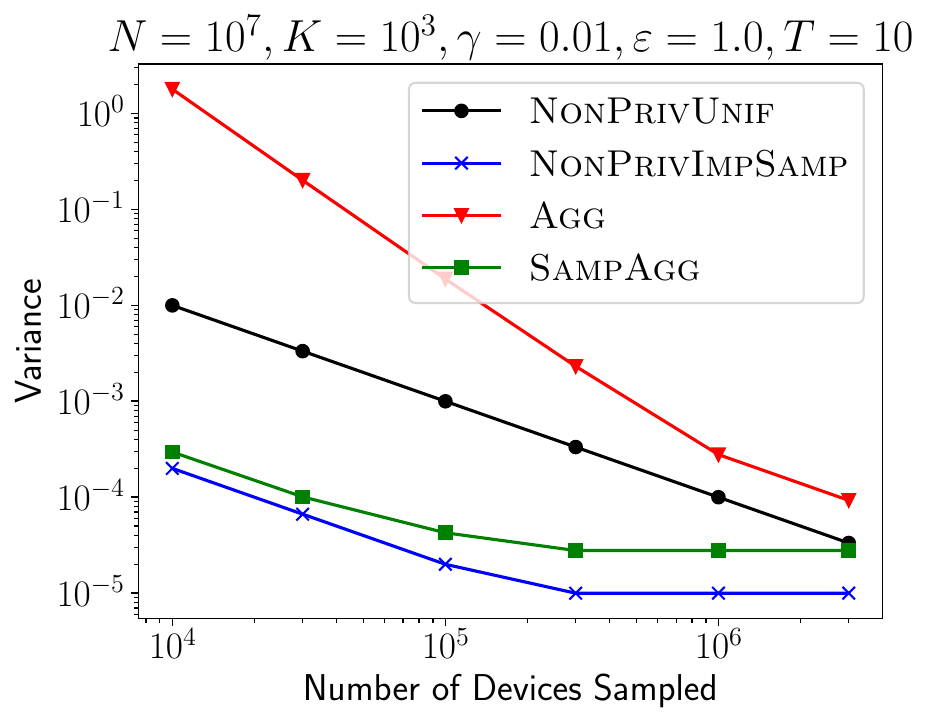}
   \includegraphics[width=0.3\linewidth]{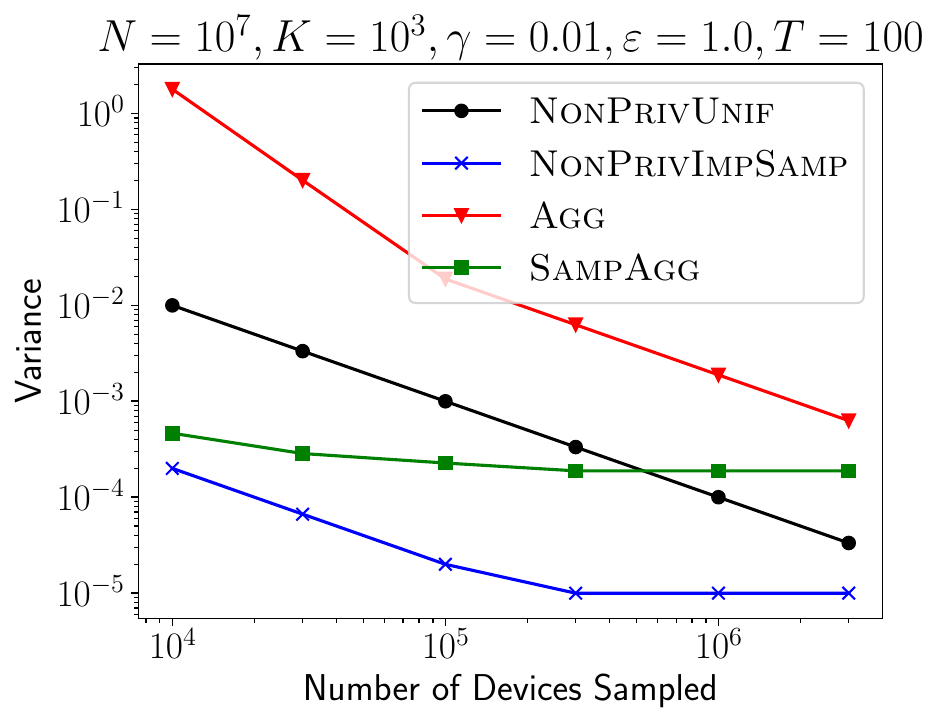}
   \includegraphics[width=0.3\linewidth]{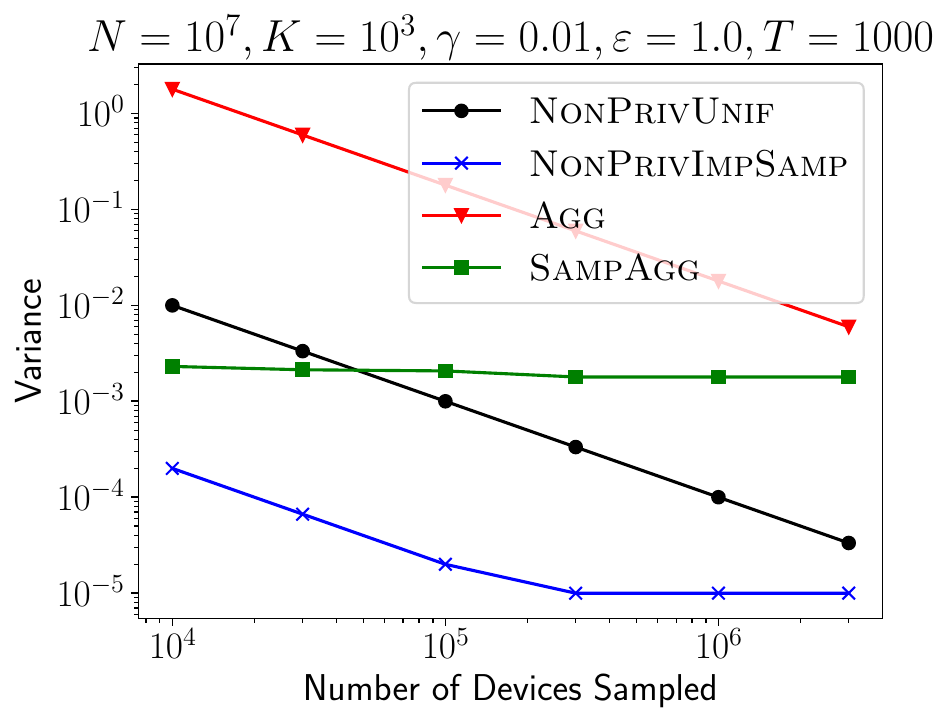}
   \caption{Expected Squared Error on the distribution of non-zero values for a sparse histogram task, for varying parameter values. The plots include a naive non-private baseline ({\sc NonPrivUnif}), non-private Importance Sampling ({\sc NonPrivImpSamp}), Aggregation model ({\sc Agg}) and Samplable Aggregation ({\sc SampAgg})  for varying number of tasks $T$.}\label{fig:sparse-hist-vary-T}
   \Description{Plots showing the benefits of Samplable Aggregation on Sparse Histograms.}
 \end{figure*}

 Similarly, in histogram tasks, one would often sample for communication cost reasons alone, and the samplable anonymity can result is significant improvements in the privacy-utility trade-off.  We demonstrate this numerically next.

 We consider the standard task of building a histogram over an alphabet of size $[K]$, where each of $N$ users has a value $i \in [K]$. To simulate practical communication constraints, we ask that at most $M$ users communicate with the server for each such task. We will consider the case where we want to do $T$ such tasks with a total privacy budget of $(1, 10^{-6})$-DP (e.g. we may want to run an iterative algorithm such as~\cite{ChadhaCDFHJMT24}, or learn multiple different histograms over the same set of users). We evaluate three algorithms, computing the expected squared error (or total variance) $\sum_{i=1}^K \|p_i - \tilde{p}_i\|_2^2$ of the an estimated distribution $\tilde{p}$. The baseline is the non-private algorithm, which only incurs error due to sampling, which is approximately\footnote{To be precise, this error is $\sum_i p_i(1-p_i)/M \leq (1-p_{max})/M$. Thus whenever $p_max << 1$, which is a uniform or power law distribution on a large $K$, this is extremely close to $1/M$.} $\frac{1}{M}$. The non-anonymous aggregation model allows adding Gaussian noise to an estimate computed from an $M/N$ fraction of the population. The noise can be calibrated so that the total privacy cost over $\lceil \frac{TM}{N} \rceil$ compositions is controlled. We use the optimal Analytical Gaussian Mechanism~\cite{Wang:2021} accounting to tune the noise here. With samplable anonymous aggregation, we can use privacy amplification by sampling and thus use the Moments Accountant~\cite{DLDP}. Figure~\ref{fig:hist-vary-T} plots the expected squared error vs. $M$ for a few different parameter settings. The plots show that \SAA allows for lower squared error, often improving significantly on the non-anonymous aggregation. We remark that the error for the two private algorithms here includes the sampling error and thus the private error is always strictly larger. Our plots show that with Samplable Aggregation, the additional error due to privacy is negligible, whereas with aggregation alone, the variance can be an order of magnitude larger.

 This impact is further amplified in the common case when there is a dominant default value, say $0$. This constraint that only a small fraction of the users have a non-default value can be thought of as a sparsity constraint. This arises naturally in many settings. For example the error codes on a device would usually denote "No Error", and we are primarily interested in the distribution of errors when an error occurs.
 Let the the universe by $[K]$ and the fraction of non-zero values by $\gamma$. The goal in this {\em counting needles} task is to estimate the distribution over the values different from $0$. Thus we measure the error of an estimate $q$ by $Err_{\ell_2^2}  =   \sum_{i=1}^{K-1} (q_i - p_i)^2$. The naive sampling approach here will have expected squared error of $\gamma / M$. The reader would notice that only a $\gamma$ fraction of the contributions provide useful (i.e. different from $0$) information. It is therefore natural to downsample the zero values and upsample the non-zero values. We consider a simple instantiation of importance sampling,  where $0$ values are downsampled at rate $M / 2N$, and other values are sampled at rate $\min(M/2\gamma N, 1)$. When $M \leq 2\gamma N$, this yield $M$ samples, roughly half of which are $0$ and the remaining half take a non-default value\footnote{We use this split for simplicity. It can be improved by a factor of two by choosing different downsampling probabilities.}. With aggregation alone, there is no way to privatize the importance sampling approach, and thus any protocol must build on uniform sampling. With anonymous aggregation, one can privatize the importance sampling algorithm, leading to better privacy-utility tradeoffs. Figure~\ref{fig:sparse-hist-vary-T} plots the expected squared error vs. $M$ for a few different parameter settings. Our plots show that with Samplable Anonymous Aggregation, the additional error due to privacy is significantly smaller than that with aggregation alone.
 In \cref{sec:additional_exp}, we present additional plots showing that our improvements continue to hold for a large range of parameters.

 \section{Architecture of a Scalable Samplable Anonymous Aggregator}
 \label{sec:architecture}
 In this section, we describe an architecture for implementing the \SAA\ primitive. Our architecture is built atop Prio~\citep{Corrigan-GibbsB17} which allows aggregation using two or more servers, and ensures that as long as one of the servers is honest, nothing except the aggregate is learnt. Briefly, Prio uses additive secret sharing, where a contribution $m \in \gf_p$ gets shared to the different servers. In the case of two aggregation servers, the first server gets a uniformly random $r \in \gf_p$, and the other server gets $m-r \in \gf_p$. In line with the IETF Distributed Aggregation Protocol draft~\cite{ietf-ppm-dap-04}, we will refer to these two aggregation servers as {\em leader} and {\em helper}. The marginal distributions of the two shares are uniformly random, and thus individually contain no information about the contribution. Each of the two servers aggregates the shares from a group of clients, and the sums are then aggregated, giving the sum of the original messages (modulo $p$) as the result. Additionally, Prio allows the clients to send zero-knowledge proofs of {\em validity}, which allow the aggregation servers to learn that a contribution is valid without learning anything more about the contributions.

 The architecture will build on a few additional components. The first component is a system for shipping {\em recipes} to devices. The recipe specifies the parameters $R, Dec, \minbatch, q$, as well as relevant information about the aggregation process. The device then tosses its own coins to decide on whether or not to contribute, and if so sends its contribution to the system. The contributions are routed through an anonymization mechanism. We also use an anonymous, rate-limited device authentication mechanism so that each device can only send one contribution to a collection. The contributions take the form of shares for the two aggregation servers, each encrypted using their respective public keys. The leader and helper implement a version of the Prio protocol, aggregating the secret shares, and only sharing aggregates once \minbatch\ many contributions are aggregated. We next describe these components in more detail.

 \subsection{Recipe Publishing and On-device Sampling}
 The recipe contains information needed to run the protocol. The randomizers $R$ itself may change from query to query; e.g. while $R$ may correspond to {\sc Rappor}$_K$, the mapping of the data on device to $[K]$, and the value $K$ itself needs to be specified in the recipe. The recipe contains the sampling rate, and the batch size, which allow the client device to verify that the protocol satisfies DP in the \SAA\ model for appropriate privacy parameters. The client device can then use its own randomness to determine whether or not to participate in the collection. If it does contribute, it runs the randomizer and computes an output. This output is then converted to a pair of secret shares as expected in Prio. The recipe also contains the public keys for the servers, which allows these shares to be encrypted so only the appropriate server can decrypt their shares. The recipes themselves can be publicly posted and logged on device, making them auditable. Additionally, the public keys can be posted on a certificate transparency log that the device checks, making them more reliable and manageable.

 \subsection{Anonymization Infrastructure}
 The privacy amplification by sampling in an \SAA\ pipeline depends on the adversary not learning the set of clients that actually contributes to the aggregate. To ensure this property, an anonymization infrastructure can be used. In addition to allowing for the strong anonymity needed in the \SAA\ primitive, this infrastructure  is also valuable in hardening the system. Further, an anonymization infrastructure can also be used by the devices to download the recipes themselves.

 A strawman approach way to ensure strong anonymity is for each device to send a contribution. Devices that choose not to contribute can instead contribute secret shares of {\em zero}, which will have no impact on the aggregate. This approach however incurs a very large communication (and computational) overhead. Indeed the communication and server-side computational cost of noisy stochastic gradient descent in this setup would be as large as {\em full-batch} gradient descent, making it impractical when the number of devices is large.

 We propose distributing the trust using a multi-hop architecture as done in Onion routing~\cite{GoldschlagRS99,DingledineMS04} and other relay-based services~\cite{rfc9298,ietf-ohttp,privaterelay}.
 Indeed these approaches have been formalized as  standard protocols~\cite{ietf-ohttp, privaterelay} that have been deployed at scale for other applications such as private browsing. These approaches can provide anonymity without relying on all devices having to send some communication, and are therefore orders of magnitude more efficient when the sampling rate is small. Our design is oblivious to the precise approach used to achieve anonymization.

 \subsection{Device Authentication and Rate Limiting}
 A device authentication method can be used to ensure that each contribution comes from an actual client. Further, one can use rate-limiting techniques to ensure that a client cannot send arbitrarily many contributions to a collection. Such an approach must build on top of a device attestation framework (e.g~\cite{devicecheck, playintegrity}), or a user attestation framework. For this discussion, we assume that each device has an attestation token that can be validated, but may leak the identity of the user or the device.

 Authentication of contributions from users/devices directly is not compatible with our anonymity goal. We thus need to authenticate devices in an anonymous way.  A simple approach uses a Trusted Execution Environment (TEE) for device authentication. Each contribution comes with an attestation token, which is validated inside a TEE. The TEE also maintains a de-duplication data structure such as a bloom filter to prevent a device from contributing multiple times. The TEE then removes the identifiable attestation token, and signs the leader and helper shares. The leader and helper verify that every share they get has been signed by the TEE. The program running in the TEE can be published and attested to by the clients, enabling the client device to be sure that its contributions can only be decrypted by this program.

 This functionality of anonymous rate-limited attestation can also be implemented using a split-trust model.  The IETF Privacy Pass proposal on rate-limited tokens~\cite{ietf-privacypass-rate-limit-tokens-01} implements precisely this functionality using blind signatures. In brief, the trust is split between two parties, one of which (the {\em attester}) can validate the (non-anonymous) device attestation and rate-limit, whereas the other (the {\em issuer}) can issue tokens meant for a certain target use case. Importantly, the attester learns nothing about the use case, nor the token, and the issuer learns nothing about the identity of the device requesting the token. The token is anonymous and unlinkable, and can be encrypted along with the leader (helper) secret-share. It can be verified by the aggregation servers (leader/helper) to ensure that a device has been authenticated/attested and rate-limited without learning the device's identity. Each of the aggregation servers can verify that the token is valid, and has not been used before.

 In our set up, the time-of-issue and time-of-use can be made uncorrelated by having each device request tokens at a time that is independent of when they need to use them, and cache these tokens. If each device requests these tokens at a random time, the token issuance time carries no information. This can make the system resilient to one of attester/issuer colluding with one of leader/helper. Finally, we note that like the anonymization infrastructure, such an authentication infrastructure is being built and deployed for other applications that require authentication~\cite{pat-cloudflare, pat-fastly}.

 \subsection{Share Aggregation}
 The contributions coming from the device consist of two shares, where each share is encrypted using the public key of the relevant server. The shares are sent to the respective server and each server aggregates their shares. Alternately, the encrypted shares can be sent to the leader, which can batch and forward the helper shares.
 The leader can collect the aggregates once the total number of contributions aggregated is at least $B$.
 Additional steps are needed to ensure that the leader and the helper are aggregating over the same set of contributions, so that a dropped message does not lead to the aggregate being arbitrarily corrupted. Note that while the leader and helper may end up selecting a subset of contributions to aggregate (e.g. dropping some contributions due to network packet loss or corruptions), the protocol still requires that the number of contributions that are successfully aggregated is at least $B$. The IETF draft on Distributed Aggregation Protocol (DAP)~\cite{ietf-ppm-dap-04} describes a specification of such a system, and one may follow the protocol specification to do this aggregation. As long as the helper (resp. leader) does not share any aggregates smaller than the minimum batch size, the privacy property of the system holds against a potentially malicious leader (resp. helper).

 \subsection{Verification of Contributions}
 \label{subsec:verification}
 The use of a Prio-based systems gives us the additional ability to ensure that malicious contributions cannot arbitrarily derail the computation. This robustness, while not needed for privacy, ensures that the aggregates are not easily corrupted by a few adversarial client contributions. Importantly, the validation preserves the zero-knowledge property: neither the leader nor the helper learns any additional information about the client contribution, except for its validity. A client following the protocol will always send a valid contribution, so that the validity predicate contains no information for an honest client. Thus this verification comes at no privacy cost for honest clients.

 While arbitrary validity predicates can in principle be verified, efficient algorithms exist for several important cases of practical interest. For example, when $R$ is the {\em randomized response} mechanism, the output is a $0$-$1$ vector with exactly one $1$. \citet{BonehBCGI21} show how to validate this property with at most $O(\sqrt{K})$ additional communication. Similarly, the output of {\sc Rappor} is a $0$-$1$ vector where the number of $1$s is bounded by $\tilde{O}(1+\frac{K}{e^{\eps_0}+1})$ with high probability\footnote{The $\tilde{O}(\cdot)$ hides factors that are logarithmic in the failure probability.}. Verifying this property, which again can be done efficiently using ~\citep{BonehBCGI21}, helps limit the impact of an adversarial client on the final aggregate. Similarly, when using the Gaussian noise mechanism as the randomizer $R$, the output, when viewed as a $d$-dimensional real vector, has bounded Euclidean norm with high probability. Several recent works~\cite{Talwar22,libprio,pine} study this question and show that this property can be efficiently verified with little overhead. Some of these protocols have been formalized as standards in the IETF draft on Verifiable Distributed Aggregation Functions~\cite{irtf-cfrg-vdaf-06}.

 \begin{figure}
   \centering
 \begin{tikzpicture}[auto, node distance=1.4cm,>=latex,block/.style={draw, fill=white, rectangle, minimum height=2em, minimum width=4em, align=center}]

 \node[block] (client) at (0,0) {Client};
 \node[block] (relay) at (6,0) {Anonymization Service};
 \node[block] (S1) at (9,3){Leader};
 \node[block] (S2) at (9, -3) {Helper};
 \node[block] (attest) at (3,-3) {\begin{tabular}{c}Rate-Limited\\ Attestation Service\end{tabular}};
 \node[block] (recipe) at (3,3) {Recipe Server};

 \draw[->] (client) --node[pos=0.7, above]{\textcircled{3}} (relay);
 \draw[->] (relay.80) -- node[pos=0.5, above, sloped]{\textcircled{4}: $Enc_{k1}(M_1, \tau)$} (S1.180);
 \draw[->] (relay.-80) -- node[pos=0.5, below, sloped]{\textcircled{4}: $ Enc_{k2}(M_2, \tau)$}(S2.180);
 \draw[->] (S1.240) -- (S2.120);
 \draw[<-] (S1.260) -- (S2.100);
 \draw[->] (S1.280) -- (S2.80);
 \draw[->] (S2.60) -- node[pos=0.5,right]{\begin{tabular}{c}\textcircled{5}: Validity Protocol\\ \\ \textcircled{6}: Aggregate\end{tabular}} (S1.300);
 \draw[->] (client.-30) -- node[pos=0.5, above, sloped]{\textcircled{2}} (attest.125);
 \draw[->] (attest.135) -- node[pos=0.5,below,sloped]{$\tau$} (client.-40);

 \draw[->] (client.40) -- node[pos=0.5, above, sloped]{\textcircled{1}} (recipe.210);
 \draw[->] (recipe.218) -- node[pos=0.5,below,sloped]{$(R,B,q,k_1,k_2)$} (client.30);

 \end{tikzpicture}
 \label{fig:schematic}
 \caption{Schematic Description of the communication pattern in the proposed protocol. The {\em Client} downloads a recipe ((Step \textcircled{1}) and gets a rate-limited anonymous attestation token $\tau$ from the {\em Rate-limited Attestation Service} (Step \textcircled{2}). It uses (Step \textcircled{3}) an {\em Anonymization Service} to send secret shares (and proof shares) $M_1$ and $M_2$ to the {\em Leader} and {\em Helper} respectively, each encrypted using their respective encryption keys (Step \textcircled{5}). The Leader and Helper run a protocol to validate the contributions((Step \textcircled{5}), and compute the aggregate over the valid shares (Step \textcircled{6}).}
 \Description{Schematic Diagram showing the communication pattern of the protocol.}
 \end{figure}

 \subsection{Security Analysis: Privacy}
 \label{subsec:security}
 In this section, we argue that the proposed system securely implements an \SAA\ primitive, under suitable assumptions on the adversary. Our privacy assurance is that an adversary learns nothing more than the output of an idealized \SAA\ functionality and the number of contributions received. We start with defining our privacy adversary model.

 \medskip\noindent{\bf Privacy Adversary Model:} We allow an adversary to control at most one of the leader or helper, and assume that the other aggregation server honestly follows the protocol. Additionally, we allow the adversary to control at most $B/2$ of the client devices.

 We next argue that the our proposed system provides the privacy guarantees of the \SAA\ primitive, assuming that anonymization and authentication infrastructures work as intended.
 \begin{theorem}
     Suppose that an adversary controls at most $k$ client devices, and at most one of leader and helper. Further assume that the device authentication and anonymous channel ensure that each device is limited to sending one contribution that is not linkable to the client device. Then for some $\tilde{B}\geq B-k$, the view of the adversary can be simulated given the number of received messages and $\sa_{\tilde{B}}^{\Dec}(M)$ where $M$ is the set of messages from clients not controlled by the adversary.
 \end{theorem}
 \medskip\noindent{\bf Proof Sketch:}
     We assume that the adversary controls the leader; the proof for the helper case is similar. First suppose that $k=0$. Then each client sends a message $m_i = R(x_i)$ to the server with probability $q$. Let $I$ be the set of clients that select themselves, and let $\hat{M}$ be the set of their messages. If there are fewer than $B$ messages received, the honest helper aborts and no information is learnt by the adversary except $|I|$ (the $|I|$ secret shares can be trivially simulated). If the number of received messages is $|I| \geq B$, then the zero knowledge property of the Prio protocol ensures that the view of the leader can be simulated from the aggregate of the messages in $\hat{M}$. The set $\hat{M}$ itself is a uniformly random subset of size $|I|$ and the anonymization infrastructure ensures this is still true from the point of view of the adversary. Thus the aggregate available to the leader is indeed distributed as $\sa_{|I|}^{\Dec}(M)$. This implies the claim for the case of $k'=0$, with $\tilde{B} = \max(B, |I|)$.

     When $k > 0$, then the adversary controls at most $k$ of the client devices. The device authentication ensures that it can control at most $\tilde{k} \leq k$ contributions, and it can simulate its view of those contributions itself. The above argument is now applied to the $|I| - \tilde{k}$ contributions not controlled by the adversary. Since the adversary now knows $\tilde{k}$ of the messages $R(x_i)$, the aggregate over the remaining $|I|-\tilde{k}$ messages from honest clients suffice to allow the simulator to compute the aggregate over all $|I|$ messages. The rest of the simulation is now identical to the $k=0$ case. Setting  $\tilde{B} = \max(B, |I|) - \tilde{k}$ now completes the proof.
 \qed

  We note that the additional leakage $B'$ here is independent of the data $\{x_i\}_{i=1}^k$ held by the clients. Further, by having each device that does not select itself send a dummy report with probability $q' \approx \frac{1}{\eps k}$, we can ensure that $B'$ is a differentially private estimate of the number of devices that actually contribute. Thus the information contained in $B'$ can be hidden, and one can argue that there is a simulator whose output is $(\eps,\delta)$-indistinguishable from the adversary's view.

 \subsection{Utility Analysis}

 The above theorem only addresses the privacy of our protocol. If one of the two servers is adversarial, the computed aggregate can be arbitrarily corrupted and the system gives no utility. Any utility results depend on the assumption that {\em both} the aggregators follow the protocol correctly. When all the clients are honest, the protocol correctly implements the desired samplable aggregation functionality. When the contributions are verified as discussed in~\cref{subsec:verification} (and the aggregation servers are honest), the system is robust to adversarial clients: the impact of $k$ adversarial clients is limited to the amount that $k$ valid contributions can influence the aggregate. This follows from the soundness of the zero knowledge proof of validity. The precise definition of validity here, and thus this robustness, depends on the specific randomizer and the validity test that is used. As an example, when we use this protocol to perform vector aggregation for vectors with Euclidean norm bounded by $1$, the robustness property says that the output of the protocol differs from the ideal functionality by at most $k$ in Euclidean distance. When $k \ll B$, this implies that the mean of the vectors is computed with $\ell_2$ error $\frac k B \ll 1$, making the estimate sufficiently useful, for example, in federated learning. Notice that the ideal functionality of aggregation itself can be corrupted by this amount by $k$ adversarial clients.

 \subsection{Additional Hardening}
 Our security analysis proves that our system implements \SAA\ as long as an attacker controls at most one of the aggregation servers, and a bounded number of devices contributing to the batch.  Indeed consider a hypothetical leader (or helper) trying to learn a specific user's message $m_{i^\star} = R(x_{i^\star})$. To mount this singling out attack, the server would have to (a) ensure that $i^\star$ decides to participate, (b) ensure no other honest user messages are in the batch, and (c) send at least $B-1$ known messages to the batch.  Our use of rate-limited authentication makes (c) very expensive for an attacker. Indeed an adversary that controls a small number of real devices can only control a small number of contributions to a collection. We next discuss various hardenings to further mitigate the risk of an attacker being able to mount such an attack.

 \subsubsection*{Public Recipes and Server Information} The parameters of each collection, including information about leader and helper parameters, can be placed on a public transparency log. This information would be auditable, and this would ensure that every client sees the same recipe.
 This hardening makes it infeasible for the server to prevent sufficiently many genuine clients from contributing to the collection. Coupled with anonymity, it helps ensure that the server cannot identify $m_{i^{\star}}$ from the set of received contributions, and thus prevents an adversarial server from singling out a single device.

 \subsubsection*{Rate validation by leader and helper} Based on the recipe, and the number of devices, the leader and the helper have an expectation on the rate at which contributions arrive. This rate can also be estimated based on historical trends for similar recipes. A noticeably higher rate of contributions may suggest a sybil attack and the leader and helper may validate that the rate is within reasonable bounds. As an example, the leader (or helper) could reject batches when the rate of contributions exceeds twice the expected rate in a certain window. This gives a fair bit of room for natural fluctuations in the rate of contributions. At the same time, if the non-attack contribution rate is close to its expected value, this ensures that at least half of the contributions come from devices not controlled by an attacker. As our privacy guarantees are robust, this would mean that the privacy parameter against the attacker that controls half the batch is worse by a factor of about $\sqrt{2}$.

 \subsubsection*{Limited Logging and Periodic Sub-Aggregation}
 In the case that an attacker can compromise both the leader and the helper, the privacy level falls to that ensurable by shuffling and sampling. In typical federated statistics applications, this is a reasonably strong guarantee; indeed the privacy parameters computed in \cref{exa:fedstats} depend on shuffling alone. In the federated learning setting, the shuffling guarantee is typically weaker and susceptible to attacks of the kind discussed in~\cite{BoenischDSSSP23a, BoenischDSSSP23b}.  The leader and helper maintain no state beyond the collection and so a security breach at the two would have to be relatively simultaneous. Further the protocol can periodically create sub-aggregates, and discard the raw shares that have been aggregated already. This can reduce the privacy risk to the already sub-aggregated contributions, as the sub-aggregate provides some privacy protection depending on the number of contributions it aggregates.

 More broadly, we aim to minimize logging at all steps. The client device does not send any telemetry that may indicate participation in a collection. While this may make it harder to debug the system, reduced logging makes the system more robust and minimizes data leakage in case of attacks.

  \subsubsection*{Reliance on Client Code}
 The implementation of the client part of the protocol is a part of the client operating system. Standard techniques for ensuring trust in this implementation include infrequent updates, auditability of the client code and signed images that ensure that each client device runs the same version of the operating system. In the proposed system, most of the privacy protection depends on the client code alone. For example, the parameter $B$ itself is part of the encrypted shares from the client, which ensures that there is agreement on this value between the client and the two aggregation servers. These hardenings ensure that the important system parameters that impact the privacy guarantee can not be altered by an adversarial leader/helper.

 \section{Discussion}
 \label{sec:discussion}
 \subsection{Federated Statistics vs. Federated Learning} Our aggregation system is designed for two somewhat different kinds of workloads. In typical federated statistics applications, one is interested in the heavy hitters over the population, and the aggregation of interest is a histogram over a noticeable fraction of the population. Thus the setup entails large batch sizes and few latency constraints. On the other hand private federated learning uses the aggregation to compute the gradient over a batch of devices, and multiple adaptive steps over small (relative to the population) batches are preferred to fewer steps over a larger batch. Thus low latency is typically desired in this setting. Our proposed system benefits from a unified design for these two settings needing the same underlying primitive. One may consider different system optimizations for the two cases to help improve the performance of the system.

 \subsection{Trust Models}

 Similar systems for private aggregation may depend on the user trusting a central curator to use the data in a manner consistent with the promised privacy assurance. The proposed system strengthens such policy-based enforcement of privacy promises, and can ensure that the privacy assurance depends primarily on the client code and an inadvertent mistake or a bad actor on the server side does not lead to a privacy violation.

 The privacy assurances of the proposed system depend on non-collusion between the leader and the helper: as long as at least one of the two servers is honest, the claimed privacy bounds hold. Such a trust model has been used for existing Prio deployments~\citep{firefox,ENPA:2021}, as well as in Private Relay~\cite{privaterelay} and Privacy Pass~\cite{ietf-privacypass-rate-limit-tokens-01}, and similar trust models have been used extensively in secure multiparty computation deployments~\cite{Wolinsky2012,Bogetoft09,Liina13,Lapets16,Abidin16,Bogdanov16,ChowB18}. In a deployment where the servers are run by different parts of the same organization, this split trust model prevents a single point of vulnerability, ensuring that any breach of the trust assumptions would be easier to detect. This can be significantly strengthened by having different entities run different servers. With standardized protocols such as DAP~\citep{ietf-ppm-dap-04}, the helper servers may be offered as a service for multiple deployments by different organizations using such a system for private federated learning and optimization. Such a market may help further improve trust in the non-collusion. We leave to future work other approaches to ensuring non-collusion and improving the trust in the system.

 Recall that our utility bounds need both servers to be honest. We believe this is a reasonable assumptions as these servers would typically be run by organizations that are incentivized to correctly compute the aggregate. Using more than two servers can relax this assumption and allow for utility results while only needing an honest majority. However, such results necessarily need to go beyond additive secret-sharing and thus are not conducive to the efficient computations and efficient validity proofs in Prio. We leave further investigation of these approaches to future work.

 \subsection{Central Noise Addition} While our aggregation primitive has the advantage of simplicity, one can also consider a richer primitive which aggregates non-noised contributions and outputs the noised aggregate. This can be useful in settings where the batch size itself may be hard to estimate. Naively implementing this primitive in the two-server framework would require both the leader and the helper to add sufficient noise, making the noise variance larger. However, there may be efficient cryptographic protocols that can allow the two servers to collaboratively generate (shares of) noise with the right variance, so that the noise is hidden from each of the two servers as long as one of them is honest. Such an approach was proposed in the fully distributed MPC setting by~\citet{ODOpaper}, and we leave the design of practical and efficient noise generation protocols in the two-server setting to future work.

 \subsection{Samplable Aggregation vs. Aggregation}

 While an aggregation functionality has been proposed in prior works, we have shown that the stronger functionality of a samplable anonymous aggregation allows us to get better privacy-utility trade-offs for private federated learning algorithms. We remark that in principle, sampling is not needed: the batched stochastic gradient descent algorithm can be replaced by a {\em full-batch} version of the algorithm which can be tuned to give the same privacy-utility trade-offs. However, this version of the algorithm will incur significantly higher cost for client and server, and lead to higher latency and thus fewer iterations in a given amount of time. The samplable aspect of the aggregation is thus crucial to be able to train accurate models in an acceptable timeframe. While some recent approaches to private learning~\cite{KairouzMSTTX21} can get around the privacy amplification by sampling, they require the use of correlated noise across iterations~\cite{DworkNPR10, ChanSS11} and hence are not compatible with our goal of adding noise in a distributed fashion. We leave secure implementations of these noise addition mechanisms to future work. We also remark that these techniques in their current form are not compatible with the use of adaptive optimizers on the server, which are often necessary for training high-accuracy models~\cite{AzamPFTSL23}.

 The sampling itself comes from two different sources of randomness. One that our analysis primarily relies on is the sampling done by a client to decide whether or not to participate in a specific batch. This randomness is derived from an on-device source of randomness, and implemented properly, would indeed be hidden from any reasonable adversary. Typically, devices only consider participating in the collection under specific circumstances; e.g. a mobile device may need to be idle and charging for it to participate. There is some inherent uncertainty in when the device considers participating and it may be reasonable to model this as being hidden from the adversary. Such modeling may allow us to prove better privacy bounds against reasonable adversaries.

 \subsection{Targeting}
 In some settings, the server may want to {\em target} a recipe to a smaller subset of client devices, e.g. we may want to learn a next-word prediction model for clients using a specific language keyboard. Thus the recipe specifies a set of criteria that a device must meet to be able to participate. Arbitrary targeting criteria may allow a server to single out a device, e.g. by crafting a recipe which will have 0 or 1 responses depending on the data on a device. To mitigate against such risks, we can restrict the set of allowed targeting criteria to be those for which we have (e.g. using a private histogram query) ascertained that the population meeting the criteria is large enough. Large populations and a sampling rate bounded away from $1$ ensure that the noise due to sampling is enough to provide a strong differential privacy guarantee for the realized batch size (e.g. \cite{ODOpaper, zhu2020federated,cormode2022sample}). Additionally, we can restrict the targeting to be based only on non-sensitive criteria. Finally, the recipes being public and inspectable further reduces the risk of undetected attacks.

 \subsection{Local Differential Privacy and Compression}
 The contributions from clients that are aggregated in the proposed framework can be high-dimensional. E.g. we may want to build a histogram using {\sc Rappor} over a large alphabet, or aggregate gradients for a large ML model. Naively, this may require the client upload to be rather large, which may be undesirable in settings where the upload bandwidth is limited. While the definition of \SAA-DP does not require the individual randomizers to satisfy a formal privacy guarantee, commonly used algorithms can be run with these randomizers satisfying differential privacy. Using local DP randomizers ensures a backup level of formal privacy. Additionally, it allows the clients to compress the contributions before sending them. Local randomizers are provably all compressible in the single-server setting~\cite{FeldmanTalwar21}, and low-communication local randomizers can often be easily adapted to the two-server setting. E.g. {\sc Rappor} can be replaced by {\sc ProjectiveGeometryResponse}~\citep{FeldmanNNT22} which reduces the communication while preserving the local DP guarantee. Similarly, a random projection can reduce the communication cost in Gaussian noise addition while preserving the accuracy~\cite{AsiFNNT23}.

 \section{Additional Related Work}
 \label{sec:related}
 Differential privacy was introduced in~\cite{Dwork:2006}, and there is by now a large literature on DP primitives, algorithms, complexity, and different models of DP. Several deployed systems use differential privacy. In addition to the private federated learning deployments cited earlier, local differential privacy has been used in~\cite{ErlingssonPK14, Apple2017,DingKY17}. In the central model, there have been DP releases by the US Census Bureau~\cite{onthemap, abowd20222020}, Google~\cite{google-mobility, aktay2020google, bavadekar2021google}, LinkedIn~\cite{rogers2020linkedins, rogers2020members}, Microsoft~\cite{global_victim_perpetrator, pereira2021us} and Facebook~\cite{FacebookURLs} amongst others~\cite{pseo, edp}.

 There has been a long line of work on the Shuffle model~\cite{Bittau17, ErlingssonFMRTT19, CheuSUZZ19}. For some tasks such as histogram estimation, the shuffle model allows for near-optimal privacy utility tradeoffs~\cite{FeldmanMT2020}. However, as we discussed in \cref{sec:experiments}, sampling allows for additional improvements when communication constraints prevent using the full population to build a histogram. For other tasks such as real summation, a single message shuffle provably leads to worse privacy-utility tradeoffs compared to aggregation~\cite{BalleBGN19a}, that can be surmounted by sending a small constant number of messages~\cite{BalleBGN20,GhaziMPV20,GhaziKMPS21}. For the vector aggregation task, which is of great interest for private federated learning, recent work~\cite{AsiFNNTZ24} shows that the Shuffle model is provably worse that aggregation: it requires sending $\Omega(d)$ messages for $d$-dimensional vectors, and is necessarily non-robust for a large class of protocols.

 There are two alternative approaches to implementing aggregation primitives that have been proposed in literature. One is to use Multi-party computation amongst the clients, facilitated by a trusted server.  Bonawitz et al.~\cite{Bonawitz17} proposed an aggregation protocol in this setting and recent works~\cite{SoGA20,BellBGLR20} have shown that this approach can scale.  It is however unclear how to ensure strong anonymity in this approach as a central server does the key distribution. Moreover, this approach requires multiple rounds of interaction involving client devices, leading to more complex protocols.
 A different approach using Trusted Execution Environments (TEEs) has been proposed for implementing shuffling~\cite{Bittau17} and aggregations~\cite{papaya}. This approach is difficult to scale, adapt and maintain and requires specialized hardware. Recent concurrent work~\cite{JinCMRYO2024, EichnerRBH+24} has proposed using TEEs for reliable execution of private federated analyses. Various attacks on existing hardware~\cite{BulckMW+18} however challenge the trust assumptions of these approaches.

 As alluded to earlier, Prio-based aggregation systems have previously been successfully deployed at scale by Mozilla for private telemetry measurment~\cite{firefox} and by several parties to enable private measurements of pandemic data in Exposure Notification Private Analytics (ENPA)~\cite{ENPA:2021}. In the case of ENPA, the system aggregated over 22 billion device aggregates in a little over a year~\cite{enpa-rwc}, with the servers being run by different organizations. While these systems demonstrate the feasibility and scalability of our proposed approach, our work argues for the additional samplability and anonymity as being part of the primitive, which can yield significant privacy benefits for PFL as discussed in~\cref{exa:pfl-sampling}.

 Several Multiparty Computation (MPC) deployments in recent years have used the split-trust model similar to ours, where the security depends on a small set of servers not colluding~\cite{Bogetoft09,Liina13,Lapets16,Abidin16,Bogdanov16,ChowB18}. These examples implement MPC for relatively complex algorithms, and are closer to the cross-silo federated learning setting as the number of parties participating in the computation is typically small.

 Another approach to federated learning that has been used in the cross-silo setting~\cite{de2020SCRAM, BlattGPG20,Froelicher21,oncological23} is to build on recent advances in homomorphic encryption. Here, the parties share encryptions of their data, typically under a secret-shared key. The computation of interest is then performed on the encrypted data, to derive an encryption of the desired result. Finally, the result can be decrypted by a sufficient number of parties collaboratively decrypting using their respective secret shares. Such an approach has been proposed for the cross-device private federated analytics setting in~\cite{unlynx,honeycrisp,orchard} and a related approach with a second non-colluding server helping with the decryption has been proposed in~\cite{crypteps}. Our approach built on Prio avoids extra rounds of interaction with the client devices, thus avoid the challenges of client dropout and churn. Additionally, the samplability in our work allows for much tighter privacy analyses.

 Efficient Anonymous Authetication protocols are an active area of research. Blind RSA signatures date back to~\cite{Chaum81}, and anonymous tokens have been recently studied in~\cite{DavidsonGSTV18,KreuterLOR20,SildeS22}. In addition to the Private Access Token work mentioned earlier~\cite{ietf-privacypass-rate-limit-tokens-01}, rate-limited anonymous tokens have also recently been studied in \cite{BenhamoudaRS23}.

 In the current work, we aimed to limit the impact of an attacker to what it could have done if it could send arbitrary but bounded gradients in proportion to the number of devices it controls.  This may still allow a variety of model poisoning attacks as discussed in several works mentioned earlier~\cite{BagdasaryanVHES20,BhagojiCMC19, BaruchBG19,WangSRVASLP20,CheuSU19}, and differentially private training can help mitigate some of these attacks~\cite{SunKSM19}. Designing practical defenses that can further limit the attacker's impact on the model is an active area of research~\cite{BlanchardEGS17,BernsteinWAA18,flame,XieKG20,WuW21,ShejwalkarHKR22,wang2023invariant}. Extending such defenses to work in our framework may require additional cryptographic primitives and is left to future work.

 \section{Conclusions}
 \label{sec:conclusions}
 In this work, we have proposed a new primitive that can allow many private federated analyses to be performed with utility guarantees close to the central setting without the strong trust assumptions it entails. In particular, the primitive supports private histograms and private federated learning with central-like utility guarantees. We propose an architecture for a large scale implementation of such a system, building on the Prio architecture and additional standard components. Our work makes it feasible to build scalable and useful private federated analysis systems that do not need to trust a single central server. Client devices in our proposed system send a single message, thus avoiding the complexity needed to handle churn in multi-round protocols. Previous deployments of Prio~\cite{firefox,ENPA:2021} attest to the scalability of such an approach.

 We have shown that the \SAA\ primitive enables central-like privacy-utility trade-offs for many problems. A natural research direction is to understand better the full power of this primitive. The number of rounds of interactivity that a practical deployment permits is typically small, motivating the study of the computational, sample, communication and round complexity of private algorithms armed with this primitive.

 An orthogonal direction is to enrich the primitive to allow a larger class of algorithms. The trust model and the two-server architecture proposed here is rich enough to enable additional functionalities in principle. Efficient algorithms to implement functionalities that are useful for differentially private algorithms can enable better utility for additional data analyses.
 Finally, it is natural to implement the \SAA\ primitive (and other richer primitives) under other trust models that further distribute the trust.

 \section{Acknowledgements}
 This work was shaped by early discussions with
 Ulfar Erlingsson, Abhishek Bhowmick, Julien Freudiger, Rogier Van Dalen, Andrew Cherakshyn, Fei Dong, Jaeman Park, Yulia Shuvkashvili, Hwasung Lee and Matt Seigel. We are grateful to Dan Boneh for valuable feedback on an earlier draft of this work. We would also like to thank Joey Meyer, Brian Lindblom, Cory Benfield, Elliot Briggs, John Duchi, Michael Hesse, Anil Katti, Cheney Lyford, Alex Palmer and Paul Pelzl for useful discussions. We also thank the anonymous CCS 2024 reviewers for their valuable feedback.

 \ifbiblatex
 \printbibliography
 \else
 \bibliography{refs}

@inproceedings{Corrigan-GibbsB17,
  author    = {Henry Corrigan{-}Gibbs and
               Dan Boneh},
  editor    = {Aditya Akella and
               Jon Howell},
  title     = {Prio: Private, Robust, and Scalable Computation of Aggregate Statistics},
  booktitle = {14th {USENIX} Symposium on Networked Systems Design and Implementation,
               {NSDI} 2017, Boston, MA, USA, March 27-29, 2017},
  pages     = {259--282},
  publisher = {{USENIX} Association},
  year      = {2017},
  bibsource = {dblp computer science bibliography, https://dblp.org}
}

@article{DworkR14,
  author    = {Cynthia Dwork and
               Aaron Roth},
  title     = {The Algorithmic Foundations of Differential Privacy},
  journal   = {Found. Trends Theor. Comput. Sci.},
  volume    = {9},
  number    = {3-4},
  pages     = {211--407},
  year      = {2014},
 }

@INPROCEEDINGS{DRV10,
  author={Dwork, Cynthia and Rothblum, Guy N. and Vadhan, Salil},
  booktitle={2010 IEEE 51st Annual Symposium on Foundations of Computer Science},
  title={Boosting and Differential Privacy},
  year={2010},
  volume={},
  number={},
  pages={51-60},}

@inproceedings{Talwar22,
  author    = {Kunal Talwar},
  editor    = {L. Elisa Celis},
  title     = {Differential Secrecy for Distributed Data and Applications to Robust
               Differentially Secure Vector Summation},
  booktitle = {3rd Symposium on Foundations of Responsible Computing, {FORC} 2022,
               June 6-8, 2022, Cambridge, MA, {USA}},
  series    = {LIPIcs},
  volume    = {218},
  pages     = {7:1--7:16},
  publisher = {Schloss Dagstuhl - Leibniz-Zentrum f{\"{u}}r Informatik},
  year      = {2022},
 }

@inproceedings{BBCGI19,
  author    = {Dan Boneh and
               Elette Boyle and
               Henry Corrigan{-}Gibbs and
               Niv Gilboa and
               Yuval Ishai},
  editor    = {Alexandra Boldyreva and
               Daniele Micciancio},
  title     = {Zero-Knowledge Proofs on Secret-Shared Data via Fully Linear PCPs},
  booktitle = {Advances in Cryptology - {CRYPTO} 2019 - 39th Annual International
               Cryptology Conference, Santa Barbara, CA, USA, August 18-22, 2019,
               Proceedings, Part {III}},
  series    = {Lecture Notes in Computer Science},
  volume    = {11694},
  pages     = {67--97},
  publisher = {Springer},
  year      = {2019},
 }

@INPROCEEDINGS{BonehBCGI21,
  author={Boneh, Dan and Boyle, Elette and Corrigan-Gibbs, Henry and Gilboa, Niv and Ishai, Yuval},
  booktitle={2021 IEEE Symposium on Security and Privacy (S \& P)},
  title={Lightweight Techniques for Private Heavy Hitters},
  year={2021},
  volume={},
  number={},
  pages={762-776},
  doi={10.1109/SP40001.2021.00048}}

@INPROCEEDINGS{Kasiviswanathan:2008,
  author={S. P. {Kasiviswanathan} and H. K. {Lee} and K. {Nissim} and S. {Raskhodnikova} and A. {Smith}},
  booktitle={2008 49th Annual IEEE Symposium on Foundations of Computer Science},
  title={What Can We Learn Privately?},
  year={2008},
  volume={},
  number={},
  pages={531-540}}

@InProceedings{Dwork:2006,
author="Dwork, Cynthia
and McSherry, Frank
and Nissim, Kobbi
and Smith, Adam",
editor="Halevi, Shai
and Rabin, Tal",
title="Calibrating Noise to Sensitivity in Private Data Analysis",
booktitle="Theory of Cryptography",
year="2006",
publisher="Springer Berlin Heidelberg",
address="Berlin, Heidelberg",
pages="265--284",
}

@InProceedings{Bun:2016,
author="Bun, Mark
and Steinke, Thomas",
editor="Hirt, Martin
and Smith, Adam",
title="Concentrated Differential Privacy: Simplifications, Extensions, and Lower Bounds",
booktitle="Theory of Cryptography",
year="2016",
publisher="Springer Berlin Heidelberg",
address="Berlin, Heidelberg",
pages="635--658",
isbn="978-3-662-53641-4"
}

@inproceedings{BassilyST14,
author = {Bassily, Raef and Smith, Adam and Thakurta, Abhradeep},
title = {Private Empirical Risk Minimization: Efficient Algorithms and Tight Error Bounds},
year = {2014},
publisher = {IEEE Computer Society},
address = {USA},
booktitle = {Proceedings of the 2014 IEEE 55th Annual Symposium on Foundations of Computer Science},
pages = {464–473},
numpages = {10},
series = {FOCS '14} }

@article{Wang:2021,
title={Subsampled Rényi Differential Privacy and Analytical Moments Accountant},
volume={10},
number={2},
journal={Journal of Privacy and Confidentiality},
author={Wang, Yu-Xiang and Balle, Borja and Kasiviswanathan, Shiva},
year={2021},
}

@article{Balle:2018,
title={Privacy Profiles and Amplification by Subsampling},
volume={10},
number={1},
journal={Journal of Privacy and Confidentiality},
author={Balle, Borja and Barthe, Gilles and Gaboardi, Marco},
year={2020},
 }

@inproceedings{DLDP,
	author    = {Mart{\'{\i}}n Abadi and
							Andy Chu and
							Ian J. Goodfellow and
							H. Brendan McMahan and
							Ilya Mironov and
							Kunal Talwar and
							Li Zhang},
	title     = {Deep Learning with Differential Privacy},
	booktitle = {Proceedings of the 2016 {ACM} {SIGSAC} Conference on Computer and Communications Security (CCS)},
	pages     = {308--318},
	year      = {2016},
}

@InProceedings{ODOpaper,
author = {Dwork, Cynthia and Kenthapadi, Krishnaram and McSherry, Frank and Mironov, Ilya and Naor, Moni},
title = {Our Data, Ourselves: Privacy Via Distributed Noise Generation},
series = {Lecture Notes in Computer Science},
booktitle = {Advances in Cryptology (EUROCRYPT 2006)},
year = {2006},
publisher = {Springer Verlag},
pages = {486-503},
volume = {4004},
edition = {Advances in Cryptology (EUROCRYPT 2006)},
}

@inproceedings{Bonawitz17,
author = {Bonawitz, Keith and Ivanov, Vladimir and Kreuter, Ben and Marcedone, Antonio and McMahan, H. Brendan and Patel, Sarvar and Ramage, Daniel and Segal, Aaron and Seth, Karn},
title = {Practical Secure Aggregation for Privacy-Preserving Machine Learning}, year = {2017},
isbn = {9781450349468},
publisher = {Association for Computing Machinery},
address = {New York, NY, USA},
doi = {10.1145/3133956.3133982},
booktitle = {Proceedings of the 2017 ACM SIGSAC Conference on Computer and Communications Security},
pages = {1175–1191},
numpages = {17},
series = {CCS '17} }

@article{CheuSU19,
title={Manipulation Attacks in Local Differential Privacy},
volume={11},
url={https://journalprivacyconfidentiality.org/index.php/jpc/article/view/754}, DOI={10.29012/jpc.754},
number={1},
journal={Journal of Privacy and Confidentiality},
author={Cheu, Albert and Smith, Adam and Ullman, Jonathan},
year={2021}, month={Feb.}
}

@InProceedings{CheuSUZZ19,
author="Cheu, Albert
and Smith, Adam
and Ullman, Jonathan
and Zeber, David
and Zhilyaev, Maxim",
editor="Ishai, Yuval
and Rijmen, Vincent",
title="Distributed Differential Privacy via Shuffling",
booktitle="Advances in Cryptology -- EUROCRYPT 2019",
year="2019",
publisher="Springer International Publishing",
address="Cham",
pages="375--403",
isbn="978-3-030-17653-2"
}

@inproceedings{ErlingssonFMRTT19,
author = {Erlingsson, \'{U}lfar and Feldman, Vitaly and Mironov, Ilya and Raghunathan, Ananth and Talwar, Kunal and Thakurta, Abhradeep},
title = {Amplification by Shuffling: From Local to Central Differential Privacy via Anonymity}, year = {2019},
publisher = {Society for Industrial and Applied Mathematics},
address = {USA},
booktitle = {Proceedings of the Thirtieth Annual ACM-SIAM Symposium on Discrete Algorithms},
pages = {2468–2479},
numpages = {12},
series = {SODA '19} }

@inproceedings{Bittau17,
author = {Bittau, Andrea and Erlingsson, \'{U}lfar and Maniatis, Petros and Mironov, Ilya and Raghunathan, Ananth and Lie, David and Rudominer, Mitch and Kode, Ushasree and Tinnes, Julien and Seefeld, Bernhard},
title = {Prochlo: Strong Privacy for Analytics in the Crowd},
year = {2017},
isbn = {9781450350853},
publisher = {Association for Computing Machinery},
address = {New York, NY, USA},
doi = {10.1145/3132747.3132769},
booktitle = {Proceedings of the 26th Symposium on Operating Systems Principles},
pages = {441–459},
numpages = {19},
series = {SOSP '17}
}

@InProceedings{BalleBGN19a,
author="Balle, Borja
and Bell, James
and Gasc{\'o}n, Adri{\`a}
and Nissim, Kobbi",
editor="Boldyreva, Alexandra
and Micciancio, Daniele",
title="The Privacy Blanket of the Shuffle Model",
booktitle="Advances in Cryptology -- CRYPTO 2019",
year="2019",
publisher="Springer International Publishing",
address="Cham",
pages="638--667",
isbn="978-3-030-26951-7"
}

@ARTICLE{SoGA20,
  author={So, Jinhyun and Güler, Başak and Avestimehr, A. Salman},
  journal={IEEE Journal on Selected Areas in Information Theory},
  title={Turbo-Aggregate: Breaking the Quadratic Aggregation Barrier in Secure Federated Learning},
  year={2021},
  volume={2},
  number={1},
  pages={479-489},
  doi={10.1109/JSAIT.2021.3054610}}

@misc{BellBGLR20,
    author = {James Bell and K. A. Bonawitz and Adrià Gascón and Tancrède Lepoint and Mariana Raykova},
    title = {Secure Single-Server Aggregation with (Poly)Logarithmic Overhead},
    howpublished = {Cryptology ePrint Archive, Report 2020/704},
    year = {2020},
    note = {\url{https://eprint.iacr.org/2020/704}},
}

@inproceedings{ErlingssonPK14,
author = {Erlingsson, \'{U}lfar and Pihur, Vasyl and Korolova, Aleksandra},
title = {RAPPOR: Randomized Aggregatable Privacy-Preserving Ordinal Response},
year = {2014},
isbn = {9781450329576},
publisher = {Association for Computing Machinery},
address = {New York, NY, USA},
booktitle = {Proceedings of the 2014 ACM SIGSAC Conference on Computer and Communications Security},
pages = {1054–1067},
numpages = {14},
series = {CCS '14}
}

@article{Apple2017,
  title={Learning with privacy at scale},
  author={{Apple's Differential Privacy Team}},
  journal={Apple Machine Learning Journal},
  year={2017},
  volume={1},
  number={9},
}

@inproceedings{DingKY17,
author = {Ding, Bolin and Kulkarni, Janardhan and Yekhanin, Sergey},
title = {Collecting Telemetry Data Privately},
year = {2017},
isbn = {9781510860964},
publisher = {Curran Associates Inc.},
address = {Red Hook, NY, USA},
booktitle = {Proceedings of the 31st International Conference on Neural Information Processing Systems},
pages = {3574–3583},
numpages = {10},
series = {NIPS'17}
}

@inproceedings{Balle2018ImprovingTG,
  title={Improving the Gaussian Mechanism for Differential Privacy: Analytical Calibration and Optimal Denoising},
  author={Borja Balle and Yu-Xiang Wang},
  booktitle={International Conference on Machine Learning},
  year={2018}
}

@inproceedings{mironov2017renyi,
  title={R{\'e}nyi differential privacy},
  author={Mironov, Ilya},
  booktitle={2017 IEEE 30th Computer Security Foundations Symposium (CSF)},
  pages={263--275},
  year={2017},
  organization={IEEE}
}

@inproceedings{FeldmanTalwar21,
  author    = {Vitaly Feldman and
               Kunal Talwar},
  editor    = {Marina Meila and
               Tong Zhang},
  title     = {Lossless Compression of Efficient Private Local Randomizers},
  booktitle = {{ICML}},
  series    = {Proceedings of Machine Learning Research},
  volume    = {139},
  pages     = {3208--3219},
  publisher = {{PMLR}},
  year      = {2021},
  bibsource = {dblp computer science bibliography, https://dblp.org}
}

@INPROCEEDINGS{FeldmanMT2020,
author={Feldman, Vitaly and McMillan, Audra and Talwar, Kunal},  booktitle={2021 IEEE 62nd Annual Symposium on Foundations of Computer Science (FOCS)},
title={Hiding Among the Clones: A Simple and Nearly Optimal Analysis of Privacy Amplification by Shuffling},
year={2022},
pages={954-964},
doi={10.1109/FOCS52979.2021.00096}}

@Misc{ENPA:2021,
  Title                    = {Exposure Notification Privacy-preserving Analytics ({ENPA}) White Paper},

  Author                   = {Apple and Google},
  HowPublished             = {\url{https://covid19-static.cdn-apple.com/applications/covid19/current/static/contact-tracing/pdf/ENPA_White_Paper.pdf}},
  Year                     = {2021}
}

@misc{enpa-rwc,
title = {Exposure Notifications Privacy Analytics},
author = {Tim Geoghegan and Mariana Raykova},
year = {2022},
note = {IACR Real World Crypto (RWC) talk}
}

@article{CanonneKS20,
	author = {Canonne, Clément L. and Kamath, Gautam and Steinke, Thomas},
	title = {The Discrete Gaussian for Differential Privacy},
	publisher = {arXiv:2004.00010},
	year = {2020}
}

@misc{Ullman:2017,
author = {Jon Ullman},
title = {Lecture Notes for CS788: Rigorous approaches to data privacy},
year = {2017},
}

@article{WangLF16b,
  title     = {Learning with Differential Privacy: Stability, Learnability and the Sufficiency and Necessity of {ERM} Principle},
  author        = {Wang, Yu-Xiang and Lei, Jing and Fienberg, Stephen E.},
  journal   = {J. Mach. Learn. Res.},
  volume    = {17},
  pages     = {183:1--183:40},
  year      = {2016},
}

@article{Mironov2019RnyiDP,
  title={R{\'e}nyi Differential Privacy of the Sampled Gaussian Mechanism},
  author={Ilya Mironov and Kunal Talwar and Li Zhang},
  journal={ArXiv},
  year={2019},
  volume={abs/1908.10530}
}

@misc{ZhangRXZZK23,
      title={Private Federated Learning in Gboard},
      author={Yuanbo Zhang and Daniel Ramage and Zheng Xu and Yanxiang Zhang and Shumin Zhai and Peter Kairouz},
      year={2023},
      eprint={2306.14793},
      archivePrefix={arXiv},
      primaryClass={cs.CR}
}

@inproceedings{CTWJ+21,
    author	  =  	{Carlini, Nicholas and Tram{\`e}r, Florian and Wallace, Eric and Jagielski, Matthew and Herbert-Voss, Ariel and Lee, Katherine and Roberts, Adam and Brown, Tom and Song, Dawn and Erlingsson, Ulfar and Oprea, Alina and Raffel, Colin},
    title	  =  	{Extracting Training Data from Large Language Models},
    booktitle	  =  	{USENIX Security Symposium},
    year	  =  	{2021},
    howpublished	  =  	{arXiv preprint arXiv:2012.07805},
    url	  =  	{https://arxiv.org/abs/2012.07805}
}

@inproceedings{CIJL+23,
author	  =  	{Carlini, Nicholas and Ippolito, Daphne and Jagielski, Matthew and Lee, Katherine and Tram{\`e}r, Florian and Zhang, Chiyuan},
title	  =  	{Quantifying Memorization Across Neural Language Models},
booktitle	  =  	{International Conference on Learning Representations (ICLR)},
year	  =  	{2023},
howpublished	  =  	{arXiv preprint arXiv:2202.07646},
url	  =  	{https://arxiv.org/abs/2202.07646}
}

@INPROCEEDINGS{ShokriSSS17,
  author={Shokri, Reza and Stronati, Marco and Song, Congzheng and Shmatikov, Vitaly},
  booktitle={2017 IEEE Symposium on Security and Privacy (S \& P)},
  title={Membership Inference Attacks Against Machine Learning Models},
  year={2017},
  volume={},
  number={},
  pages={3-18},
  doi={10.1109/SP.2017.41}}

@INPROCEEDINGS{NasrSH19,
  author={Nasr, Milad and Shokri, Reza and Houmansadr, Amir},
  booktitle={2019 IEEE Symposium on Security and Privacy (S \& P)},
  title={Comprehensive Privacy Analysis of Deep Learning: Passive and Active White-box Inference Attacks against Centralized and Federated Learning},
  year={2019},
  volume={},
  number={},
  pages={739-753},
  doi={10.1109/SP.2019.00065}}

@inproceedings{CarliniLEKS19,
author = {Carlini, Nicholas and Liu, Chang and Erlingsson, \'{U}lfar and Kos, Jernej and Song, Dawn},
title = {The Secret Sharer: Evaluating and Testing Unintended Memorization in Neural Networks},
year = {2019},
isbn = {9781939133069},
publisher = {USENIX Association},
address = {USA},
booktitle = {Proceedings of the 28th USENIX Conference on Security Symposium},
pages = {267–284},
numpages = {18},
series = {SEC'19}
}

@inproceedings{Feldman20,
author = {Feldman, Vitaly},
title = {Does Learning Require Memorization? A Short Tale about a Long Tail},
year = {2020},
isbn = {9781450369794},
publisher = {Association for Computing Machinery},
address = {New York, NY, USA},
doi = {10.1145/3357713.3384290},
booktitle = {Proceedings of the 52nd Annual ACM SIGACT Symposium on Theory of Computing},
pages = {954–959},
numpages = {6},
series = {STOC 2020}
}

@inproceedings{FeldmanZ20,
 author = {Feldman, Vitaly and Zhang, Chiyuan},
 booktitle = {Advances in Neural Information Processing Systems},
 editor = {H. Larochelle and M. Ranzato and R. Hadsell and M.F. Balcan and H. Lin},
 pages = {2881--2891},
 publisher = {Curran Associates, Inc.},
 title = {What Neural Networks Memorize and Why: Discovering the Long Tail via Influence Estimation},
 volume = {33},
 year = {2020}
}

@article{ChaudhuriMS11,
  author  = {Kamalika Chaudhuri and Claire Monteleoni and Anand D. Sarwate},
  title   = {Differentially Private Empirical Risk Minimization},
  journal = {Journal of Machine Learning Research},
  year    = {2011},
  volume  = {12},
  number  = {29},
  pages   = {1069--1109},
  url     = {http://jmlr.org/papers/v12/chaudhuri11a.html}
}

@inproceedings{BoenischDSSSP23b,
      title={Reconstructing Individual Data Points in Federated Learning Hardened with Differential Privacy and Secure Aggregation},
      author={Franziska Boenisch and Adam Dziedzic and Roei Schuster and Ali Shahin Shamsabadi and Ilia Shumailov and Nicolas Papernot},
      year={2023},
      booktitle={8th IEEE European Symposium on Security and Privacy}
}

@inproceedings{BoenischDSSSP23a,
      title={When the Curious Abandon Honesty: Federated Learning Is Not Private},
      author={Franziska Boenisch and Adam Dziedzic and Roei Schuster and Ali Shahin Shamsabadi and Ilia Shumailov and Nicolas Papernot},
      year={2023},
      booktitle={8th IEEE European Symposium on Security and Privacy}
}

@inproceedings{BellGGKMRS22,
author = {Bell, James and Gasc\'{o}n, Adri\`{a} and Ghazi, Badih and Kumar, Ravi and Manurangsi, Pasin and Raykova, Mariana and Schoppmann, Phillipp},
title = {Distributed, Private, Sparse Histograms in the Two-Server Model},
year = {2022},
isbn = {9781450394505},
publisher = {Association for Computing Machinery},
address = {New York, NY, USA},
doi = {10.1145/3548606.3559383},
booktitle = {Proceedings of the 2022 ACM SIGSAC Conference on Computer and Communications Security},
pages = {307–321},
numpages = {15},
series = {CCS '22}
}

@InProceedings{CastroP22,
author="de Castro, Leo
and Polychroniadou, Anitgoni",
editor="Dunkelman, Orr
and Dziembowski, Stefan",
title="Lightweight, Maliciously Secure Verifiable Function Secret Sharing",
booktitle="Advances in Cryptology -- EUROCRYPT 2022",
year="2022",
publisher="Springer International Publishing",
address="Cham",
pages="150--179",
isbn="978-3-031-06944-4"
}

@article{RatheeSWP22,
  author       = {Mayank Rathee and
                  Conghao Shen and
                  Sameer Wagh and
                  Raluca Ada Popa},
  title        = {{ELSA:} Secure Aggregation for Federated Learning with Malicious Actors},
  journal      = {{IACR} Cryptol. ePrint Arch.},
  pages        = {1695},
  year         = {2022},
  url          = {https://eprint.iacr.org/2022/1695},
  bibsource    = {dblp computer science bibliography, https://dblp.org}
}

@inproceedings{AddankiGJOP22,
  author       = {Surya Addanki and
                  Kevin Garbe and
                  Eli Jaffe and
                  Rafail Ostrovsky and
                  Antigoni Polychroniadou},
  editor       = {Clemente Galdi and
                  Stanislaw Jarecki},
  title        = {Prio+: Privacy Preserving Aggregate Statistics via Boolean Shares},
  booktitle    = {Security and Cryptography for Networks - 13th International Conference,
                  {SCN} 2022, Amalfi, Italy, September 12-14, 2022, Proceedings},
  series       = {Lecture Notes in Computer Science},
  volume       = {13409},
  pages        = {516--539},
  publisher    = {Springer},
  year         = {2022},
  doi          = {10.1007/978-3-031-14791-3\_23},
}

@misc{firefox,
author="Robert Helmer and Anthony Miyaguchi and Eric Rescorla",
title="Testing Privacy-Preserving Telemetry with Prio",
year="2018",
howpublished="\url{https://hacks.mozilla.org/2018/10/testing-privacy-preserving-telemetry-with-prio/}"
}

@inproceedings{papaya,
 author = {Huba, Dzmitry and Nguyen, John and Malik, Kshitiz and Zhu, Ruiyu and Rabbat, Mike and Yousefpour, Ashkan and Wu, Carole-Jean and Zhan, Hongyuan and Ustinov, Pavel and Srinivas, Harish and Wang, Kaikai and Shoumikhin, Anthony and Min, Jesik and Malek, Mani},
 booktitle = {Proceedings of Machine Learning and Systems},
 editor = {D. Marculescu and Y. Chi and C. Wu},
 pages = {814--832},
 title = {PAPAYA: Practical, Private, and Scalable Federated Learning},
 volume = {4},
 year = {2022}
}

@InProceedings{BagdasaryanVHES20,
  title = 	 {How To Backdoor Federated Learning},
  author =       {Bagdasaryan, Eugene and Veit, Andreas and Hua, Yiqing and Estrin, Deborah and Shmatikov, Vitaly},
  booktitle = 	 {Proceedings of the Twenty Third International Conference on Artificial Intelligence and Statistics},
  pages = 	 {2938--2948},
  year = 	 {2020},
  editor = 	 {Chiappa, Silvia and Calandra, Roberto},
  volume = 	 {108},
  series = 	 {Proceedings of Machine Learning Research},
  publisher =    {PMLR},
}

@article{SunKSM19,
  author       = {Ziteng Sun and
                  Peter Kairouz and
                  Ananda Theertha Suresh and
                  H. Brendan McMahan},
  title        = {Can You Really Backdoor Federated Learning?},
  journal      = {CoRR},
  volume       = {abs/1911.07963},
  year         = {2019},
  url          = {http://arxiv.org/abs/1911.07963},
  eprinttype    = {arXiv},
  eprint       = {1911.07963},
  bibsource    = {dblp computer science bibliography, https://dblp.org}
}

@InProceedings{BhagojiCMC19,
  title = 	 {Analyzing Federated Learning through an Adversarial Lens},
  author =       {Bhagoji, Arjun Nitin and Chakraborty, Supriyo and Mittal, Prateek and Calo, Seraphin},
  booktitle = 	 {Proceedings of the 36th International Conference on Machine Learning},
  pages = 	 {634--643},
  year = 	 {2019},
  editor = 	 {Chaudhuri, Kamalika and Salakhutdinov, Ruslan},
  volume = 	 {97},
  series = 	 {Proceedings of Machine Learning Research},
  publisher =    {PMLR},
 }

@inproceedings{BaruchBG19,
 author = {Baruch, Gilad and Baruch, Moran and Goldberg, Yoav},
 booktitle = {Advances in Neural Information Processing Systems},
 editor = {H. Wallach and H. Larochelle and A. Beygelzimer and F. d\textquotesingle Alch\'{e}-Buc and E. Fox and R. Garnett},
 pages = {},
 publisher = {Curran Associates, Inc.},
 title = {A Little Is Enough: Circumventing Defenses For Distributed Learning},
 volume = {32},
 year = {2019}
}

@inproceedings{WangSRVASLP20,
 author = {Wang, Hongyi and Sreenivasan, Kartik and Rajput, Shashank and Vishwakarma, Harit and Agarwal, Saurabh and Sohn, Jy-yong and Lee, Kangwook and Papailiopoulos, Dimitris},
 booktitle = {Advances in Neural Information Processing Systems},
 editor = {H. Larochelle and M. Ranzato and R. Hadsell and M.F. Balcan and H. Lin},
 pages = {16070--16084},
 publisher = {Curran Associates, Inc.},
 title = {Attack of the Tails: Yes, You Really Can Backdoor Federated Learning},
 volume = {33},
 year = {2020}
}

@article{yousefpour2021opacus,
	title={Opacus: User-friendly differential privacy library in {P}y{T}orch},
	author={Yousefpour, Ashkan and Shilov, Igor and Sablayrolles, Alexandre and Testuggine, Davide and Prasad, Karthik and Malek, Mani and Nguyen, John and Ghosh, Sayan and Bharadwaj, Akash and Zhao, Jessica and others},
	journal={arXiv preprint arXiv:2109.12298},
	year={2021},
  note = {PriML Workshop at NeurIPS}
}

@misc{tfprivacy,
title = {Tensorflow {P}rivacy},
url = {https://github.com/tensorflow/privacy},
note = {Retreived June, 2023}
}

@misc{libprio,
author = {ISRG},
title = {DivviUp LibPrio Rust},
url = {https://github.com/divviup/libprio-rs},
year = {2023},
note = {Retreived June 2023}
}

@misc{pine,
title = {PINE: Efficient Verification of a Euclidean Norm Bound of a Secret-Shared Vector},
author = {Guy Rothblum and Eran Omri and Junye Chen and Kunal Talwar},
note = {Usenix Security 2024. To Appear.}
}

@InProceedings{KairouzMSTTX21,
  title = 	 {Practical and Private (Deep) Learning Without Sampling or Shuffling},
  author =       {Kairouz, Peter and Mcmahan, Brendan and Song, Shuang and Thakkar, Om and Thakurta, Abhradeep and Xu, Zheng},
  booktitle = 	 {Proceedings of the 38th International Conference on Machine Learning},
  pages = 	 {5213--5225},
  year = 	 {2021},
  editor = 	 {Meila, Marina and Zhang, Tong},
  volume = 	 {139},
  series = 	 {Proceedings of Machine Learning Research},
  publisher =    {PMLR},
}

@inproceedings{DworkNPR10,
author = {Dwork, Cynthia and Naor, Moni and Pitassi, Toniann and Rothblum, Guy N.},
title = {Differential Privacy under Continual Observation},
year = {2010},
isbn = {9781450300506},
publisher = {Association for Computing Machinery},
doi = {10.1145/1806689.1806787},
booktitle = {Proceedings of the Forty-Second ACM Symposium on Theory of Computing},
pages = {715–724},
numpages = {10},
location = {Cambridge, Massachusetts, USA},
series = {STOC '10}
}

@article{ChanSS11,
author = {Chan, T.-H. Hubert and Shi, Elaine and Song, Dawn},
title = {Private and Continual Release of Statistics},
year = {2011},
issue_date = {November 2011},
publisher = {Association for Computing Machinery},
address = {New York, NY, USA},
volume = {14},
number = {3},
issn = {1094-9224},
doi = {10.1145/2043621.2043626},
journal = {ACM Trans. Inf. Syst. Secur.},
articleno = {26},
numpages = {24}
}

@InProceedings{FeldmanNNT22,
  title = 	 {Private Frequency Estimation via Projective Geometry},
  author =       {Feldman, Vitaly and Nelson, Jelani and Nguyen, Huy and Talwar, Kunal},
  booktitle = 	 {Proceedings of the 39th International Conference on Machine Learning},
  pages = 	 {6418--6433},
  year = 	 {2022},
  editor = 	 {Chaudhuri, Kamalika and Jegelka, Stefanie and Song, Le and Szepesvari, Csaba and Niu, Gang and Sabato, Sivan},
  volume = 	 {162},
  series = 	 {Proceedings of Machine Learning Research},
  publisher =    {PMLR},
}

@misc{AsiFNNT23,
      title={Fast Optimal Locally Private Mean Estimation via Random Projections},
      author={Hilal Asi and Vitaly Feldman and Jelani Nelson and Huy L. Nguyen and Kunal Talwar},
      year={2023},
      eprint={2306.04444},
      archivePrefix={arXiv},
      primaryClass={cs.LG}
}

@article{Chaum81,
author = {Chaum, David L.},
title = {Untraceable Electronic Mail, Return Addresses, and Digital Pseudonyms},
year = {1981},
issue_date = {Feb. 1981},
publisher = {Association for Computing Machinery},
address = {New York, NY, USA},
volume = {24},
number = {2},
issn = {0001-0782},
doi = {10.1145/358549.358563},
journal = {Commun. ACM},
pages = {84–90},
numpages = {7}
}

@inproceedings {DingledineMS04,
author = {Roger Dingledine and Nick Mathewson and Paul Syverson},
title = {Tor: The {Second-Generation} Onion Router},
booktitle = {13th USENIX Security Symposium (USENIX Security 04)},
year = {2004},
address = {San Diego, CA},
publisher = {USENIX Association},
}

@article{GoldschlagRS99,
author = {Goldschlag, David and Reed, Michael and Syverson, Paul},
title = {Onion Routing},
year = {1999},
issue_date = {Feb. 1999},
publisher = {Association for Computing Machinery},
address = {New York, NY, USA},
volume = {42},
number = {2},
issn = {0001-0782},
doi = {10.1145/293411.293443},
journal = {Commun. ACM},
pages = {39–41},
numpages = {3}
}

@misc{privaterelay,
title = "i{C}loud {P}rivate {R}elay {O}verview",
url = "https://www.apple.com/icloud/docs/iCloud_Private_Relay_Overview_Dec2021.pdf",
note = "Retreived June, 2023",
}

@misc{devicecheck,
title = "Apple {D}evice{C}heck {F}ramework",
url = "https://developer.apple.com/documentation/devicecheck",
note = "Retreived June, 2023"
}

@misc{playintegrity,
title = "Google {P}lay {I}ntegrity {A}{P}{I}",
url="https://developer.android.com/google/play/integrity",
note = "Retreived June, 2023"}

@techreport{ietf-ppm-dap-04,
    number =    {draft-ietf-ppm-dap-04},
    type =      {Internet-Draft},
    institution =   {Internet Engineering Task Force},
    publisher = {Internet Engineering Task Force},
    note =      {Work in Progress},
    url =       {https://datatracker.ietf.org/doc/draft-ietf-ppm-dap/04/},
        author =    {Tim Geoghegan and Christopher Patton and Eric Rescorla and Christopher A. Wood},
    title =     {{Distributed Aggregation Protocol for Privacy Preserving Measurement}},
    pagetotal = 74,
    year =      2023,
    month =     mar,
    day =       13,
}

@techreport{ietf-privacypass-rate-limit-tokens-01,
    number =    {draft-ietf-privacypass-rate-limit-tokens-01},
    type =      {Internet-Draft},
    institution =   {Internet Engineering Task Force},
    publisher = {Internet Engineering Task Force},
    note =      {Work in Progress},
    url =       {https://datatracker.ietf.org/doc/draft-ietf-privacypass-rate-limit-tokens/01/},
        author =    {Scott Hendrickson and Jana Iyengar and Tommy Pauly and Steven Valdez and Christopher A. Wood},
    title =     {{Rate-Limited Token Issuance Protocol}},
    pagetotal = 49,
    year =      2023,
    month =     mar,
    day =       3,
}

@techreport{irtf-cfrg-vdaf-06,
    number =    {draft-irtf-cfrg-vdaf-06},
    type =      {Internet-Draft},
    institution =   {Internet Engineering Task Force},
    publisher = {Internet Engineering Task Force},
    note =      {Work in Progress},
    url =       {https://datatracker.ietf.org/doc/draft-irtf-cfrg-vdaf/06/},
        author =    {Richard Barnes and David Cook and Christopher Patton and Phillipp Schoppmann},
    title =     {{Verifiable Distributed Aggregation Functions}},
    pagetotal = 108,
    year =      2023,
    month =     jun,
    day =       15,
}

@techreport{ietf-ohttp,
    number =    {draft-ietf-ohai-ohttp-08},
    type =      {Internet-Draft},
    institution =   {Internet Engineering Task Force},
    publisher = {Internet Engineering Task Force},
    note =      {Work in Progress},
    url =       {https://ietf-wg-ohai.github.io/oblivious-http/draft-ietf-ohai-ohttp.html},
        author =    {Martin Thomson and Christopher A. Wood},
    title =     {{Oblivious HTTP}},
    year =      2023,
    month =     mar,
    day =       13,
}

@inproceedings{zhu2020federated,
  title={Federated heavy hitters discovery with differential privacy},
  author={Zhu, Wennan and Kairouz, Peter and McMahan, Brendan and Sun, Haicheng and Li, Wei},
  booktitle={International Conference on Artificial Intelligence and Statistics},
  pages={3837--3847},
  year={2020},
  organization={PMLR}
}

@inproceedings{cormode2022sample,
  title={Sample-and-threshold differential privacy: Histograms and applications},
  author={Cormode, Graham and Bharadwaj, Akash},
  booktitle={International Conference on Artificial Intelligence and Statistics},
  pages={1420--1431},
  year={2022},
  organization={PMLR}
}

@misc{FacebookURLs,
author = {Messing, Solomon and DeGregorio, Christina and Hillenbrand, Bennett and King, Gary and Mahanti, Saurav and Mukerjee, Zagreb and Nayak, Chaya and Persily, Nate and State, Bogdan and Wilkins, Arjun},
publisher = {Harvard Dataverse},
title = {{Facebook Privacy-Protected Full URLs Data Set}},
year = {2020},
version = {V10},
doi = {10.7910/DVN/TDOAPG},
url = {https://doi.org/10.7910/DVN/TDOAPG}
}

@misc{aktay2020google,
      title={Google COVID-19 Community Mobility Reports: Anonymization Process Description (version 1.1)},
      author={Ahmet Aktay and Shailesh Bavadekar and Gwen Cossoul and John Davis and Damien Desfontaines and Alex Fabrikant and Evgeniy Gabrilovich and Krishna Gadepalli and Bryant Gipson and Miguel Guevara and Chaitanya Kamath and Mansi Kansal and Ali Lange and Chinmoy Mandayam and Andrew Oplinger and Christopher Pluntke and Thomas Roessler and Arran Schlosberg and Tomer Shekel and Swapnil Vispute and Mia Vu and Gregory Wellenius and Brian Williams and Royce J Wilson},
      year={2020},
      eprint={2004.04145},
      archivePrefix={arXiv},
      primaryClass={cs.CR}
}

@article{google-mobility,
author={Aleix Bassolas and Hugo Barbosa-Filho and Brian Dickinson and Xerxes Dotiwalla and Paul Eastham and Riccardo Gallotti and Gourab Ghoshal and Bryant Gipson and Surendra A. Hazarie and Henry Kautz and Onur Kucuktunc and Allison Lieber and Adam Sadilek and José J. Ramasco},
title = {Hierarchical organization of urban mobility and its connection with city livability},
journal = {Nat Commun},
volume = {10},
number = {4817},
year = {2019},
}

@misc{bavadekar2021google,
      title={Google COVID-19 Vaccination Search Insights: Anonymization Process Description},
      author={Shailesh Bavadekar and Adam Boulanger and John Davis and Damien Desfontaines and Evgeniy Gabrilovich and Krishna Gadepalli and Badih Ghazi and Tague Griffith and Jai Gupta and Chaitanya Kamath and Dennis Kraft and Ravi Kumar and Akim Kumok and Yael Mayer and Pasin Manurangsi and Arti Patankar and Irippuge Milinda Perera and Chris Scott and Tomer Shekel and Benjamin Miller and Karen Smith and Charlotte Stanton and Mimi Sun and Mark Young and Gregory Wellenius},
      year={2021},
      eprint={2107.01179},
      archivePrefix={arXiv},
      primaryClass={cs.CR}
}

@misc{pereira2021us,
      title={U.S. Broadband Coverage Data Set: A Differentially Private Data Release},
      author={Mayana Pereira and Allen Kim and Joshua Allen and Kevin White and Juan Lavista Ferres and Rahul Dodhia},
      year={2021},
      eprint={2103.14035},
      archivePrefix={arXiv},
      primaryClass={cs.CR}
}

@misc{global_victim_perpetrator,
author = {Counter-Trafficking Data Collaborative (CTDC)},
title = {Global Victim-Perpetrator Synthetic Dataset},
url = {https://www.ctdatacollaborative.org/global-victim-perpetrator-synthetic-dataset},
year = {2022},
note = {Retreived June, 2023}
}

@misc{rogers2020members,
      title={A Members First Approach to Enabling LinkedIn's Labor Market Insights at Scale},
      author={Ryan Rogers and Adrian Rivera Cardoso and Koray Mancuhan and Akash Kaura and Nikhil Gahlawat and Neha Jain and Paul Ko and Parvez Ahammad},
      year={2020},
      eprint={2010.13981},
      archivePrefix={arXiv},
      primaryClass={cs.CR}
}

@misc{rogers2020linkedins,
      title={LinkedIn's Audience Engagements API: A Privacy Preserving Data Analytics System at Scale},
      author={Ryan Rogers and Subbu Subramaniam and Sean Peng and David Durfee and Seunghyun Lee and Santosh Kumar Kancha and Shraddha Sahay and Parvez Ahammad},
      year={2020},
      eprint={2002.05839},
      archivePrefix={arXiv},
      primaryClass={cs.CR}
}

@misc{pseo,
title = {"Post-Secondary Employment Outcomes},
author = {Andrew Foote and Joyce Key Hahn and Stephen Tibbets and Larry Warren},
url = {https://lehd.ces.census.gov/doc/PSEOTechnicalDocumentation.pdf},
year = {2023},
note = {Retreived June, 2023}
}

@misc{edp,
author = {Marc Paré and  Mariano Teehan and Stephen Suffian and
Joe Glass and Adam Scheer and McGee Young and Matt Golden},
title = {Applying Energy Differential Privacy To Enable Measurement of the OhmConnect Virtual Power Plant},
url = {https://edp.recurve.com},
note = {Retreived June 2023}
}

@INPROCEEDINGS{onthemap,
  author={Machanavajjhala, Ashwin and Kifer, Daniel and Abowd, John and Gehrke, Johannes and Vilhuber, Lars},
  booktitle={2008 IEEE 24th International Conference on Data Engineering},
  title={Privacy: Theory meets Practice on the Map},
  year={2008},
  volume={},
  number={},
  pages={277-286},
  doi={10.1109/ICDE.2008.4497436}}

@misc{abowd20222020,
      title={The 2020 Census Disclosure Avoidance System TopDown Algorithm},
      author={John M. Abowd and Robert Ashmead and Ryan Cumings-Menon and Simson Garfinkel and Micah Heineck and Christine Heiss and Robert Johns and Daniel Kifer and Philip Leclerc and Ashwin Machanavajjhala and Brett Moran and William Sexton and Matthew Spence and Pavel Zhuravlev},
      year={2022},
      eprint={2204.08986},
      archivePrefix={arXiv},
      primaryClass={cs.CR}
}

@misc{BonehBCGI23,
      author = {Dan Boneh and Elette Boyle and Henry Corrigan-Gibbs and Niv Gilboa and Yuval Ishai},
      title = {Arithmetic Sketching},
      howpublished = {Cryptology ePrint Archive, Paper 2023/1012},
      year = {2023},
      note = {\url{https://eprint.iacr.org/2023/1012}},
      url = {https://eprint.iacr.org/2023/1012}
}

@inproceedings {BulckMW+18,
author = {Jo Van Bulck and Marina Minkin and Ofir Weisse and Daniel Genkin and Baris Kasikci and Frank Piessens and Mark Silberstein and Thomas F. Wenisch and Yuval Yarom and Raoul Strackx},
title = {Foreshadow: Extracting the Keys to the Intel {SGX} Kingdom with Transient {Out-of-Order} Execution},
booktitle = {27th USENIX Security Symposium (USENIX Security 18)},
year = {2018},
isbn = {978-1-939133-04-5},
address = {Baltimore, MD},
pages = {991{\textendash}1008},
publisher = {USENIX Association},
}

@misc{xu2023federated,
      title={Federated Learning of Gboard Language Models with Differential Privacy},
      author={Zheng Xu and Yanxiang Zhang and Galen Andrew and Christopher A. Choquette-Choo and Peter Kairouz and H. Brendan McMahan and Jesse Rosenstock and Yuanbo Zhang},
      year={2023},
      eprint={2305.18465},
      archivePrefix={arXiv},
      primaryClass={cs.LG}
}

@misc{wang2023invariant,
      title={Invariant Aggregator for Defending against Federated Backdoor Attacks},
      author={Xiaoyang Wang and Dimitrios Dimitriadis and Sanmi Koyejo and Shruti Tople},
      year={2023},
      eprint={2210.01834},
      archivePrefix={arXiv},
      primaryClass={cs.LG}
}

@inproceedings {flame,
author = {Thien Duc Nguyen and Phillip Rieger and Huili Chen and Hossein Yalame and Helen M{\"o}llering and Hossein Fereidooni and Samuel Marchal and Markus Miettinen and Azalia Mirhoseini and Shaza Zeitouni and Farinaz Koushanfar and Ahmad-Reza Sadeghi and Thomas Schneider},
title = {{FLAME}: Taming Backdoors in Federated Learning},
booktitle = {31st USENIX Security Symposium (USENIX Security 22)},
year = {2022},
isbn = {978-1-939133-31-1},
address = {Boston, MA},
pages = {1415--1432},
publisher = {USENIX Association},
}

@inproceedings{WuW21,
 author = {Wu, Dongxian and Wang, Yisen},
 booktitle = {Advances in Neural Information Processing Systems},
 editor = {M. Ranzato and A. Beygelzimer and Y. Dauphin and P.S. Liang and J. Wortman Vaughan},
 pages = {16913--16925},
 publisher = {Curran Associates, Inc.},
 title = {Adversarial Neuron Pruning Purifies Backdoored Deep Models},
 volume = {34},
 year = {2021}
}

@InProceedings{BernsteinWAA18,
  title = 	 {sign{SGD}: Compressed Optimisation for Non-Convex Problems},
  author =       {Bernstein, Jeremy and Wang, Yu-Xiang and Azizzadenesheli, Kamyar and Anandkumar, Animashree},
  booktitle = 	 {Proceedings of the 35th International Conference on Machine Learning},
  pages = 	 {560--569},
  year = 	 {2018},
  editor = 	 {Dy, Jennifer and Krause, Andreas},
  volume = 	 {80},
  series = 	 {Proceedings of Machine Learning Research},
  month = 	 {10--15 Jul},
  publisher =    {PMLR},
}

@InProceedings{XieKG20,
author="Xie, Cong
and Koyejo, Oluwasanmi
and Gupta, Indranil",
editor="Brefeld, Ulf
and Fromont, Elisa
and Hotho, Andreas
and Knobbe, Arno
and Maathuis, Marloes
and Robardet, C{\'e}line",
title="{SLSGD}: Secure and Efficient Distributed On-device Machine Learning",
booktitle="Machine Learning and Knowledge Discovery in Databases",
year="2020",
publisher="Springer International Publishing",
address="Cham",
pages="213--228",
isbn="978-3-030-46147-8"
}

@INPROCEEDINGS{ShejwalkarHKR22,
  author={Shejwalkar, Virat and Houmansadr, Amir and Kairouz, Peter and Ramage, Daniel},
  booktitle={2022 IEEE Symposium on Security and Privacy (S \& P)},
  title={Back to the Drawing Board: A Critical Evaluation of Poisoning Attacks on Production Federated Learning},
  year={2022},
  volume={},
  number={},
  pages={1354-1371},
  doi={10.1109/SP46214.2022.9833647}}

@inproceedings{BlanchardEGS17,
 author = {Blanchard, Peva and El Mhamdi, El Mahdi and Guerraoui, Rachid and Stainer, Julien},
 booktitle = {Advances in Neural Information Processing Systems},
 editor = {I. Guyon and U. Von Luxburg and S. Bengio and H. Wallach and R. Fergus and S. Vishwanathan and R. Garnett},
 pages = {},
 publisher = {Curran Associates, Inc.},
 title = {Machine Learning with Adversaries: Byzantine Tolerant Gradient Descent},
  volume = {30},
 year = {2017}
}

@misc{oncological23,
title = {Collaborative Privacy-Preserving Analysis of Oncological Data},
author = {Zohar Duchin and Marcelo Blatt and Yuriy Polyakov},
howpublished = {\url{https://dualitytech.com/blog/collaborative-analysis-oncological-data/}},
note = {Retreived July 2023}
}

@article{Froelicher21,
author ={David Froelicher and Juan R Troncoso-Pastoriza  and Jean Louis Raisaro  and Michel A Cuendet  and Joao Sa Sousa  and Hyunghoon Cho  and Bonnie Berger  and Jacques Fellay  and Jean-Pierre Hubaux},
title = {Truly privacy-preserving federated analytics for precision medicine with multiparty homomorphic encryption},
journal ={Nat Commun.},
year = {2021},
volume = {12},
number = {1}
}

@article{ChoWB18,
title = {Secure genome-wide association analysis using multiparty computation},
author = {Hyunghoon Cho and David J Wu and Bonnie Berger},
journal = {Nature Biotechnology},
volume = {36},
pages ={547–-551},
year = {2018}
}

@article{BlattGPG20,
author = {Marcelo Blatt  and Alexander Gusev  and Yuriy Polyakov  and Shafi Goldwasser },
title = {Secure large-scale genome-wide association studies using homomorphic encryption},
journal = {Proceedings of the National Academy of Sciences},
volume = {117},
number = {21},
pages = {11608-11613},
year = {2020},
doi = {10.1073/pnas.1918257117},
URL = {https://www.pnas.org/doi/abs/10.1073/pnas.1918257117},
eprint = {https://www.pnas.org/doi/pdf/10.1073/pnas.1918257117},
}

@article{de2020SCRAM,
	author = {de Castro, Leo and Lo, Andrew W. and Reynolds, Taylor and Susan, Fransisca and Vaikuntanathan, Vinod and Weitzner, Daniel and Zhang, Nicolas},
	journal = {Harvard Data Science Review},
	number = {3},
	year = {2020},
	month = {sep 16},
	note = {https://hdsr.mitpress.mit.edu/pub/gylaxji4},
	publisher = {},
	title = {SCRAM: A {Platform} for {Securely} {Measuring} {Cyber} {Risk}},
	volume = {2},
}

@InProceedings{Abidin16,
author="Abidin, Aysajan
and Aly, Abdelrahaman
and Cleemput, Sara
and Mustafa, Mustafa A.",
editor="Foresti, Sara
and Persiano, Giuseppe",
title="An {M}{P}{C}-Based Privacy-Preserving Protocol for a Local Electricity Trading Market",
booktitle="Cryptology and Network Security",
year="2016",
publisher="Springer International Publishing",
address="Cham",
pages="615--625",
isbn="978-3-319-48965-0"
}

@inproceedings{Bogdanov16,
title = {Students and Taxes: a Privacy-Preserving Study Using Secure Computation},
author = {Dan Bogdanov and Liina Kamm and Baldur Kubo and Reimo Rebane and Ville Sokk and Riivo Talviste},
booktitle ={Proceedings on Privacy Enhancing Technologies},
volume = {2016},
number = {3},
pages= {117–-135},
}

@article{Liina13,
    author = {Kamm, Liina and Bogdanov, Dan and Laur, Sven and Vilo, Jaak},
    title = "{A new way to protect privacy in large-scale genome-wide association studies}",
    journal = {Bioinformatics},
    volume = {29},
    number = {7},
    pages = {886-893},
    year = {2013},
    month = {02},
    issn = {1367-4803},
    doi = {10.1093/bioinformatics/btt066},
}

@INPROCEEDINGS{Lapets16,
  author={Lapets, Andrei and Volgushev, Nikolaj and Bestavros, Azer and Jansen, Frederick and Varia, Mayank},
  booktitle={2016 IEEE Cybersecurity Development (SecDev)},
  title={Secure MPC for Analytics as a Web Application},
  year={2016},
  volume={},
  number={},
  pages={73-74},
  doi={10.1109/SecDev.2016.027}}

@inbook{Bogetoft09,
author = {Bogetoft, Peter and Christensen, Dan Lund and Damg\r{a}rd, Ivan and Geisler, Martin and Jakobsen, Thomas and Kr\o{}igaard, Mikkel and Nielsen, Janus Dam and Nielsen, Jesper Buus and Nielsen, Kurt and Pagter, Jakob and Schwartzbach, Michael and Toft, Tomas},
title = {Secure Multiparty Computation Goes Live},
year = {2009},
isbn = {9783642035487},
publisher = {Springer-Verlag},
address = {Berlin, Heidelberg},
booktitle = {Financial Cryptography and Data Security: 13th International Conference, FC 2009, Accra Beach, Barbados, February 23-26, 2009. Revised Selected Papers},
pages = {325–343},
numpages = {19}
}

@misc{rfc9298,
    series =    {Request for Comments},
    number =    9298,
    howpublished =  {RFC 9298},
    publisher = {RFC Editor},
    doi =       {10.17487/RFC9298},
    url =       {https://www.rfc-editor.org/info/rfc9298},
        author =    {David Schinazi},
    title =     {{Proxying UDP in HTTP}},
    pagetotal = 16,
    year =      2022,
    month =     aug,
}

@inproceedings{crypteps,
author = {Roy Chowdhury, Amrita and Wang, Chenghong and He, Xi and Machanavajjhala, Ashwin and Jha, Somesh},
title = {Crypt$\eps$: Crypto-Assisted Differential Privacy on Untrusted Servers}, year = {2020},
isbn = {9781450367356},
publisher = {Association for Computing Machinery},
address = {New York, NY, USA},
doi = {10.1145/3318464.3380596},
booktitle = {Proceedings of the 2020 ACM SIGMOD International Conference on Management of Data},
pages = {603–619},
numpages = {17},
location = {Portland, OR, USA},
series = {SIGMOD '20} }

@inproceedings{unlynx,
title = {{UnLynx}: A Decentralized System for Privacy-Conscious Data Sharing},
author={David Froelicher and Patricia Egger and João Sá Sousa and Jean Louis Raisaro and Zhicong Huang and Christian Mouchet and Bryan Ford and Jean-Pierre Hubaux},
volume ={2017},
pages = {232–-250},
booktitle ={Proceedings on Privacy Enhancing Technologies},
}

@inproceedings{DavidsonGSTV18,
title = {Privacy Pass: Bypassing Internet Challenges Anonymously},
author = {Alex Davidson and Ian Goldberg and Nick Sullivan and George Tankersley and Filippo Valsorda},
volume={2018},
pages={164–-180},
booktitle ={Proceedings on Privacy Enhancing Technologies},
}

@inproceedings {orchard,
author = {Edo Roth and Hengchu Zhang and Andreas Haeberlen and Benjamin C. Pierce},
title = {Orchard: Differentially Private Analytics at Scale},
booktitle = {14th USENIX Symposium on Operating Systems Design and Implementation (OSDI 20)},
year = {2020},
isbn = {978-1-939133-19-9},
pages = {1065--1081},
publisher = {USENIX Association},
month = nov,
}

@inproceedings{honeycrisp,
author = {Roth, Edo and Noble, Daniel and Falk, Brett Hemenway and Haeberlen, Andreas},
title = {Honeycrisp: Large-Scale Differentially Private Aggregation without a Trusted Core},
year = {2019},
isbn = {9781450368735},
publisher = {Association for Computing Machinery},
address = {New York, NY, USA},
doi = {10.1145/3341301.3359660},
booktitle = {Proceedings of the 27th ACM Symposium on Operating Systems Principles},
pages = {196–210},
numpages = {15},
location = {Huntsville, Ontario, Canada},
series = {SOSP '19}
}

@inproceedings{Wolinsky2012,
  title={Scalable Anonymous Group Communication in the Anytrust Model},
  author={David Wolinsky and Henry Corrigan-Gibbs and Bryan Ford and Aaron Johnson},
  year={2012},
  booktitle = {European Symposium on System Security (EuroSec)}
}

@misc{BenhamoudaRS23,
      author = {Fabrice Benhamouda and Mariana Raykova and Karn Seth},
      title = {Anonymous Counting Tokens},
      howpublished = {Cryptology ePrint Archive, Paper 2023/320},
      year = {2023},
      note = {\url{https://eprint.iacr.org/2023/320}},
      url = {https://eprint.iacr.org/2023/320}
}

@inproceedings{KreuterLOR20,
  title={Anonymous Tokens with Private Metadata Bit},
  publisher={Springer-Verlag},
  doi={10.1007/978-3-030-56784-2_11},
  author={Ben Kreuter and Tancrède Lepoint and Michele Orrù and Mariana P. Raykova},
  year=2020
}

@inproceedings{SildeS22,
author = {Silde, Tjerand and Strand, Martin},
title = {Anonymous Tokens With Public Metadata And Applications ToPrivate Contact Tracing},
year = {2022},
isbn = {978-3-031-18282-2},
publisher = {Springer-Verlag},
address = {Berlin, Heidelberg},
doi = {10.1007/978-3-031-18283-9_9},
booktitle = {Financial Cryptography and Data Security: 26th International Conference, FC 2022, Grenada, May 2–6, 2022, Revised Selected Papers},
pages = {179–199},
numpages = {21},
location = {Grenada, Grenada}
}

@misc{federated-personalization,
title = {Federated Evaluation and Tuning for On-Device Personalization: System Design and Applications},
author = {Matthias Paulik and Matt Seigel and Henry Mason and Dominic Telaar and Joris Kluivers and Rogier van Dalen and Chi Wai Lau and Luke Carlson and Filip Granqvist and Chris Vandevelde and Sudeep Agarwal and Julien Freudiger and Andrew Byde and Abhishek Bhowmick and Gaurav Kapoor and Si Beaumont and Áine Cahill and Dominic Hughes and Omid Javidbakht and Fei Dong and Rehan Rishi and Stanley Hung},
year = {2022},
URL = {https://arxiv.org/pdf/2102.08503.pdf}
}

@inproceedings{improving-on-device-speaker,
title = {Improving On-Device Speaker Verification Using Federated Learning With Privacy},
booktitle = {Interspeech},
author = {Filip Granqvist and Matt Seigel and Rogier van Dalen and Áine Cahill and Stephen Shum and Matthias Paulik},
year = {2020},
URL = {https://arxiv.org/pdf/2008.02651.pdf}
}

@inproceedings{
ChadhaCDFHJMT24,
title={Differentially Private Heavy Hitter Detection using Federated Analytics},
author={Karan Chadha and Junye Chen and John Duchi and Vitaly Feldman and Hanieh Hashemi and Omid Javidbakht and Audra McMillan and Kunal Talwar},
booktitle={2nd IEEE Conference on Secure and Trustworthy Machine Learning},
year={2024},
url={https://openreview.net/forum?id=1HF8tLztGX}
}

@misc{AsiFNNTZ24,
      title={Private Vector Mean Estimation in the Shuffle Model: Optimal Rates Require Many Messages}, 
      author={Hilal Asi and Vitaly Feldman and Jelani Nelson and Huy L. Nguyen and Kunal Talwar and Samson Zhou},
      year={2024},
      eprint={2404.10201},
      archivePrefix={arXiv},
      primaryClass={cs.DS}
}

@misc{EichnerRBH+24,
      title={Confidential Federated Computations}, 
      author={Hubert Eichner and Daniel Ramage and Kallista Bonawitz and Dzmitry Huba and Tiziano Santoro and Brett McLarnon and Timon Van Overveldt and Nova Fallen and Peter Kairouz and Albert Cheu and Katharine Daly and Adria Gascon and Marco Gruteser and Brendan McMahan},
      year={2024},
      eprint={2404.10764},
      archivePrefix={arXiv},
      primaryClass={cs.CR}
}

@misc{JinCMRYO2024,
      title={Elephants Do Not Forget: Differential Privacy with State Continuity for Privacy Budget}, 
      author={Jiankai Jin and Chitchanok Chuengsatiansup and Toby Murray and Benjamin I. P. Rubinstein and Yuval Yarom and Olga Ohrimenko},
      year={2024},
      eprint={2401.17628},
      archivePrefix={arXiv},
      primaryClass={cs.CR}
}

@InProceedings{BartheO13,
author="Barthe, Gilles
and Olmedo, Federico",
editor="Fomin, Fedor V.
and Freivalds, R{\={u}}si{\c{n}}{\v{s}}
and Kwiatkowska, Marta
and Peleg, David",
title="Beyond Differential Privacy: Composition Theorems and Relational Logic for f-divergences between Probabilistic Programs",
booktitle="Automata, Languages, and Programming",
year="2013",
publisher="Springer Berlin Heidelberg",
address="Berlin, Heidelberg",
pages="49--60"
}

@misc{pat-cloudflare,
author = {Reid Tatoris and Maxime Guerreiro},
title = {Private Access Tokens: eliminating CAPTCHAs on iPhones and Macs with open standards},
url = {https://blog.cloudflare.com/eliminating-captchas-on-iphones-and-macs-using-new-standard/},
Note = {Blog Post, 2022},
}

@misc{pat-fastly,
author = {Jana Iyengar and Jonathan Foot},
title = {Private Access Tokens: stepping into the privacy-respecting, CAPTCHA-less future we were promised},
url = {https://www.fastly.com/blog/private-access-tokens-stepping-into-the-privacy-respecting-captcha-less},
note = {Blog Post, 2022}
}

@inproceedings{
AzamPFTSL23,
title={Federated Learning for Speech Recognition: Revisiting Current Trends Towards Large-Scale {ASR}},
author={Sheikh Shams Azam and Martin Pelikan and Vitaly Feldman and Kunal Talwar and Jan Silovsky and Tatiana Likhomanenko},
booktitle={International Workshop on Federated Learning in the Age of Foundation Models in Conjunction with NeurIPS 2023},
year={2023},
url={https://openreview.net/forum?id=ozN92d7CHX}
}

@inproceedings{BalleBGN20,
  author       = {Borja Balle and
                  James Bell and
                  Adri{\`{a}} Gasc{\'{o}}n and
                  Kobbi Nissim},
  title        = {Private Summation in the Multi-Message Shuffle Model},
  booktitle    = {{CCS} '20: 2020 {ACM} {SIGSAC} Conference on Computer and Communications Security},
  pages        = {657--676},
  year         = {2020}
}

@inproceedings{GhaziMPV20,
  author       = {Badih Ghazi and
                  Pasin Manurangsi and
                  Rasmus Pagh and
                  Ameya Velingker},
  title        = {Private Aggregation from Fewer Anonymous Messages},
  booktitle    = {Advances in Cryptology - {EUROCRYPT} 2020 - 39th Annual International Conference on the Theory and Applications of Cryptographic Techniques, Proceedings, Part {II}},
  pages        = {798--827},
  year         = {2020}
}

@inproceedings{GhaziKMPS21,
  author       = {Badih Ghazi and
                  Ravi Kumar and
                  Pasin Manurangsi and
                  Rasmus Pagh and
                  Amer Sinha},
  title        = {Differentially Private Aggregation in the Shuffle Model: Almost Central
                  Accuracy in Almost a Single Message},
  booktitle    = {Proceedings of the 38th International Conference on Machine Learning,
                  {ICML}},
  pages        = {3692--3701},
  year         = {2021}
}

@inproceedings{PasquiniFA22,
  author       = {Dario Pasquini and
                  Danilo Francati and
                  Giuseppe Ateniese},
  editor       = {Heng Yin and
                  Angelos Stavrou and
                  Cas Cremers and
                  Elaine Shi},
  title        = {Eluding Secure Aggregation in Federated Learning via Model Inconsistency},
  booktitle    = {Proceedings of the 2022 {ACM} {SIGSAC} Conference on Computer and
                  Communications Security, {CCS} 2022, Los Angeles, CA, USA, November
                  7-11, 2022},
  pages        = {2429--2443},
  publisher    = {{ACM}},
  year         = {2022},
  url          = {https://doi.org/10.1145/3548606.3560557},
  doi          = {10.1145/3548606.3560557},
  bibsource    = {dblp computer science bibliography, https://dblp.org}
}

@INPROCEEDINGS{MelisSCS19,
  author={Melis, Luca and Song, Congzheng and De Cristofaro, Emiliano and Shmatikov, Vitaly},
  booktitle={2019 IEEE Symposium on Security and Privacy (SP)}, 
  title={Exploiting Unintended Feature Leakage in Collaborative Learning}, 
  year={2019},
  volume={},
  number={},
  pages={691-706},
  doi={10.1109/SP.2019.00029}}

@inproceedings{FowlGCGG22,
title={Robbing the Fed:  Directly Obtaining Private Data in Federated Learning with Modified Models},
author={Liam H Fowl and Jonas Geiping and Wojciech Czaja and Micah Goldblum and Tom Goldstein},
booktitle={International Conference on Learning Representations},
year={2022},
url={https://openreview.net/forum?id=fwzUgo0FM9v}
}

@inproceedings{GeipingBDM20,
author = {Geiping, Jonas and Bauermeister, Hartmut and Dr\"{o}ge, Hannah and Moeller, Michael},
title = {Inverting gradients - how easy is it to break privacy in federated learning?},
year = {2020},
isbn = {9781713829546},
publisher = {Curran Associates Inc.},
address = {Red Hook, NY, USA},
booktitle = {Proceedings of the 34th International Conference on Neural Information Processing Systems},
articleno = {1421},
numpages = {11},
location = {Vancouver, BC, Canada},
series = {NIPS '20}
}

@inproceedings{WangSZSWQ19,
author = {Wang, Zhibo and Song, Mengkai and Zhang, Zhifei and Song, Yang and Wang, Qian and Qi, Hairong},
title = {Beyond Inferring Class Representatives: User-Level Privacy Leakage From Federated Learning},
year = {2019},
publisher = {IEEE Press},
url = {https://doi.org/10.1109/INFOCOM.2019.8737416},
doi = {10.1109/INFOCOM.2019.8737416},
booktitle = {IEEE INFOCOM 2019 - IEEE Conference on Computer Communications},
pages = {2512–2520},
numpages = {9},
location = {Paris, France}
}

@INPROCEEDINGS{YinMVAKM21,
  author={Yin, Hongxu and Mallya, Arun and Vahdat, Arash and Alvarez, Jose M. and Kautz, Jan and Molchanov, Pavlo},
  booktitle={2021 IEEE/CVF Conference on Computer Vision and Pattern Recognition (CVPR)}, 
  title={See through Gradients: Image Batch Recovery via GradInversion}, 
  year={2021},
  volume={},
  number={},
  pages={16332-16341},
  doi={10.1109/CVPR46437.2021.01607}}

@article{ZhaoMB20,
  author       = {Bo Zhao and
                  Konda Reddy Mopuri and
                  Hakan Bilen},
  title        = {iDLG: Improved Deep Leakage from Gradients},
  journal      = {CoRR},
  volume       = {abs/2001.02610},
  year         = {2020},
  url          = {http://arxiv.org/abs/2001.02610},
  eprinttype    = {arXiv},
  eprint       = {2001.02610},
  bibsource    = {dblp computer science bibliography, https://dblp.org}
}

@inproceedings{ZhuLH19,
 author = {Zhu, Ligeng and Liu, Zhijian and Han, Song},
 booktitle = {Advances in Neural Information Processing Systems},
 editor = {H. Wallach and H. Larochelle and A. Beygelzimer and F. d\textquotesingle Alch\'{e}-Buc and E. Fox and R. Garnett},
 pages = {},
 publisher = {Curran Associates, Inc.},
 title = {Deep Leakage from Gradients},
 url = {https://proceedings.neurips.cc/paper_files/paper/2019/file/60a6c4002cc7b29142def8871531281a-Paper.pdf},
 volume = {32},
 year = {2019}
}

@InProceedings{WenGFGG22,
  title = 	 {Fishing for User Data in Large-Batch Federated Learning via Gradient Magnification},
  author =       {Wen, Yuxin and Geiping, Jonas A. and Fowl, Liam and Goldblum, Micah and Goldstein, Tom},
  booktitle = 	 {Proceedings of the 39th International Conference on Machine Learning},
  pages = 	 {23668--23684},
  year = 	 {2022},
  editor = 	 {Chaudhuri, Kamalika and Jegelka, Stefanie and Song, Le and Szepesvari, Csaba and Niu, Gang and Sabato, Sivan},
  volume = 	 {162},
  series = 	 {Proceedings of Machine Learning Research},
  month = 	 {17--23 Jul},
  publisher =    {PMLR},
  url = 	 {https://proceedings.mlr.press/v162/wen22a.html},
}
 \fi
 \appendix

\section{Privacy Amplification by Randomized Donation Time}

In Section~\ref{subsampling}, we discussed how sampling a subset of the available users at any time provides additional privacy by giving each user plausible deniability about whether their data was included in the computation. In this section, we will discuss how added uncertainty in when a user participates in a computation can be used to further amplify privacy guarantees in iterative algorithms like DP-SGD.
Devices typically need to satisfy specific conditions in order to consider participating, for example a device may need to charging. Since there is often uncertainty in when any particular device becomes available for participation, there is uncertainty in donation time inherent in the system. Relying on this randomness to provide formal privacy guarantees would require careful modeling of this randomness. However,
we can also enable devices to deliberately (and according to a specific distribution) randomize the times at which they consider participating.
In this section, we will discuss how a specific algorithm for randomizing donation time can be used to amplify privacy guarantees in iterative algorithms like DP-SGD.

Suppose that an analyst is running an iterative algorithm in the following manner. When a device becomes available, it chooses whether or not to participate by flipping a biased coin and participating with probability $q$. If the device chooses to participate it then chooses an amount of time to wait before participating. When the wait time is complete, if the device is available, it downloads the recipe for the currently occurring round of the iterative algorithm and sends a privatized report to the aggregator. The aggregator aggregates the local reports for some amount of time and sends the aggregate to the data analyst. The analyst then begins the next round by selecting the next round's query. This process continues until the iterative algorithm is complete. By randomizing the wait time, we can inject uncertainty into which iteration the device participates in, and use this uncertainty to amplify the privacy guarantee. However, selecting a wait time that is greater than 0 runs the risk of the device not being available at the subsequent donation time, increasing latency and/or reducing utility. Thus, we need to carefully balance any privacy amplification with latency/utility concerns.

The key intuition we will use for defining an algorithm for choosing the wait time is that we want to build uncertainty into which iteration a device contributes to. Ideally, we would have devices randomly select an iteration to participate in, but this is difficult to implement since iterations may last for variable lengths of time making it difficult for devices to predict when a specific iteration may occur. Our algorithm uses parameters $K>0$ and $m\in\mathbb{N}$. For all $i\in\{0,\cdots,m-1\}$, the algorithm samples from a Bernoulli random variable $\chi_i$ with bias $1/m$. If $\max\chi_i=0$, then the device does not participate. Otherwise, the wait time is set to set to $K\cdot\arg\min\{i\;|\; \chi_i=1\}$. If we can ensure each iteration lasts for at \emph{most} $K$ minutes, then this algorithm either selects not to participate, or selects randomly among $m$ future iterations. In order to formalize the privacy amplification of this mechanism, we need one additional functionality of our aggregator. In addition to implementing the \SAA\ primitive, we will require that the aggregator does not release the result if the length of collection time is longer than $K$ minutes.

\begin{align*}
    \agg_{\minbatch, K}^{\Dec}&(t_0, \{(m_1, t_1), \ldots, (m_k, t_k)\}) \\&= \left\{\begin{array}{ll} \sum_{i=1}^k Dec(m_i) & \mbox{if } k \geq \minbatch \text{ and } t_k\in[t_0,t_0+K]\\ \bot &\mbox{otherwise} \end{array} \right.
\end{align*}

Suppose each device participates at most once in the iterative algorithm and each round is $(\eps, \delta)$-DP in the \SAA\ model. If this algorithm is run with the above randomized donation time algorithm in the \SAA\ model equipped with an aggregator with this additional functionality then it is $(\epsilon', \delta')$-DP where
\[\epsilon'=O\left(\epsilon\sqrt{\frac{\log(1/\delta)}{m}}\right) \text{ and } \delta'=O(m\delta).\]

To see this, note that the privacy loss of this algorithm is upper bounded by the privacy loss of the related algorithm where for all $i\in\{0,\cdots, m-1\}$, if $\chi_i=1$, then the device participates in the round occurring after a wait time of $i\cdot K$. For each round, each device is either not eligible, or participates with probability $1/m$. Thus, using Lemma~\ref{samplingWOR}, we can see that each round is $(\log(1+\frac{1}{m}(e^{\epsilon}-1)), \delta/m)$-DP. Since each device only considers participating in at most $m$ rounds, we can obtain the final privacy guarantee by using the advanced composition theorem \cite{DRV10} over $m$ rounds, so \[\epsilon' \le \log\left(1+\frac{1}{m}(e^{\epsilon}-1)\right)\sqrt{2\log(1/\delta)m}+\frac{1}{2}m \log^2\left(1+\frac{1}{m}(e^{\epsilon}-1)\right) \]\[ \text{ and } \delta'\le (m+1)\delta.\]

We note that for small $\eps, \frac{1}{m}$,  this $\eps' \approx \eps \sqrt{\frac{2\log 1/\delta}{m}}$, and the bounds can be improved by using numerical accounting techniques based on Renyi DP.

To maximize utility and minimize latency, the constant $K$ should be chosen so that most iterations are expected to receive enough donations within $K$ minutes. The constant $m$ should be chosen so that the device is likely to be available at the assigned donation time. This may be achieved by choosing $m$ small enough that $m\cdot K$ is a lower bound on the length of time that devices typically satisfy the conditions required in order for them to be available.

\section{Additional Experiments}
\label{sec:additional_exp}
\begin{figure*}[h]
  \centering
  \includegraphics[width=0.3\linewidth]{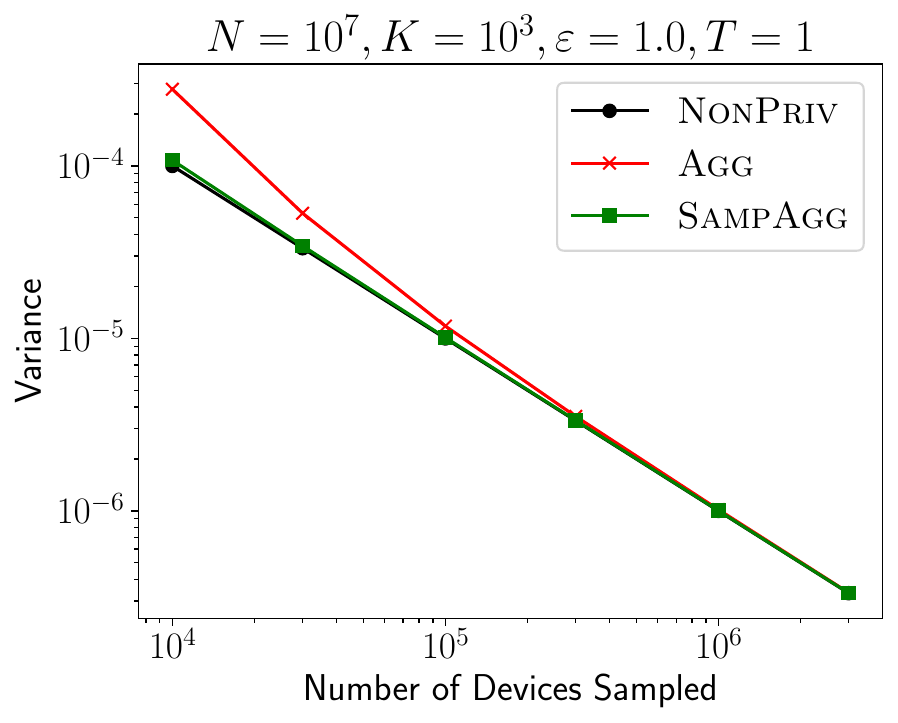}
  \includegraphics[width=0.3\linewidth]{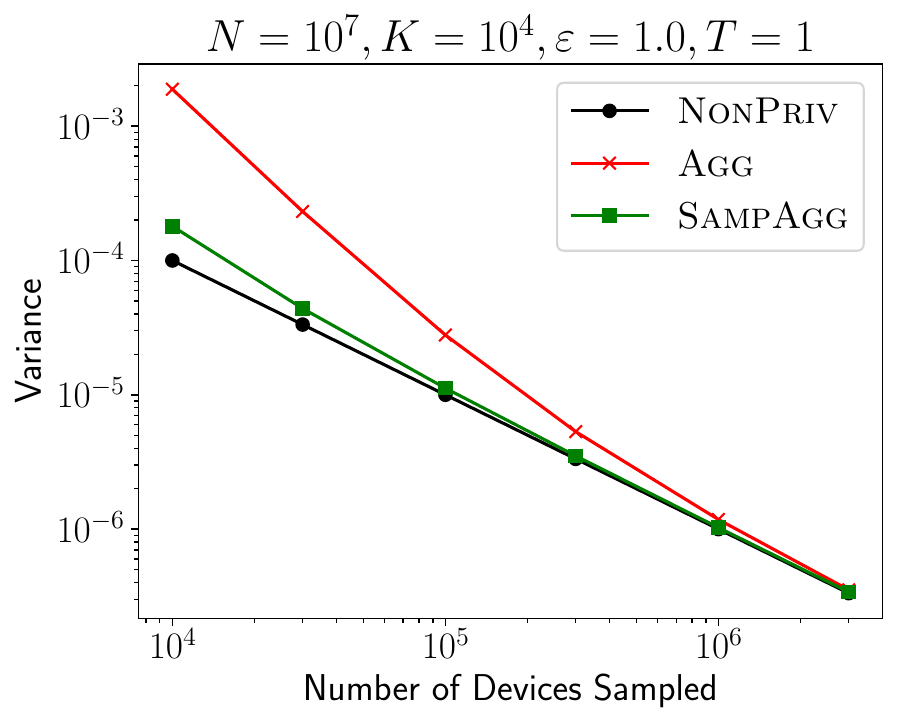}
  \includegraphics[width=0.3\linewidth]{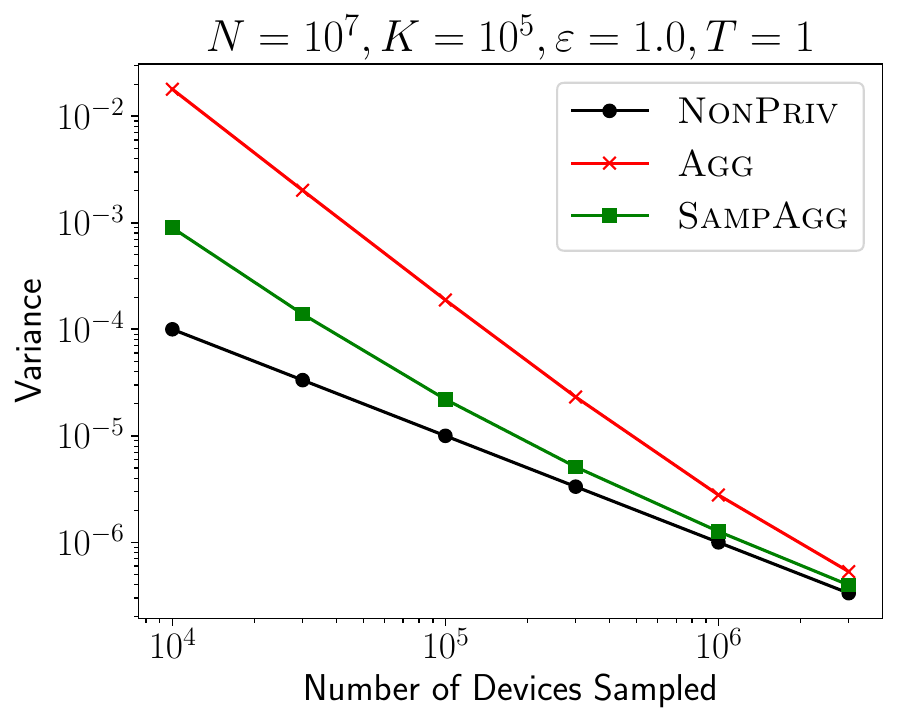}
  \includegraphics[width=0.3\linewidth]{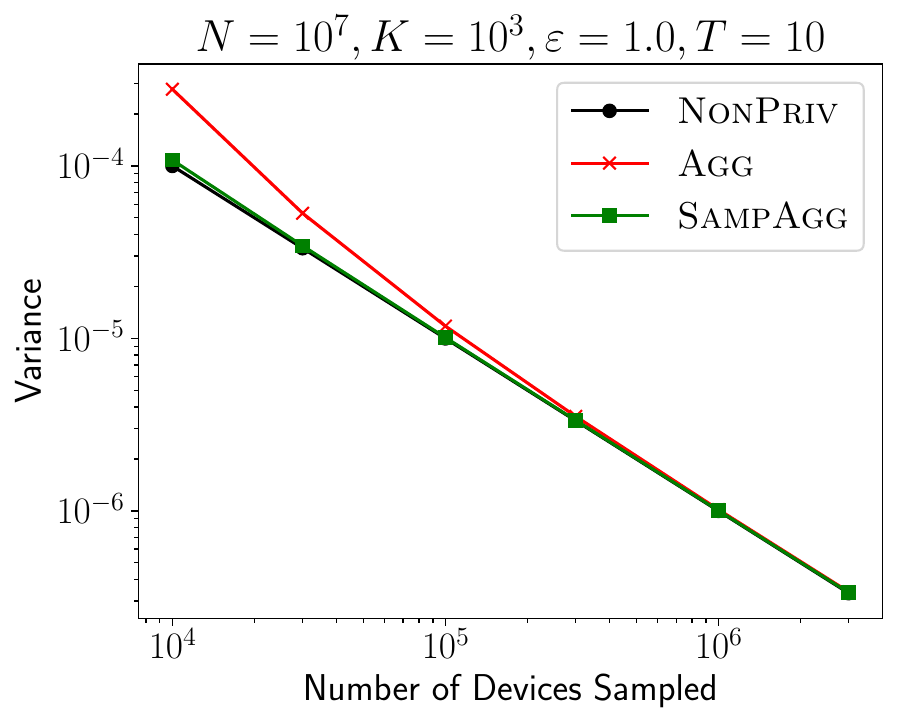}
  \includegraphics[width=0.3\linewidth]{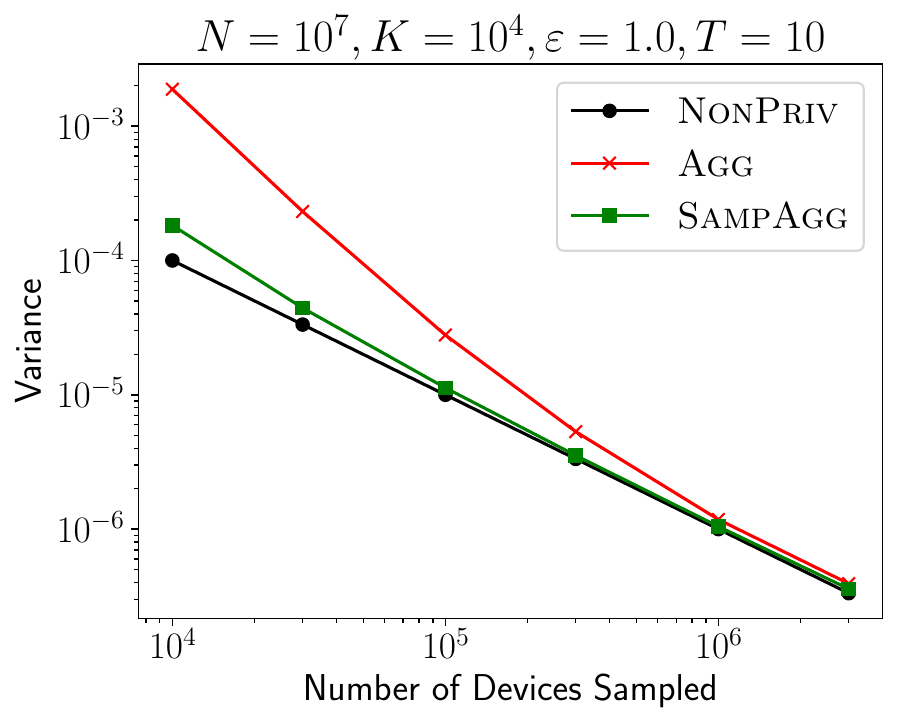}
  \includegraphics[width=0.3\linewidth]{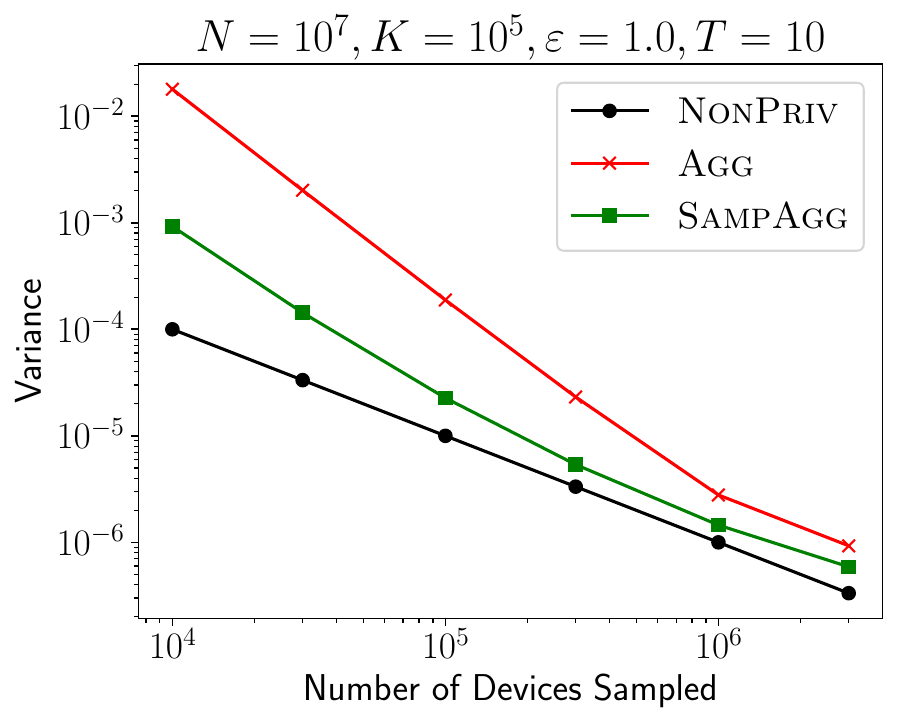}
  \includegraphics[width=0.3\linewidth]{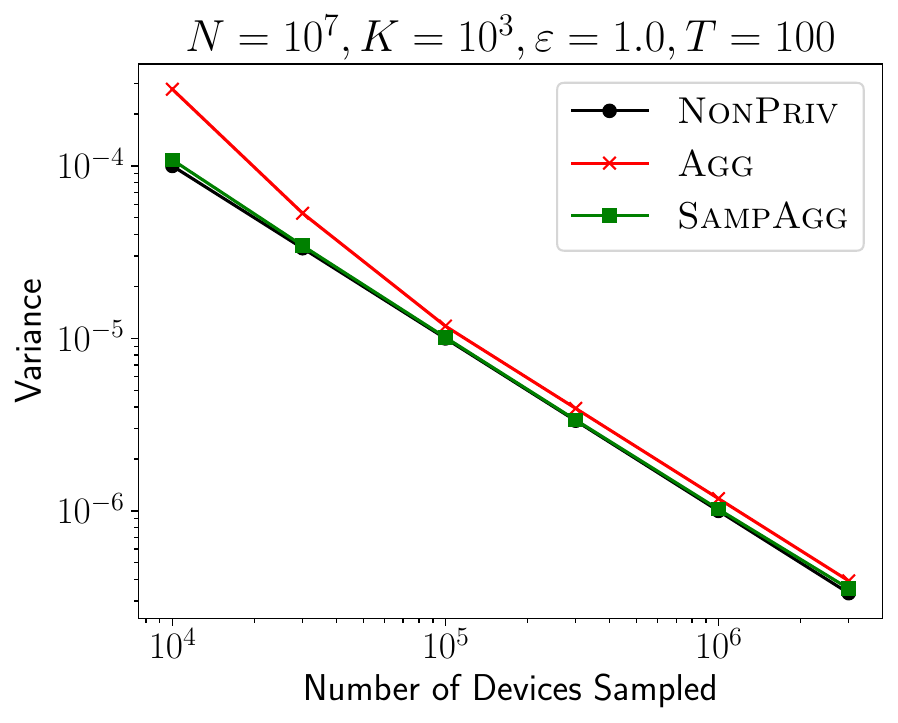}
  \includegraphics[width=0.3\linewidth]{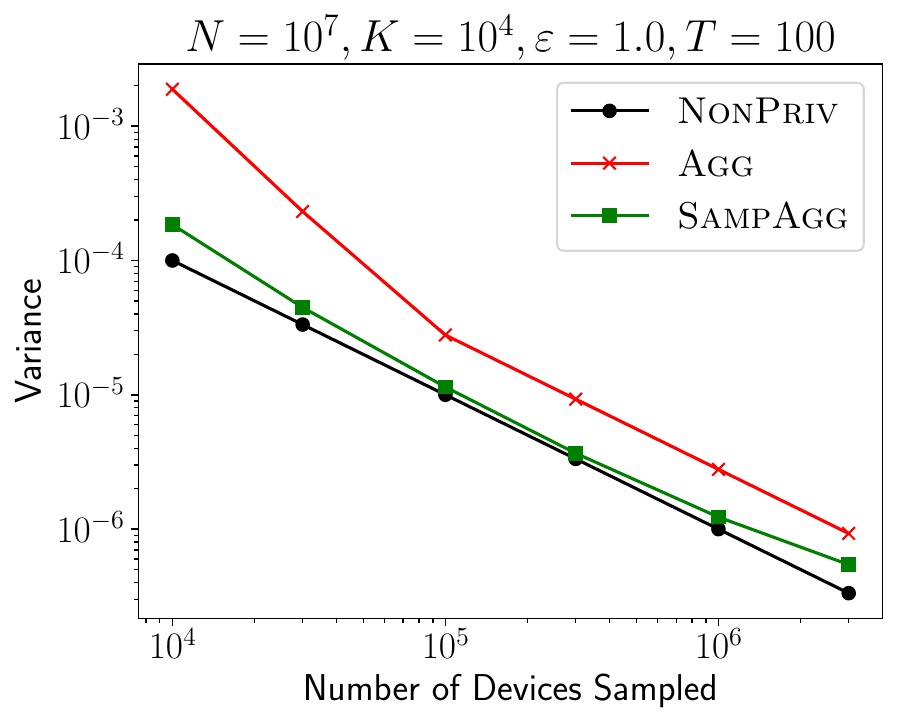}
  \includegraphics[width=0.3\linewidth]{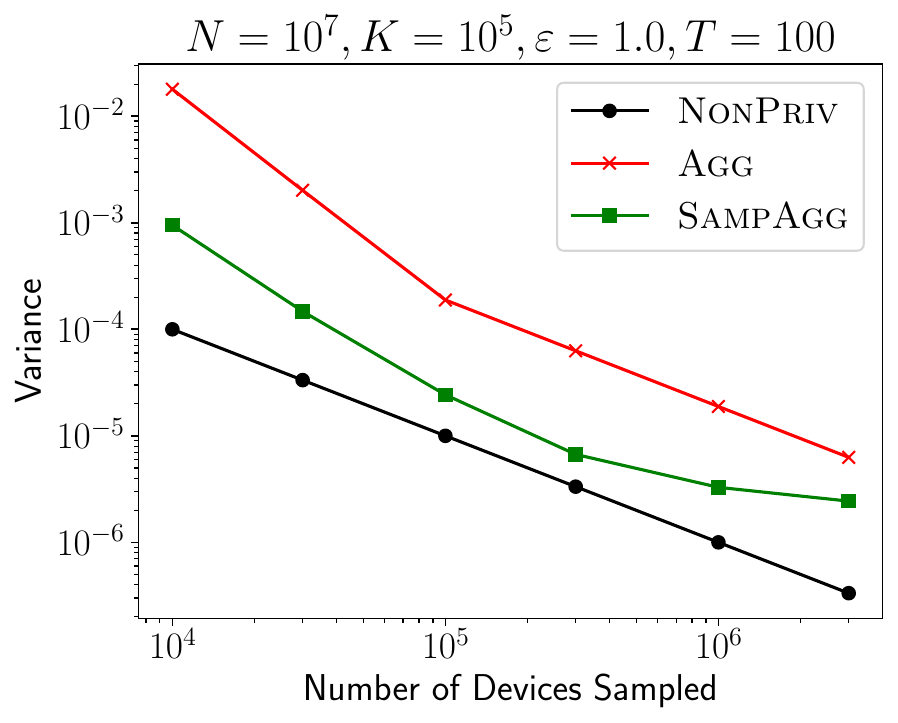}
  \includegraphics[width=0.3\linewidth]{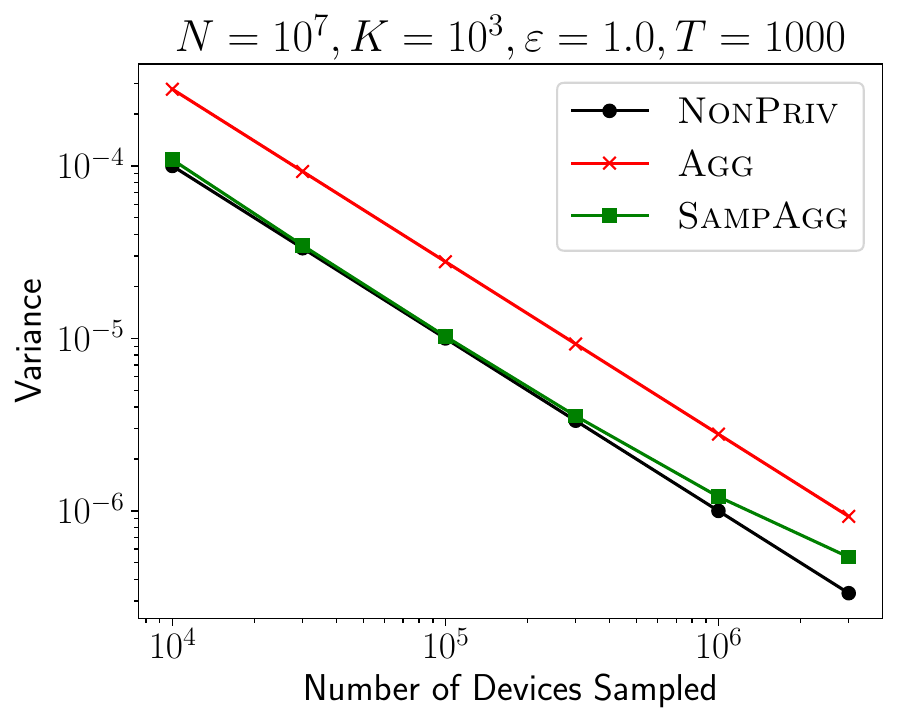}
  \includegraphics[width=0.3\linewidth]{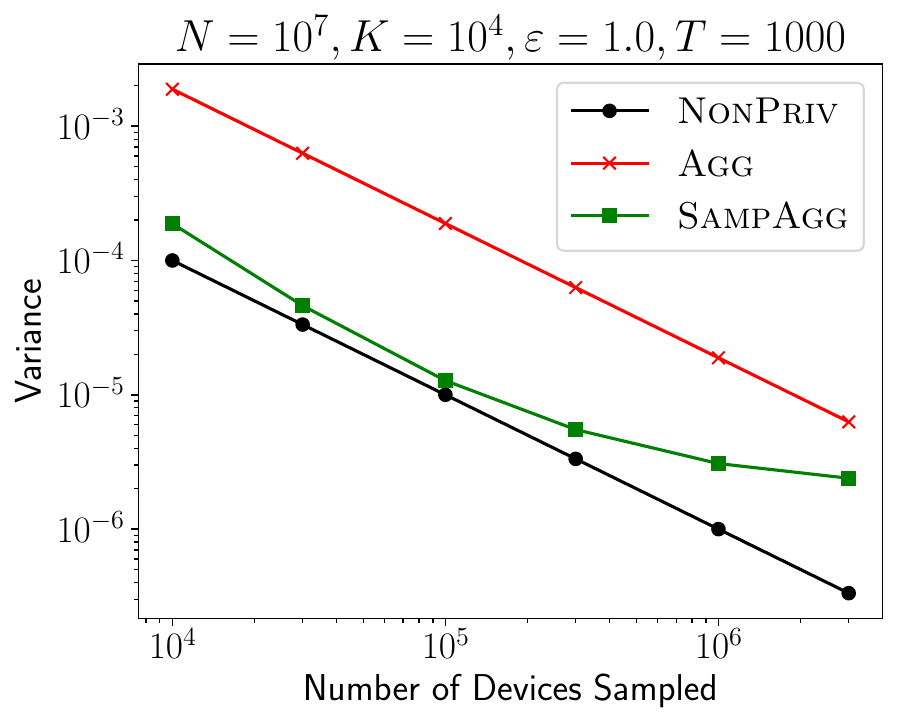}
  \includegraphics[width=0.3\linewidth]{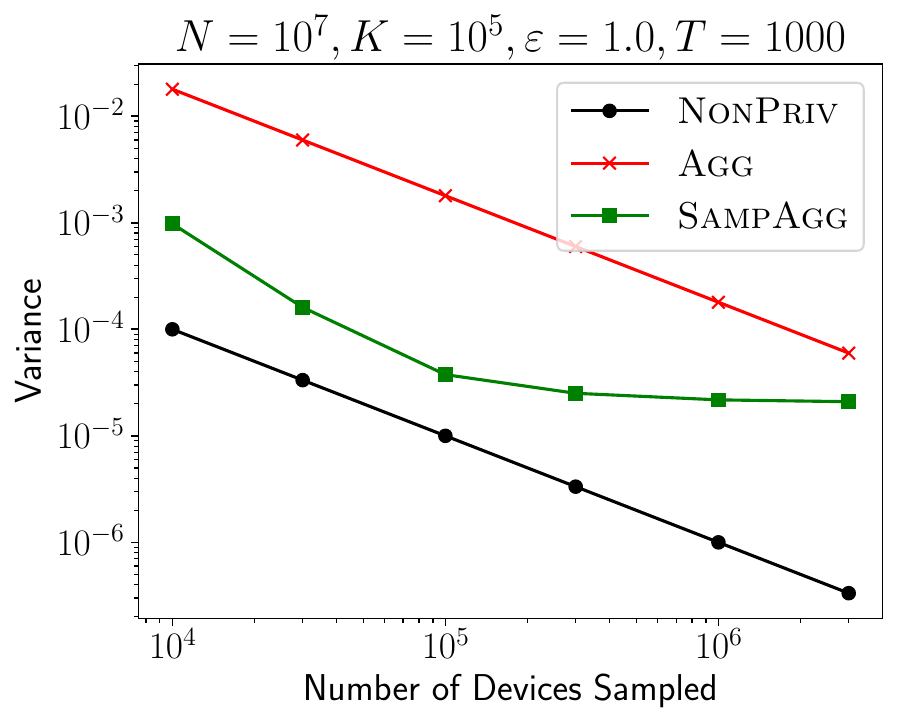}
  \caption{Expected Squared Error of a non-private baseline ({\sc NonPriv}), Aggregation model ({\sc Agg}) and Samplable Aggregation ({\sc SampAgg}) on a histogram task, for varying values of vocabulary size $K$ and tasks $T$ for a fixed total privacy budget.}\label{fig:hist-vary-K-T}
  \Description{Plots showing the benefits of Samplable Aggregation on Histograms.}
\end{figure*}

In this section, we present additional experiments that vary parameters. For the case of histogram, the relevant parameters are $K$ and $T$. The population size $N$ has no impact on the calculation, as long as it is at least $M$. We have considered a large range for $M$ in our plots. We show all results for $\varepsilon=1.0$ and $\delta=1e-6$. Our results in \cref{fig:hist-vary-K-T} show a range of vocabulary sizes $K$ (varying left to right from 1000 to 100,000), and number of tasks solved $T$ (varying top to bottom from $T=1$ to $T=1000$). As we increase $K$, the statistical squared error of $\approx 1/M$ is unaffected, whereas the error due to privacy noise increases as $K\sigma^2 / M^2$.
As we increase $T$, the privacy budget available per task decreases as $\approx 1/\sqrt{T}$, thus leading $\sigma \approx \sqrt{\log M\frac 1 \delta} / N\eps_{\mbox{task}} \approx \sqrt{MT\log \frac 1 \delta}{N}$. The precise value of $\sigma$ is computed using tight numerical accounting tools, as these formulae are only valid for a certain range of parameters. Our plots show that the improvement over an aggregator alone continues to hold over a large range of parameter and is specially pronounced for large $T$ and $K$.

\begin{figure*}[h]
  \centering
  \includegraphics[width=0.3\linewidth]{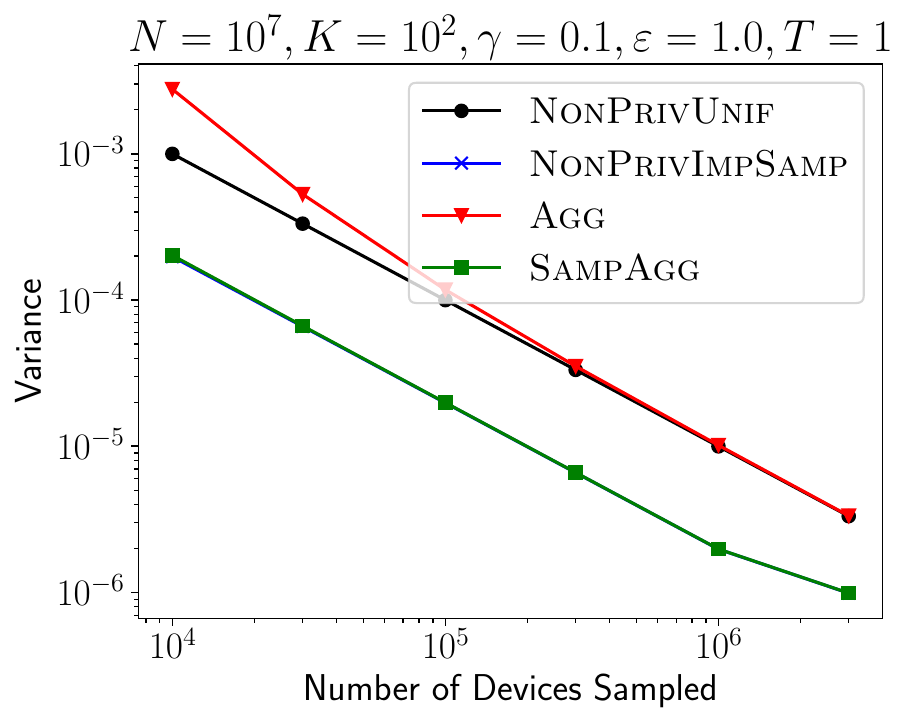}
  \includegraphics[width=0.3\linewidth]{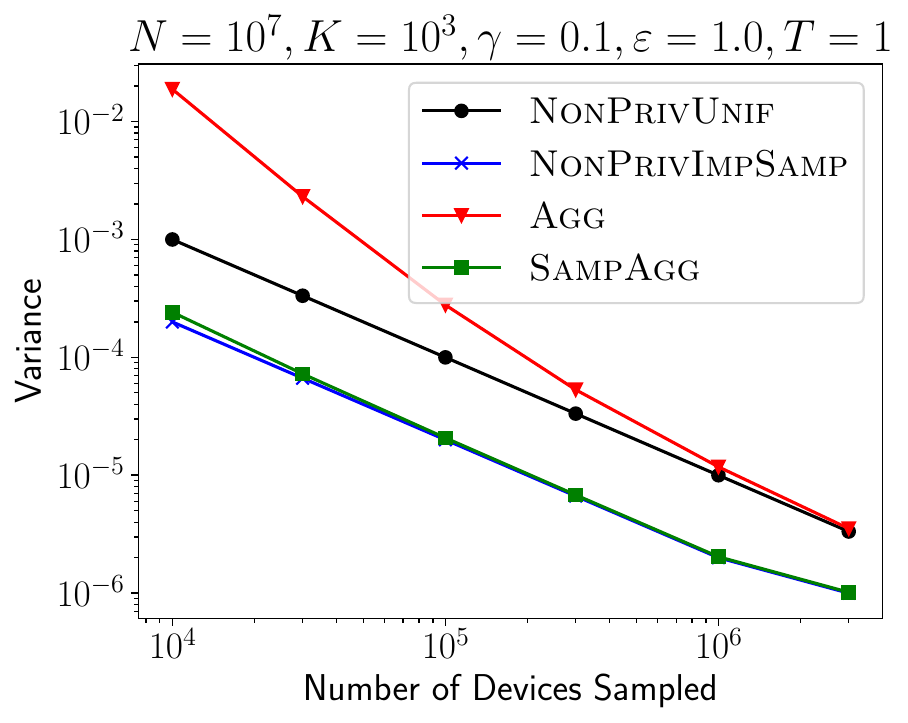}
  \includegraphics[width=0.3\linewidth]{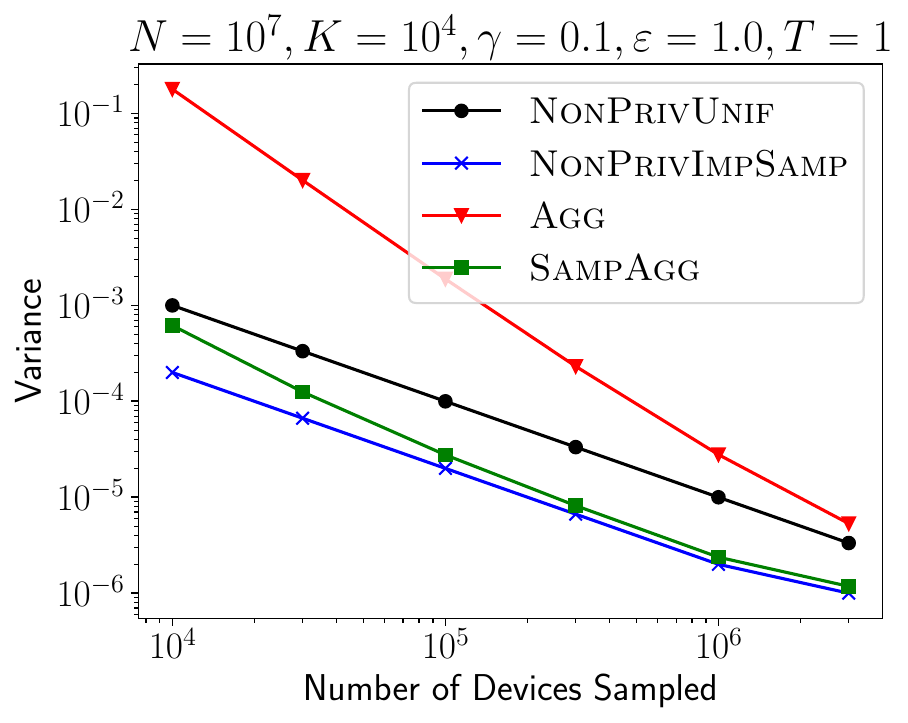}
  \includegraphics[width=0.3\linewidth]{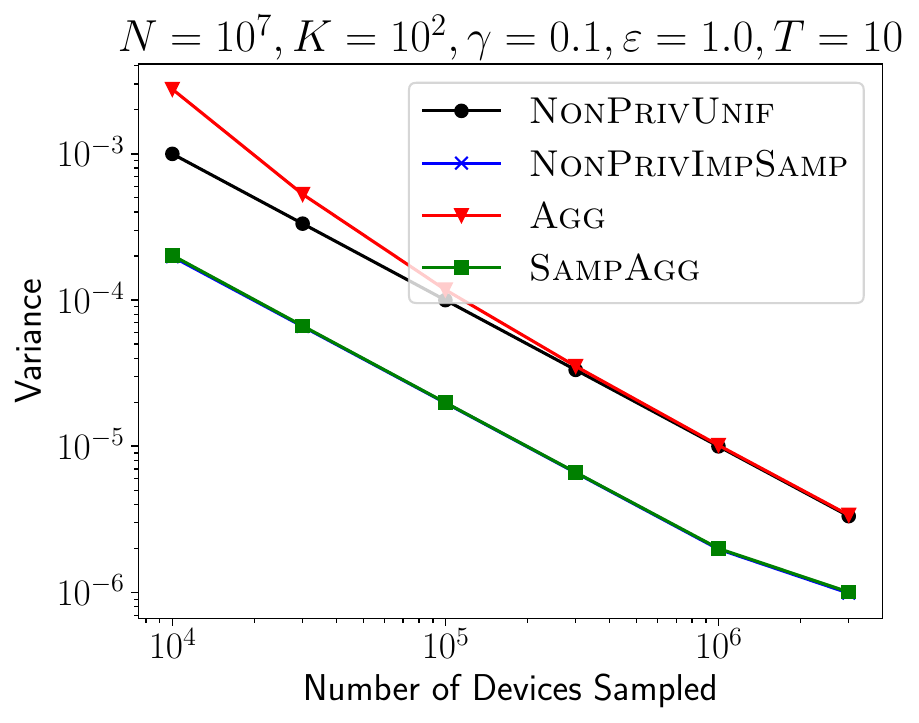}
  \includegraphics[width=0.3\linewidth]{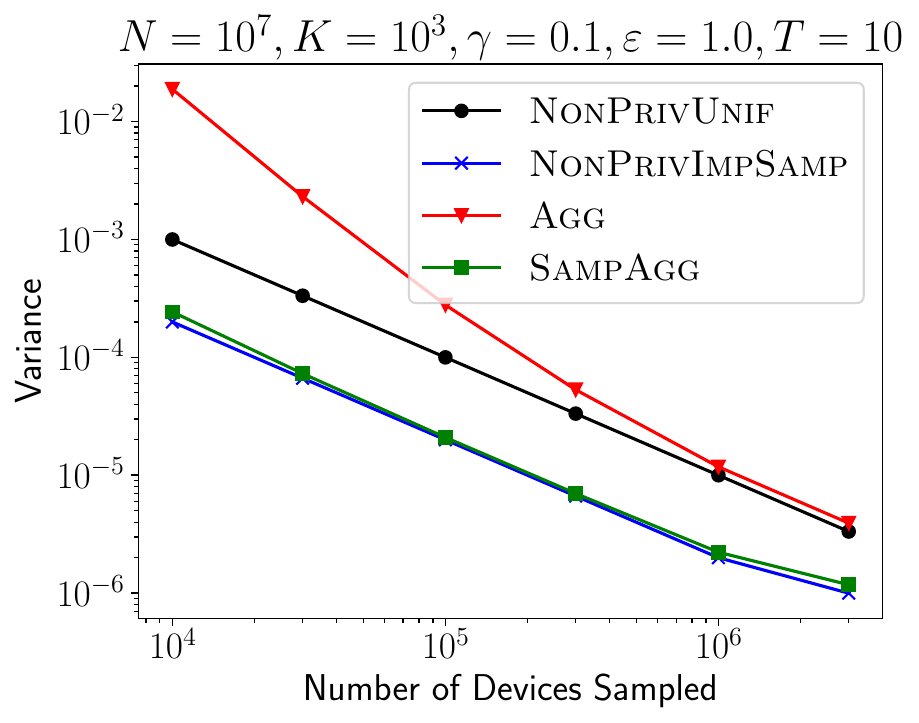}
  \includegraphics[width=0.3\linewidth]{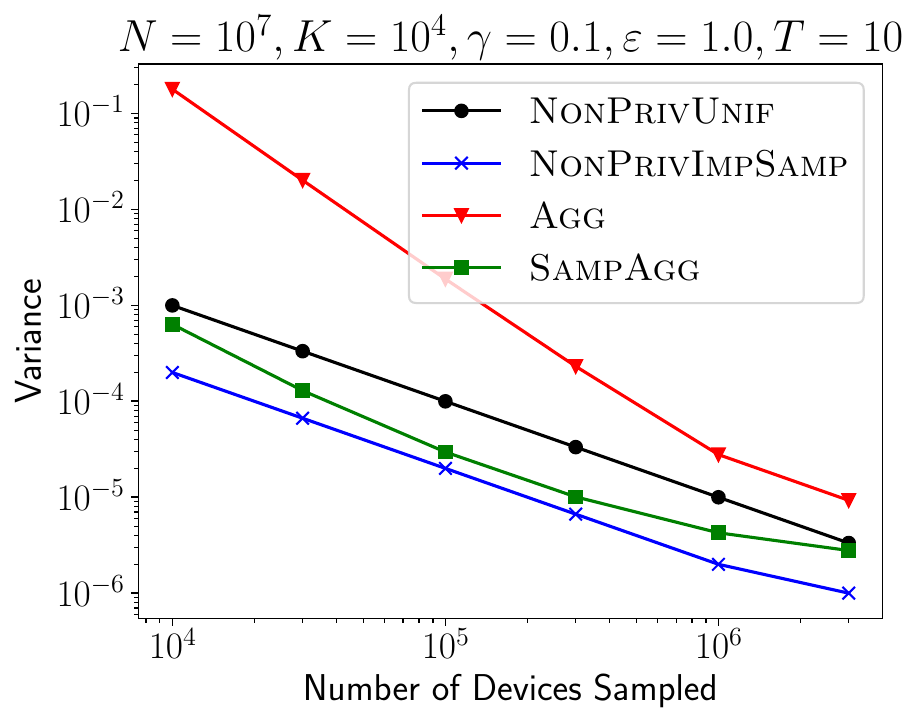}
  \includegraphics[width=0.3\linewidth]{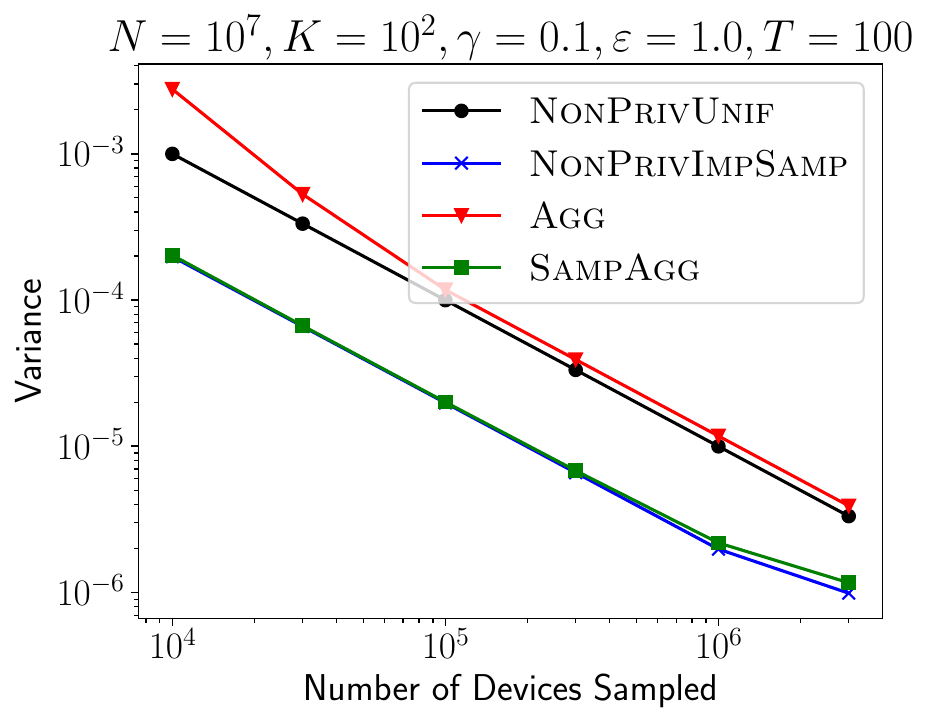}
  \includegraphics[width=0.3\linewidth]{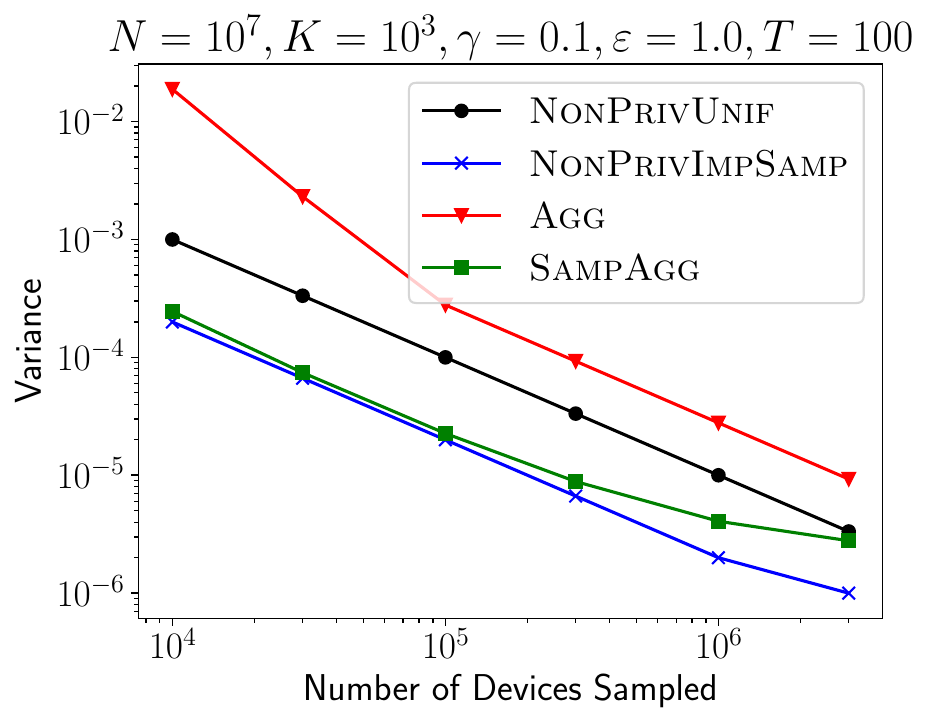}
  \includegraphics[width=0.3\linewidth]{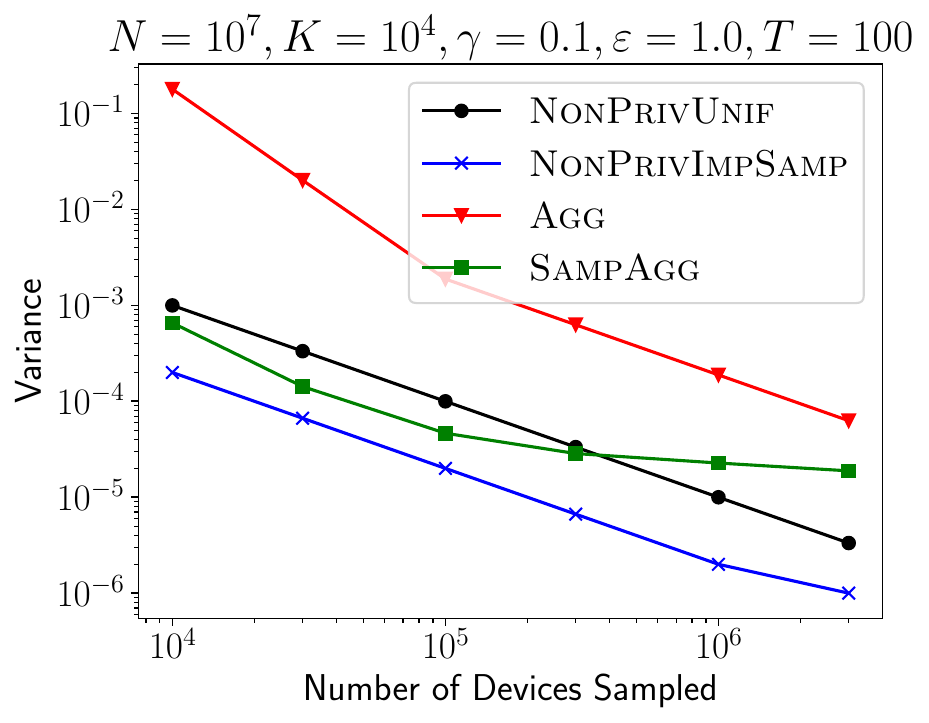}
  \includegraphics[width=0.3\linewidth]{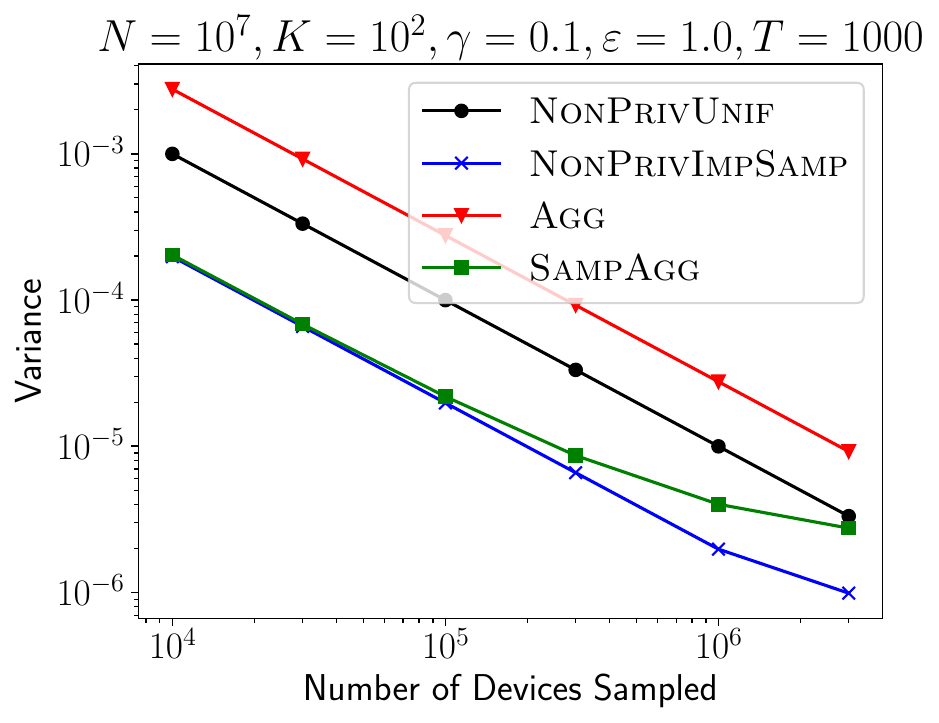}
  \includegraphics[width=0.3\linewidth]{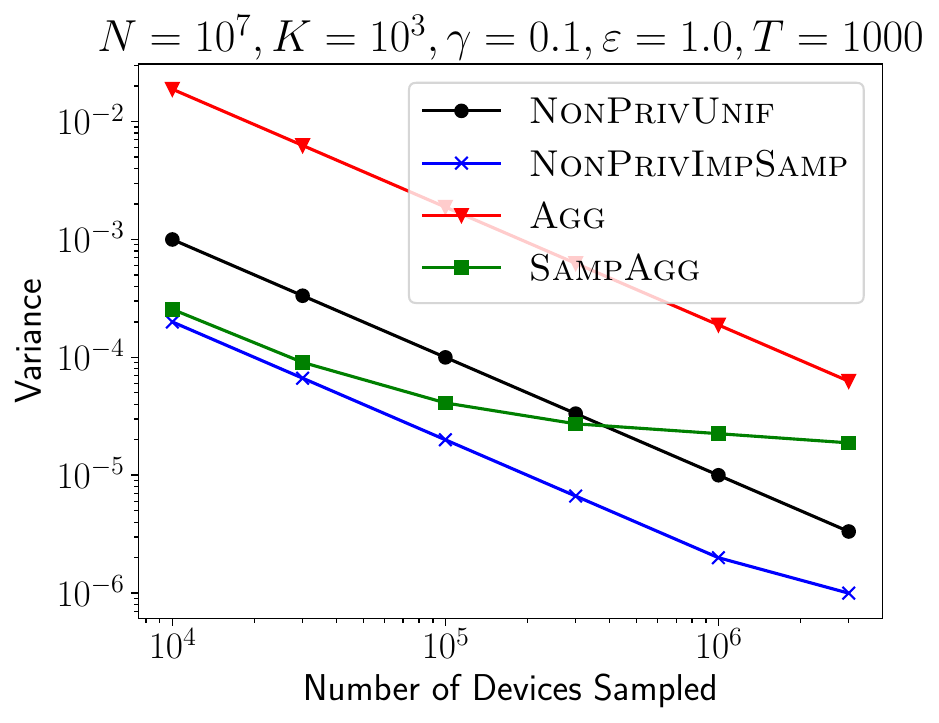}
  \includegraphics[width=0.3\linewidth]{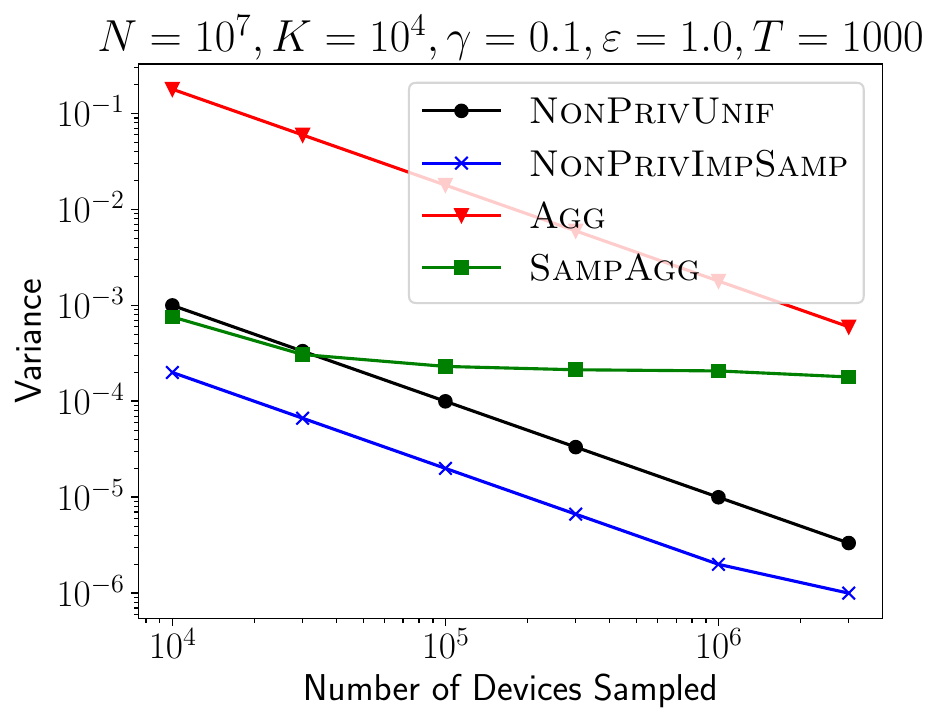}
  \caption{Expected Squared Error on the distribution of non-zero values for a sparse histogram task, for varying parameter values $K$ and $T$, for $\gamma = 0.1$. The plots include a naive non-private baseline ({\sc NonPrivUnif}), non-private Importance Sampling ({\sc NonPrivImpSamp}), Aggregation model ({\sc Agg}) and Samplable Aggregation ({\sc SampAgg})  for varying number of tasks $T$.}\label{fig:sparse-hist-vary-K-T-0.1}
  \Description{Plots showing the benefits of Samplable Aggregation on Sparse Histograms.}
\end{figure*}

\begin{figure*}[h]
  \centering
  \includegraphics[width=0.3\linewidth]{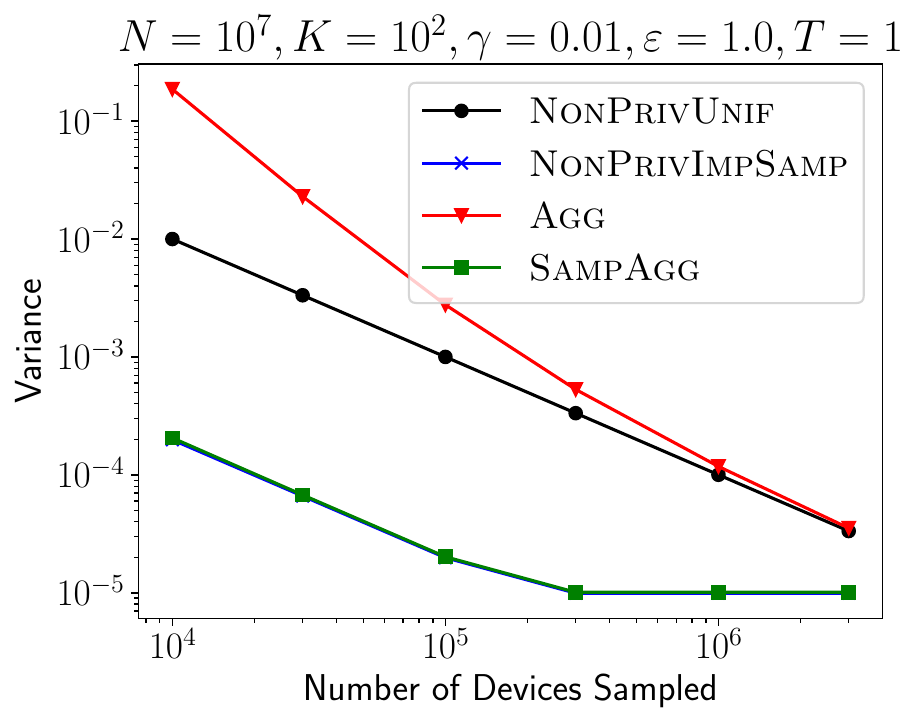}
  \includegraphics[width=0.3\linewidth]{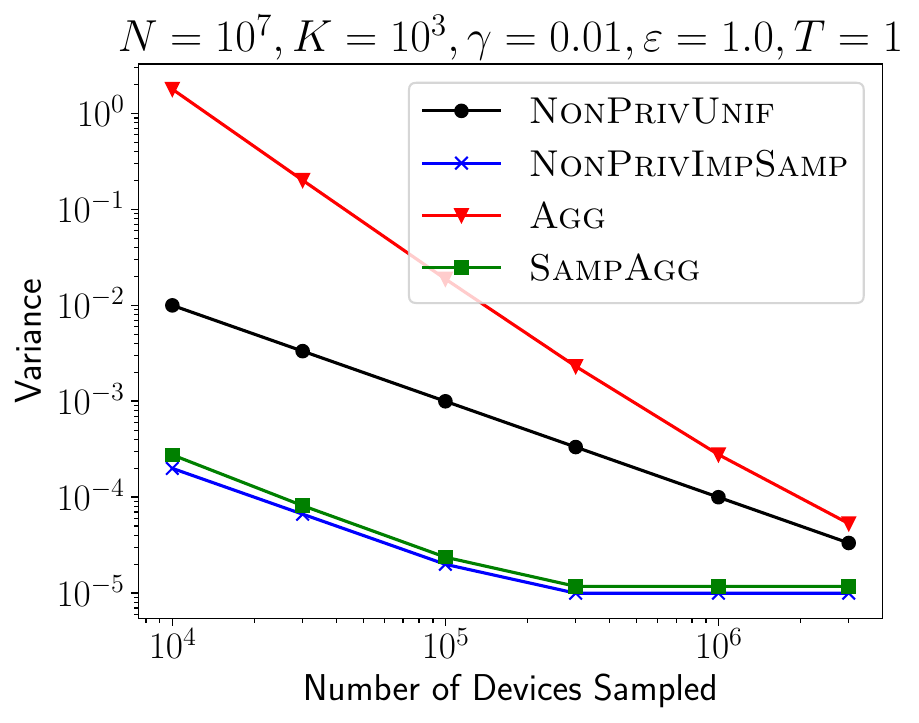}
  \includegraphics[width=0.3\linewidth]{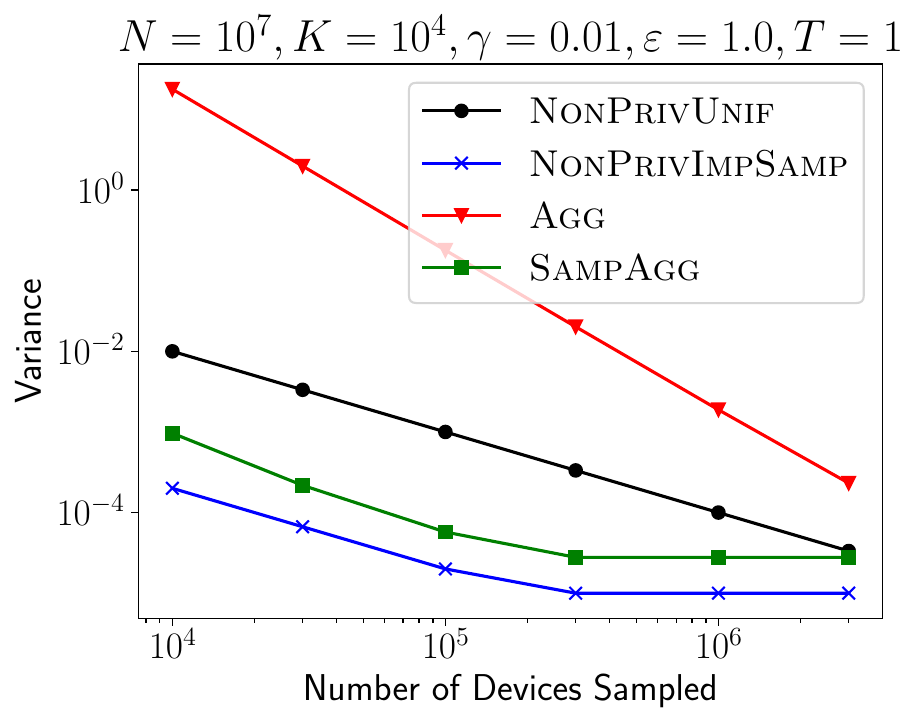}
  \includegraphics[width=0.3\linewidth]{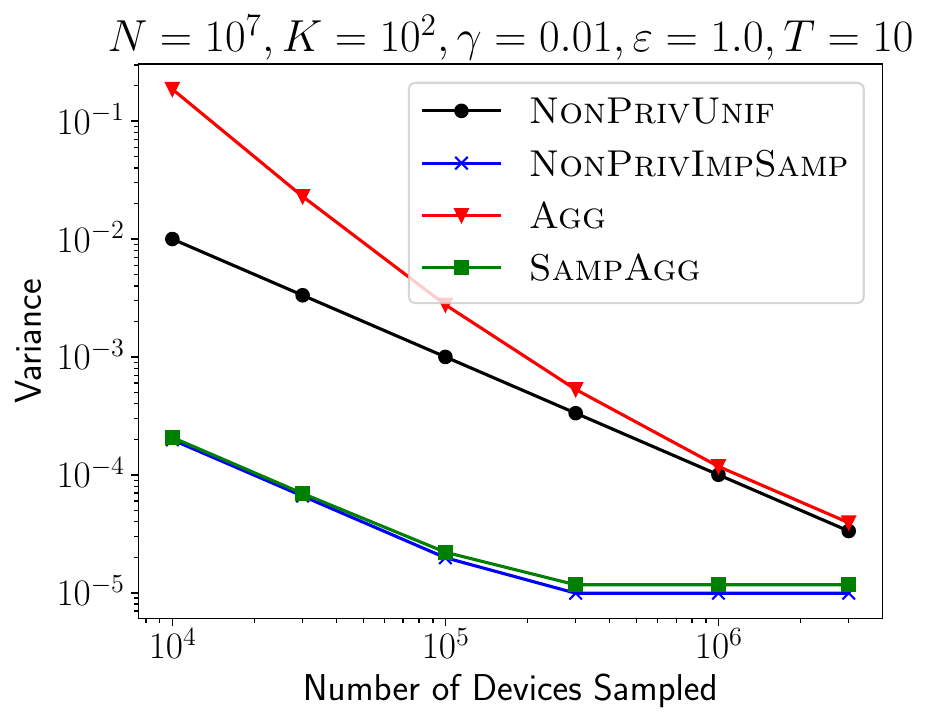}
  \includegraphics[width=0.3\linewidth]{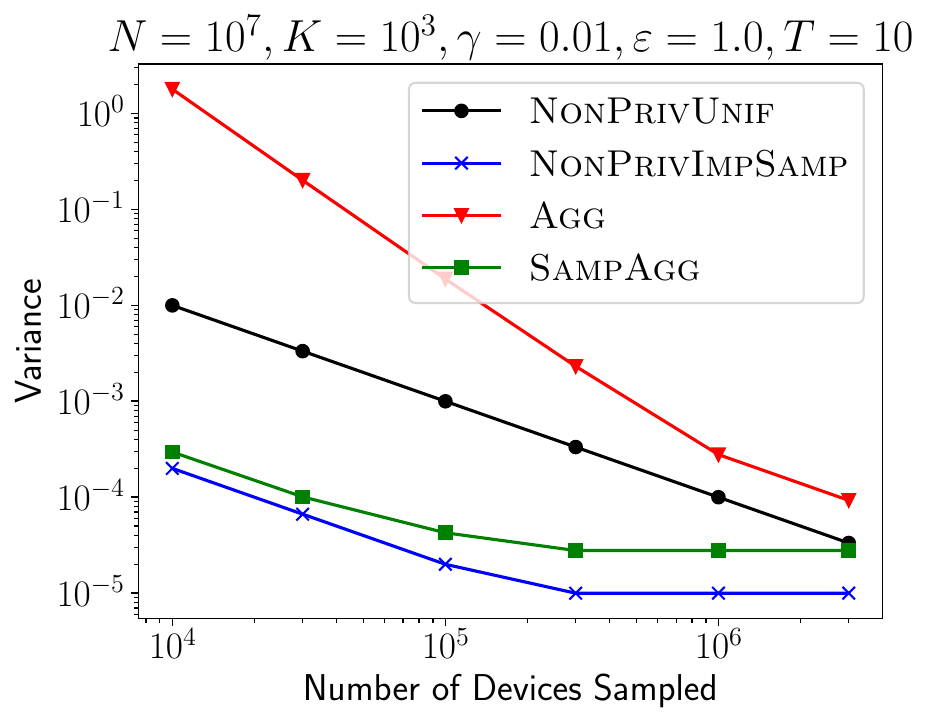}
  \includegraphics[width=0.3\linewidth]{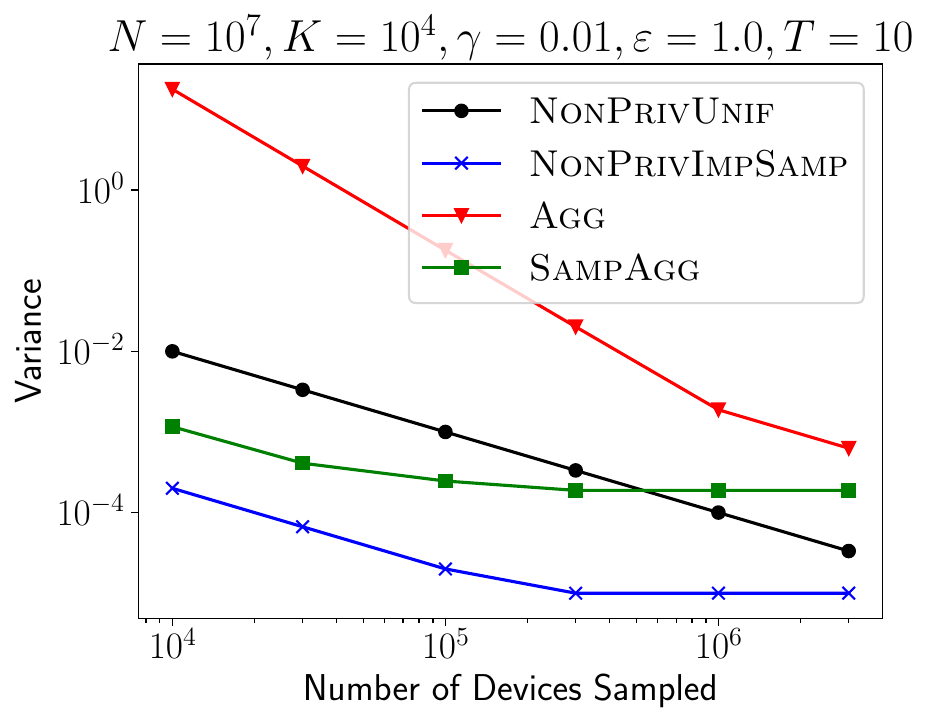}
  \includegraphics[width=0.3\linewidth]{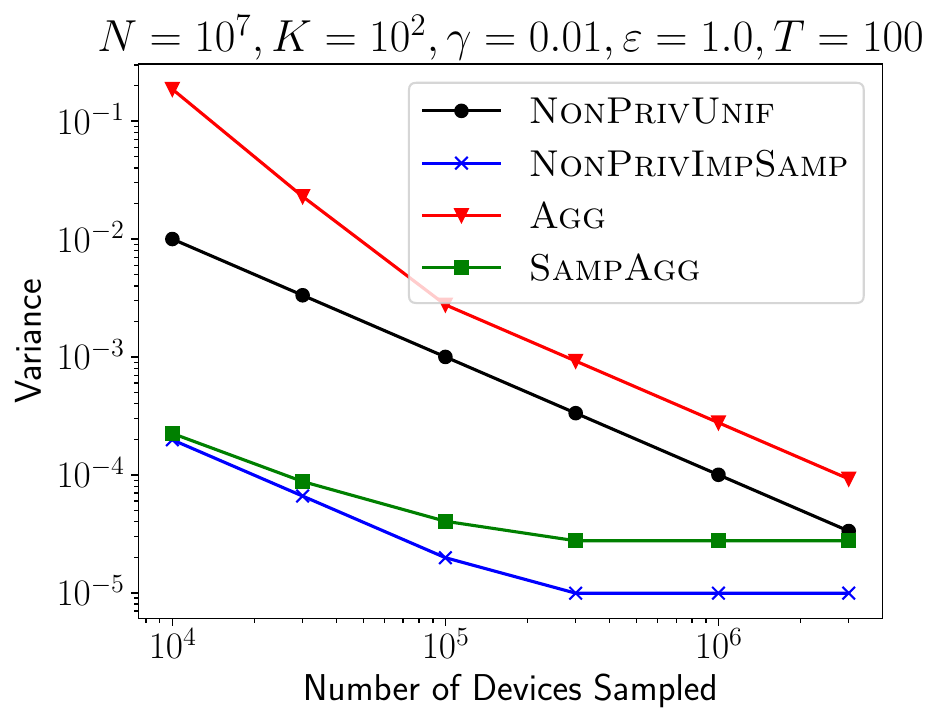}
  \includegraphics[width=0.3\linewidth]{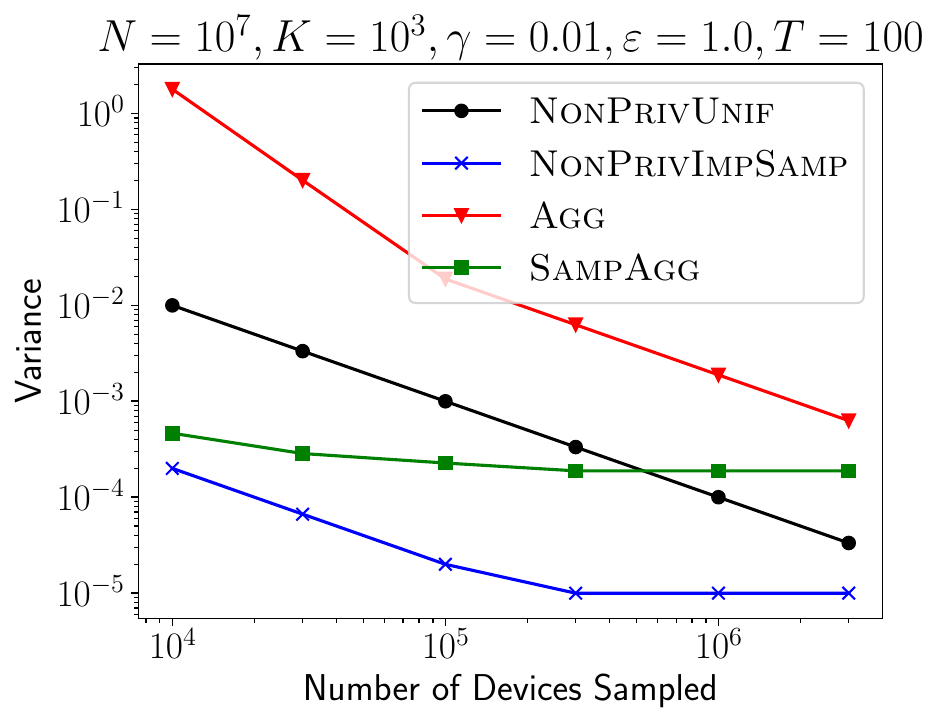}
  \includegraphics[width=0.3\linewidth]{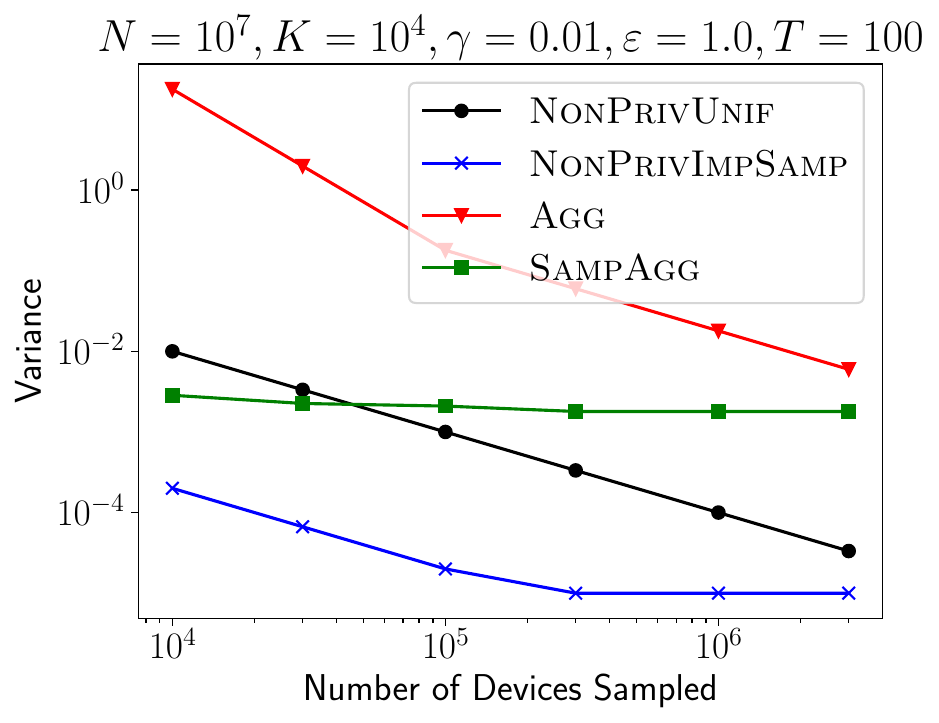}
  \includegraphics[width=0.3\linewidth]{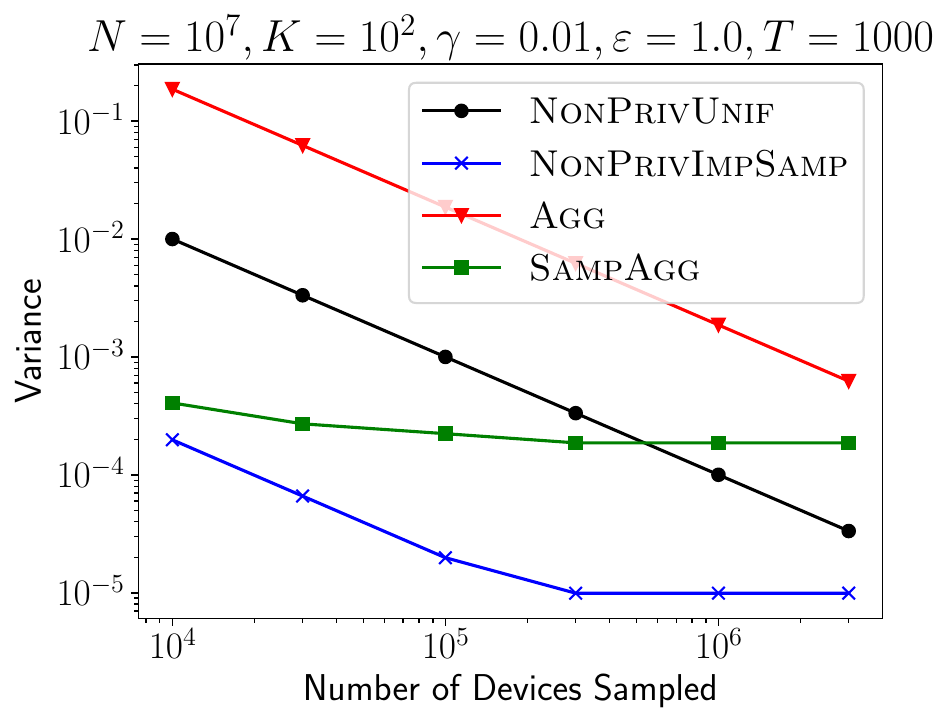}
  \includegraphics[width=0.3\linewidth]{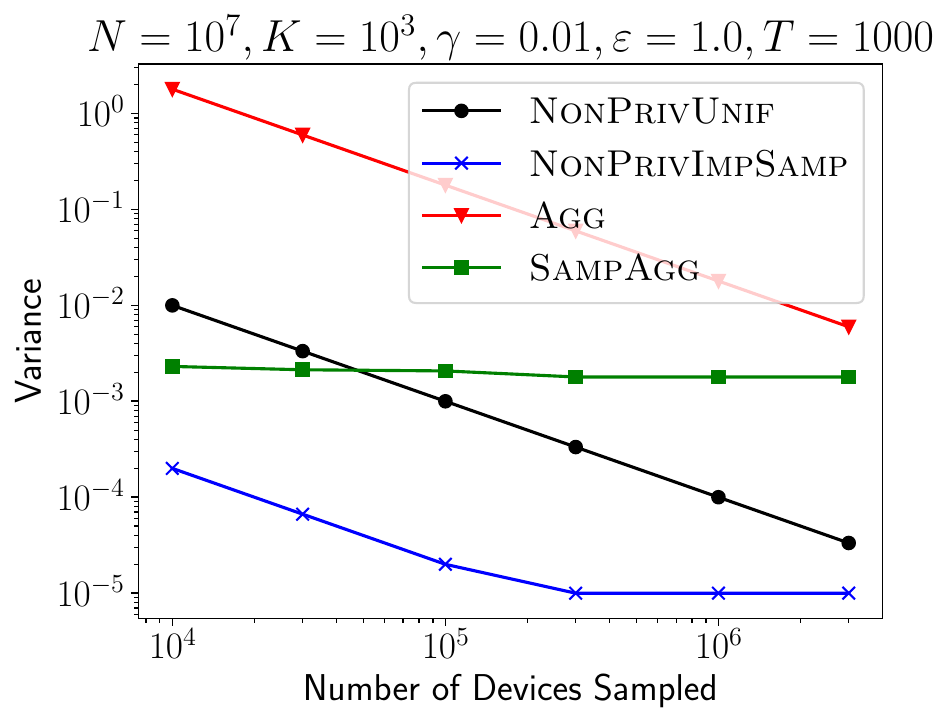}
  \includegraphics[width=0.3\linewidth]{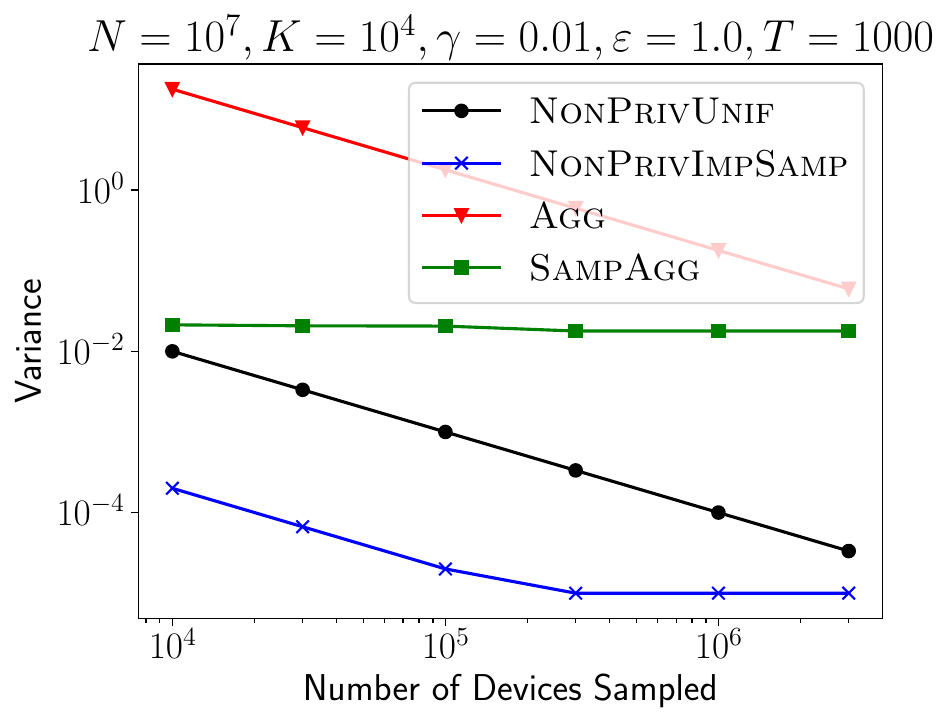}
  \caption{Expected Squared Error on the distribution of non-zero values for a sparse histogram task, for varying parameter values $K$ and $T$, for $\gamma = 0.01$. The plots include a naive non-private baseline ({\sc NonPrivUnif}), non-private Importance Sampling ({\sc NonPrivImpSamp}), Aggregation model ({\sc Agg}) and Samplable Aggregation ({\sc SampAgg})  for varying number of tasks $T$.}\label{fig:sparse-hist-vary-K-T-0.01}
  \Description{Plots showing the benefits of Samplable Aggregation on Sparse Histograms.}
\end{figure*}

\begin{figure*}[h]
  \centering
  \includegraphics[width=0.3\linewidth]{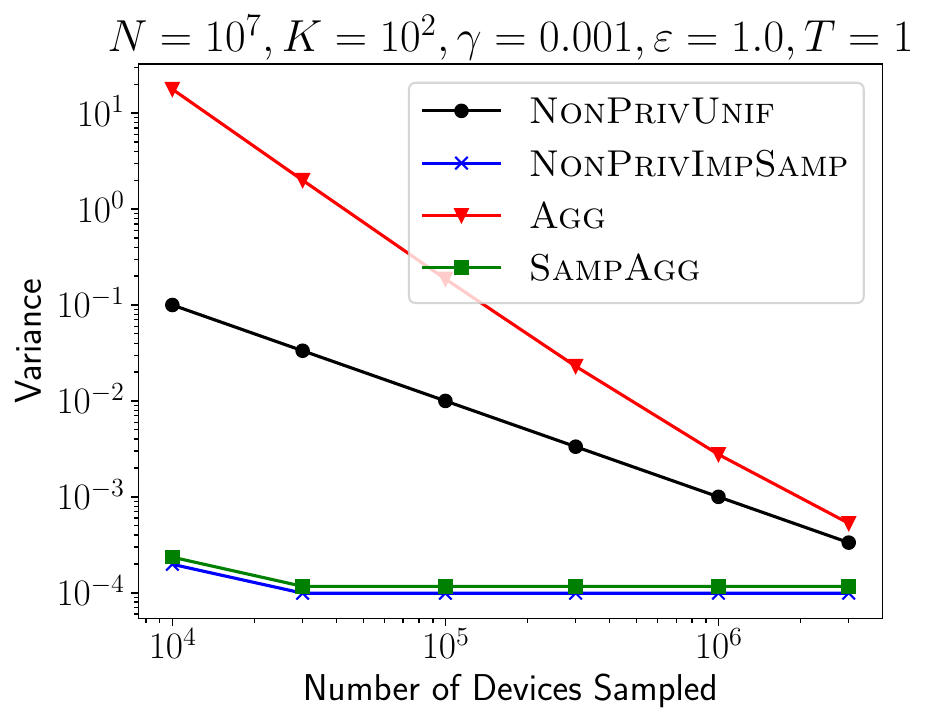}
  \includegraphics[width=0.3\linewidth]{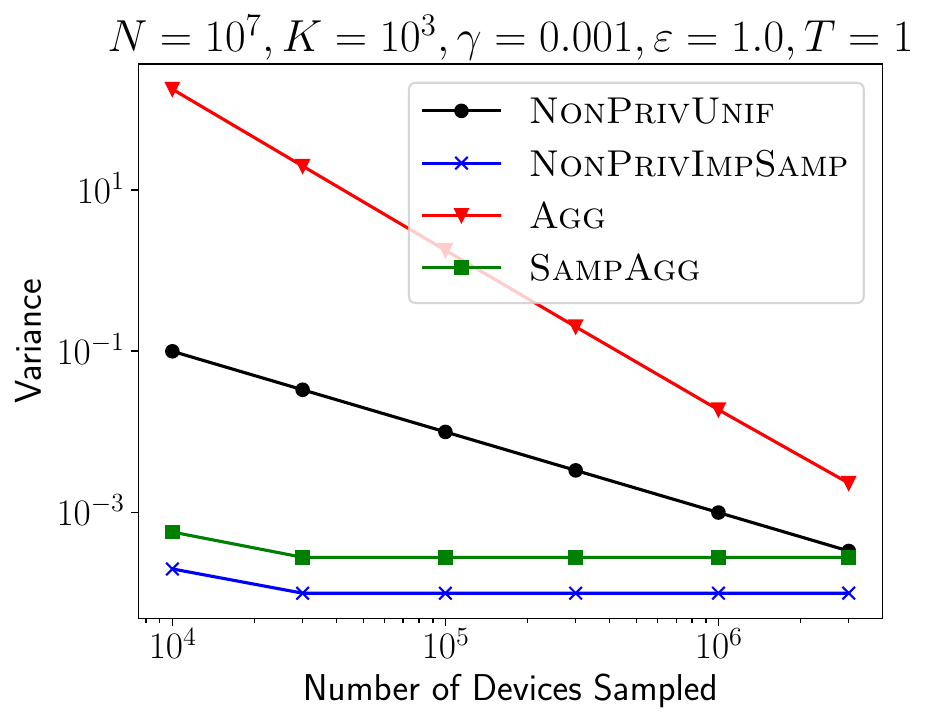}
  \includegraphics[width=0.3\linewidth]{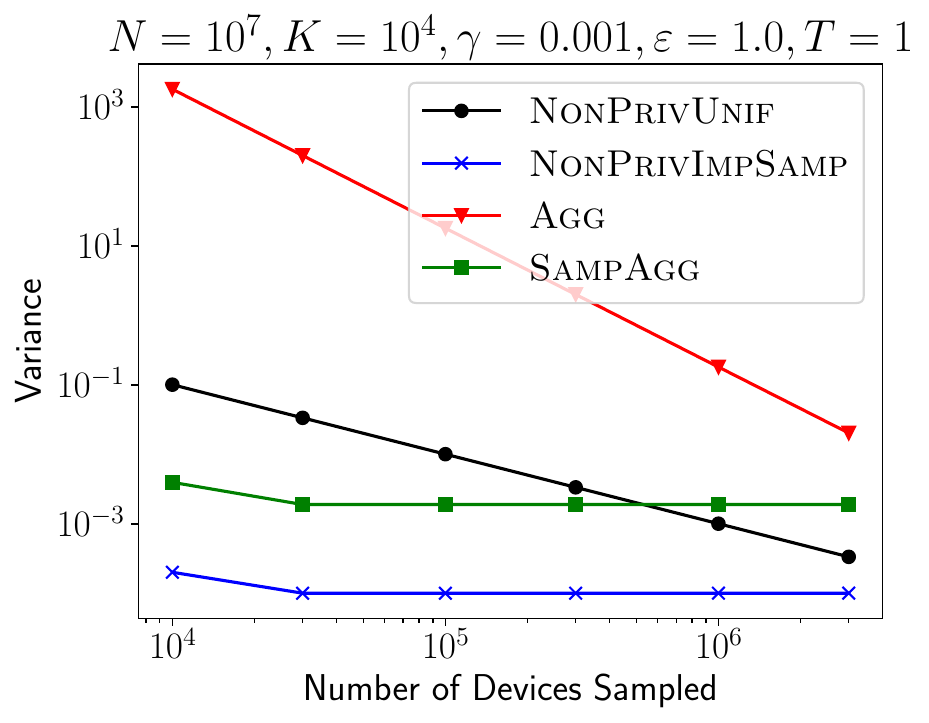}
  \includegraphics[width=0.3\linewidth]{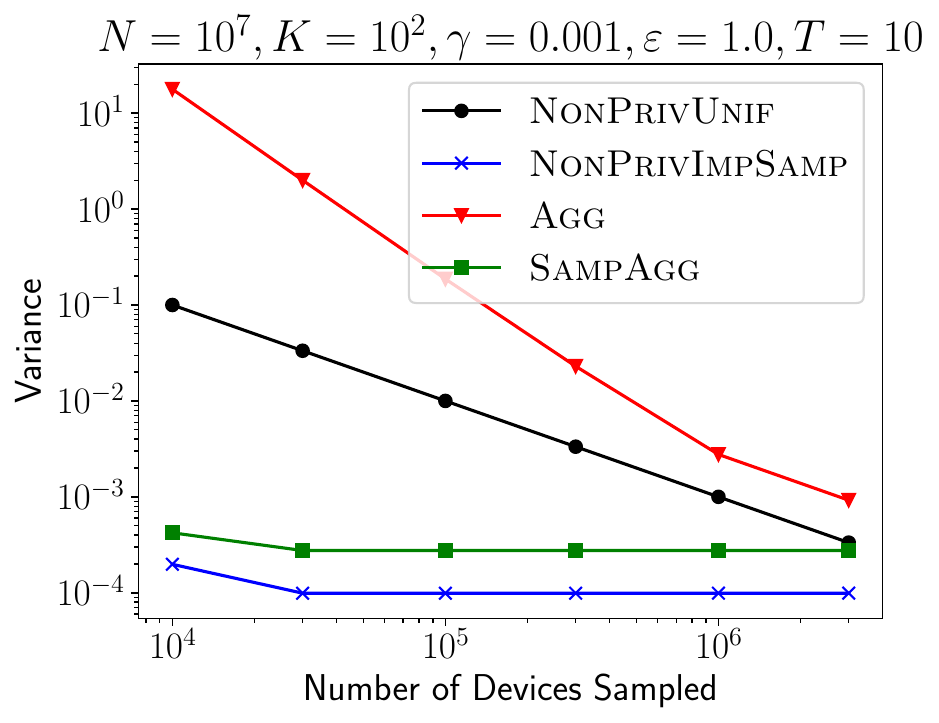}
  \includegraphics[width=0.3\linewidth]{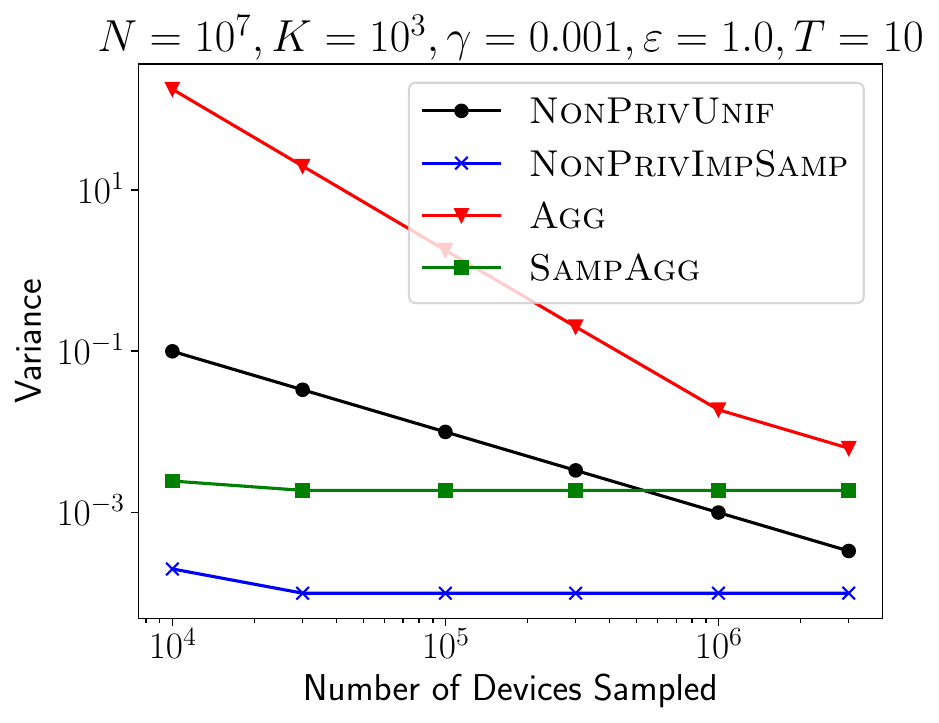}
  \includegraphics[width=0.3\linewidth]{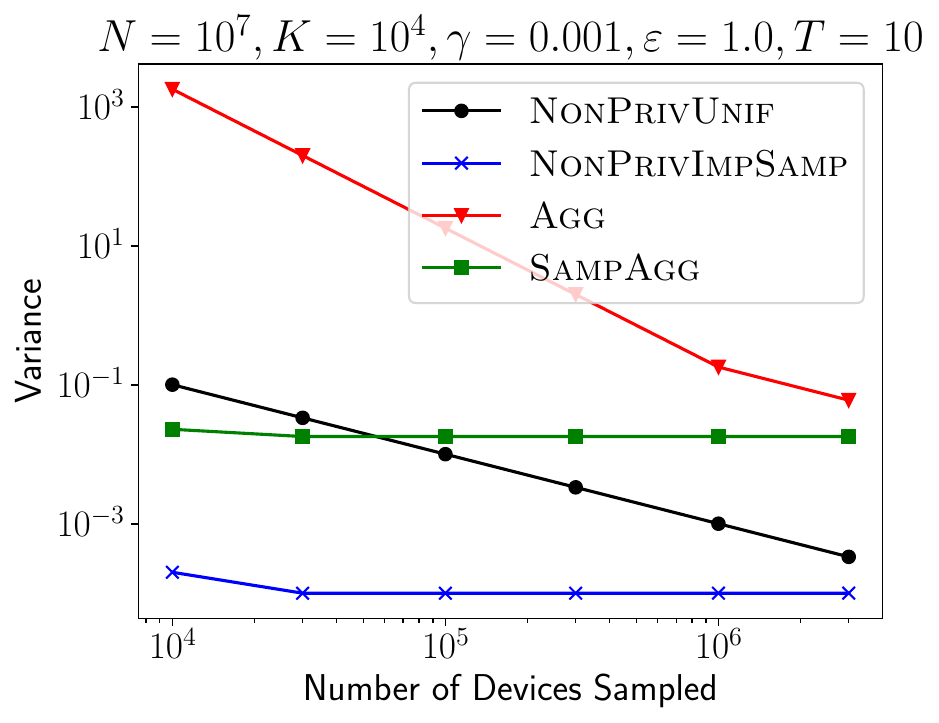}
  \includegraphics[width=0.3\linewidth]{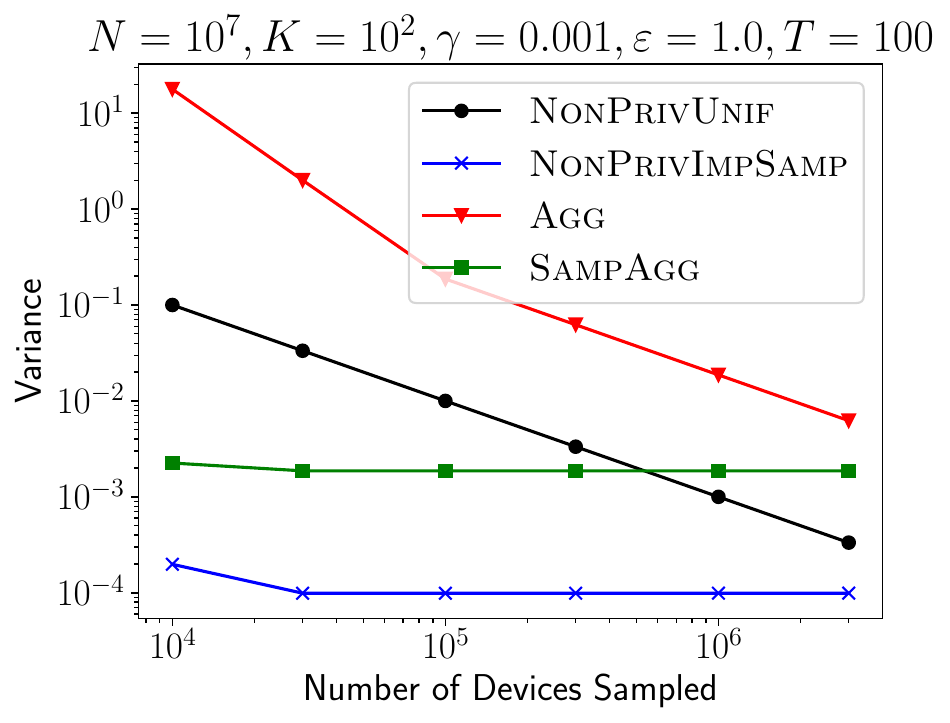}
  \includegraphics[width=0.3\linewidth]{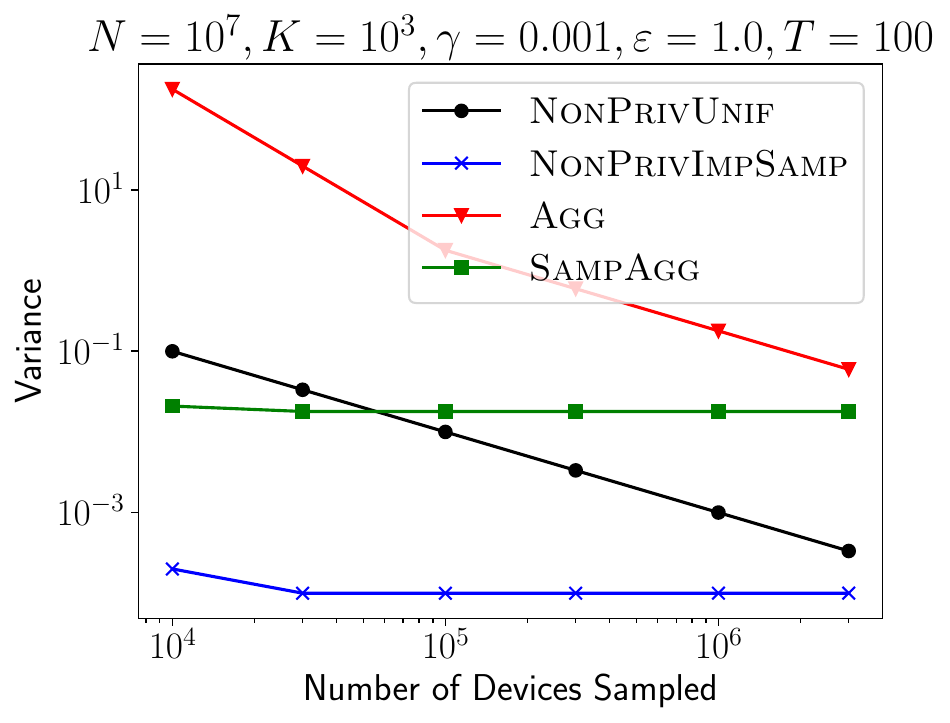}
  \includegraphics[width=0.3\linewidth]{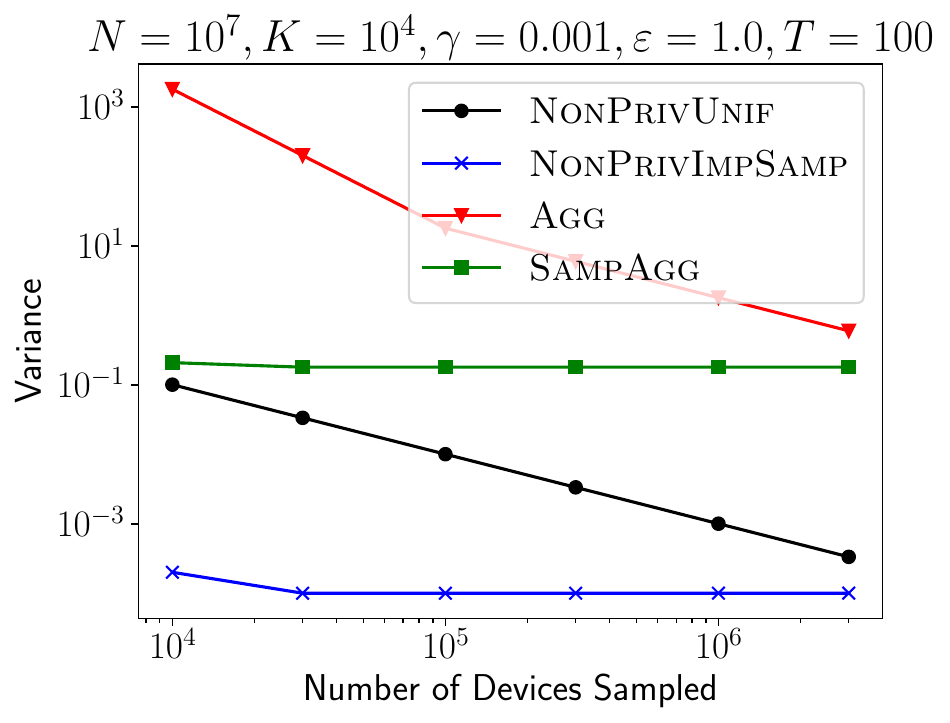}
  \includegraphics[width=0.3\linewidth]{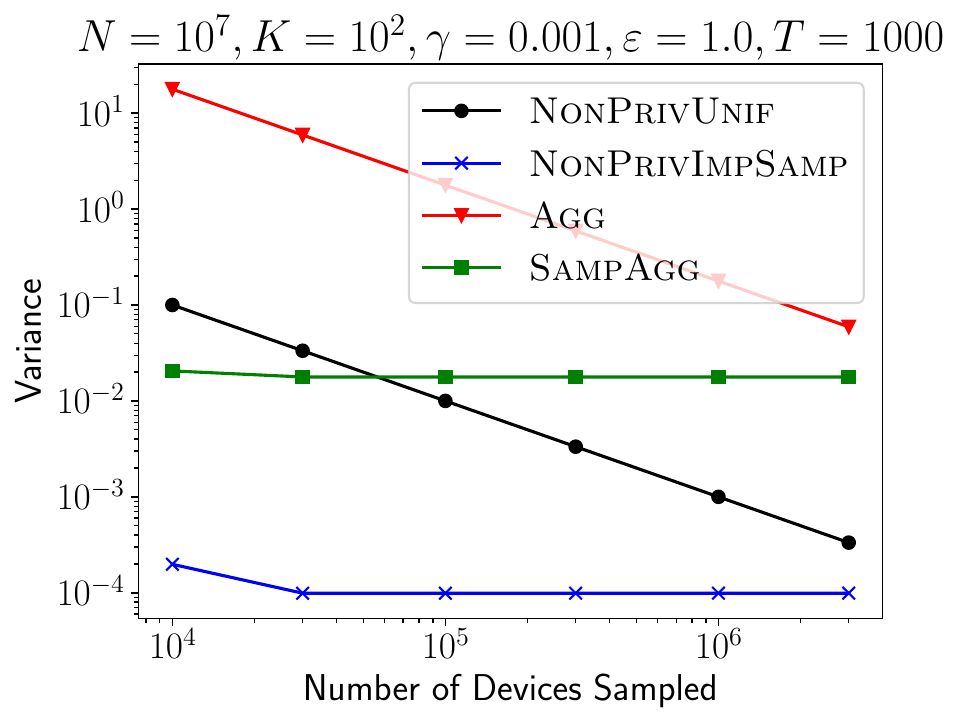}
  \includegraphics[width=0.3\linewidth]{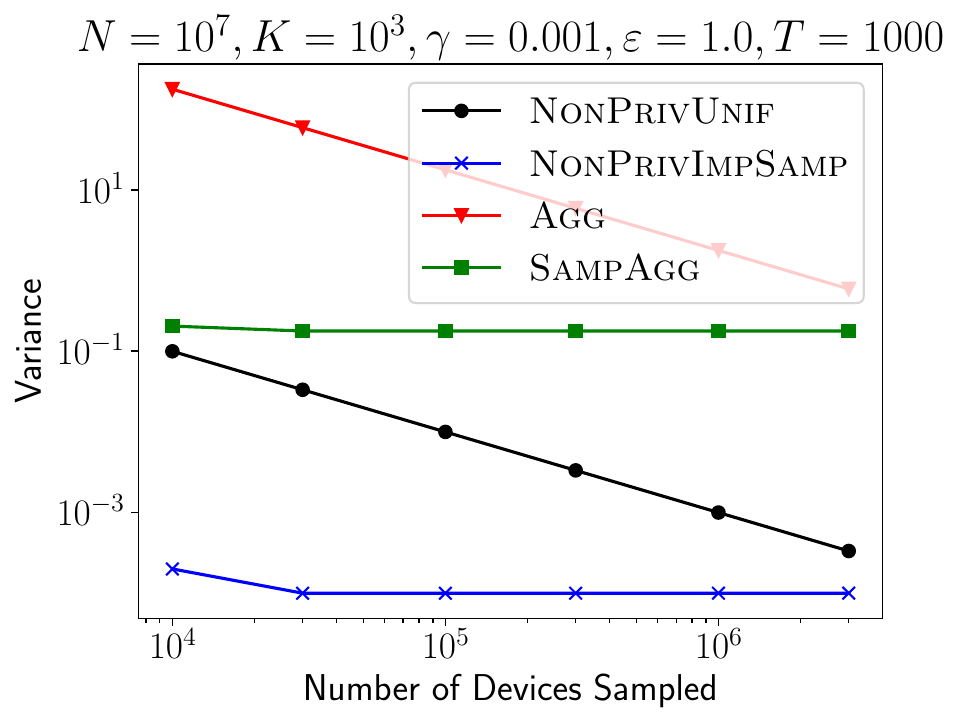}
  \includegraphics[width=0.3\linewidth]{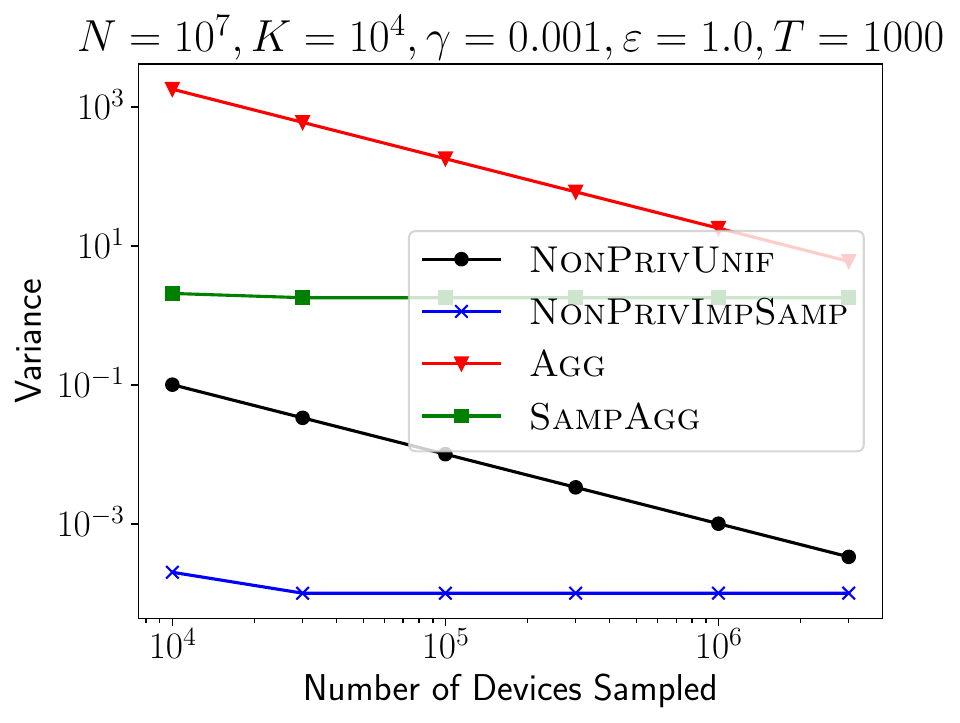}
  \caption{Expected Squared Error on the distribution of non-zero values for a sparse histogram task, for varying parameter values $K$ and $T$, for $\gamma = 0.001$. The plots include a naive non-private baseline ({\sc NonPrivUnif}), non-private Importance Sampling ({\sc NonPrivImpSamp}), Aggregation model ({\sc Agg}) and Samplable Aggregation ({\sc SampAgg})  for varying number of tasks $T$.}\label{fig:sparse-hist-vary-K-T-0.001}
  \Description{Plots showing the benefits of Samplable Aggregation on Sparse Histograms.}
\end{figure*}

For the case of sparse histogram, the fraction of values that are non-default is an additional parameter. We show the same plots as the histograms case above, for $\gamma=0.1$ (\cref{fig:sparse-hist-vary-K-T-0.1}), $\gamma=0.01$ (\cref{fig:sparse-hist-vary-K-T-0.01}) and $\gamma=0.001$ (\cref{fig:sparse-hist-vary-K-T-0.001}). Once again, these plots demonstrate that the gain from our improved primitive are significant across a large range of parameters. When $\gamma$ is very small, there is fewer users to learn from, leading to larger privacy costs.

\end{document}